\documentclass[12pt,preprint]{aastex}

\begin{document}

\title{Nascent starbursts in synchrotron-deficient galaxies
       with hot dust\footnote{Based on observations with the 100-m telescope
       of the Max-Planck Institut f\"ur Radioastronomie at Effelsberg.}}

\author{H. Roussel\altaffilmark{1}, G. Helou\altaffilmark{1,5},
R. Beck\altaffilmark{2}, J.J. Condon\altaffilmark{3}, A. Bosma\altaffilmark{4},
K. Matthews\altaffilmark{1}, and T.H. Jarrett\altaffilmark{5}}
\affil{1) California Institute of Technology, Pasadena, CA 91125}
\affil{2) Max-Planck Institut f\"ur Radioastronomie, 53121 Bonn, Germany}
\affil{3) National Radio Astronomy Observatory, Charlottesville, VA 22903}
\affil{4) Observatoire de Marseille, 2 place Le Verrier, 13248 Marseille cedex 4, France}
\affil{5) Infrared Processing and Analysis Center, Pasadena, CA 91125}

\email{hroussel@irastro.caltech.edu}

\begin{abstract}
Three nearby galaxies which have abnormally high infrared to radio
continuum ratios, NGC\,1377, IC\,1953 and NGC\,4491, are investigated
with a view to understanding the physical origin of their peculiarity.
We review the existing data and present new radio continuum
measurements along with near-infrared integral-field spectroscopy
and molecular gas observations.
The three galaxies have low luminosities but starburst-like infrared colors;
in NGC\,1377, no synchrotron emission is detected at any wavelength;
in IC\,1953, the observed synchrotron component is attributable to
the spiral disk alone, and is lacking in the central regions; the
radio spectrum of NGC\,4491 is unusually flat.
We also compare and contrast them with NGC\,4418, a heavily
extinguished galaxy which shares some attributes with them.
After examining various scenarios, we conclude that these galaxies
are most likely observed within a few Myr of the onset
of an intense star formation episode after being quiescent for
at least $\approx 100$\,Myr. This starburst, while heating the dust,
has not produced optical signatures nor a normal amount of cosmic
rays yet. We briefly discuss the statistics of such galaxies
and what they imply for star formation surveys.
\end{abstract}


\section{Introduction}

One of the implications from the data collected by the IRAS mission,
combined with extensive radio surveys, is a nearly universal and tight
correlation between the far-infrared thermal dust emission and the
optically-thin total radio continuum emission (mixing thermal and
synchrotron components) in normal star-forming galaxies. \citet{Jong}
have reached this conclusion comparing 60\,$\mu$m fluxes with 6.3\,cm
fluxes from the Effelsberg dish, \citet{Helou_letter} comparing
40--120\,$\mu$m fluxes with 21\,cm fluxes from Westerbork. Although
the correlation is generally expressed between luminosities, it holds
when they are normalized to cancel size and distance effects, and is
valid for a very broad range of star formation rate densities (review
by \citet{Condon_review}). At centimeter wavelengths, the radio
emission is dominated by non-thermal processes.

All models accounting for the constancy of the infrared to radio flux
ratios invoke the star formation process as the indirect source of both
types of emission, via a production rate of cosmic rays by type-II
supernov\ae\ proportional to the production rate of heating photons.
Additional coupling mechanisms are needed, such as between the interstellar
medium density and the magnetic field intensity \citep{Helou_model},
or between the star formation rate and the magnetic field amplification
\citep{Beck, Niklas_model}. Various physical mechanisms can lead to
these couplings, among which differential rotation, or turbulence driven
by massive star formation amplifying the magnetic field. In addition,
given that the various timescales involved in the underlying mechanisms
are very different (heating of dust by massive and then intermediate-mass
stars, explosion of supernov\ae, acceleration, diffusion and decay of
the relativistic electrons, variations of the magnetic field strength),
some variation in infrared to radio flux ratios with the star formation
recent history is to be expected.

There are naturally known exceptions to the general correlation.
A fraction of galaxies hosting non-stellar nuclear activity have excess radio
continuum emission, from cosmic rays accelerated in jets from the nucleus
\citep{Sopp, Roy}. Star-forming galaxies in cluster environments can also
have radio excess \citep{Miller}, perhaps triggered by external magnetic
field compression due to the intracluster gas pressure or to interactions with
neighbor galaxies. Two pairs of colliding galaxies with excess synchrotron
emission originating from a bridge between the disks were studied by
\citet{Condon_taffy1} and \citet{Condon_taffy2}, to illustrate the
importance to the correlation of regulated cosmic ray escape.
\citet{Hummel_compression} also discovered excess synchrotron in an
interacting galaxy, due to external compression of the magnetic field.

Some statistical studies found high far-infrared to radio ratios in compact
galaxy groups \citep{Menon} and in lenticular galaxies \citep{Walsh}, but
may suffer from photometric errors (confusion problem for IRAS fluxes
in galaxy groups, as pointed out by \citet{Sulentic}, and extended radio
emission missed by interferometric observations).

For dwarf galaxies, metallicity effects can lower the dust emission,
and the radio spectra of many are dominated by thermal emission.
\citet{Wunderlich} found that the far-infrared to radio ratio tends
to be lower in blue compact galaxies than in spirals, while \citet{Klein}
found similar ratios in both. Using the radio fluxes
of \citet{Klein} and new IRAS fluxes measured with {\it Scanpi}
\citep{scanpi}, it seems in fact that there is a continuum of infrared
to radio ratios from very low to very high values, i.e. with a much
larger dispersion than in spirals, perhaps depending on two competing
effects: reduction of the dust emission by metal deficiency, and reduction
of the synchrotron emission by a lack of supernov\ae\ or by fast escape
of cosmic rays.

Variations also exist inside spiral galaxies: \citet{Beck} found
higher 60\,$\mu$m to 11\,cm flux ratios in the central regions
of large angular size spirals than in their disks, following the
$F_{60}/F_{100}$ ratio, an average dust temperature indicator.
This is likely because the scale length of the synchrotron emission
is larger than in the infrared, due to cosmic ray propagation far
from the site of their acceleration, located in star formation regions
hosting hot dust \citep{Bicay, Marsh}.

In this paper, we study three nearby galaxies (an amorphous galaxy,
an early-type spiral and a late-type spiral) which are strongly
deficient in radio emission with respect to their dust emission,
trying to identify the cause of their peculiarity. In normal
galaxies, the synchrotron radiation comes mainly from electrons
having left the supernova remnants which have accelerated them, and
propagating in the interstellar magnetic field \citep{Condon_model}.
A lack of synchrotron emission may thus have its origin in either
an unusually low strength of the magnetic field, or a reduced
supernova rate in the recent past. With non-detections in
the synchrotron emission, the magnetic field strength is extremely
difficult to estimate. However, we address the star formation properties
using an array of new observations, and thereby approach indirectly
the issue of magnetic fields. We will argue that these galaxies
have a high star formation rate associated with a low supernova rate,
reflecting a starburst in a very early stage occurring after a
long period of quiescence, since supernov\ae\ begin to explode
a few Myr after the birth of their progenitors.

If this interpretation is valid, these galaxies could provide the
opportunity to observe the properties of the earliest stages of a
starburst, and to understand better the regulation of the galactic
infrared and radio emission. Although such galaxies are apparently
quite rare, they may exist in greater numbers at more primitive epochs
of star formation. They must therefore be taken into account when
using high-redshift radio surveys to study the star formation history,
and when using high-resolution radio observations to identify 
faint infrared or submillimeter sources.

The paper is structured as follows: in Section~\ref{review} we discuss
existing data and assemble comprehensive information about NGC\,1377,
IC\,1953 and NGC\,4491; Section~\ref{effelsberg} presents and analyses
new radio continuum observations; in Section~\ref{mid_ir} we describe
and exploit near-infrared spectroscopic observations, extract estimates
of the star formation rate from the dust emission and discuss the excitation
and apparent cooling rate of the interstellar medium;
Section~\ref{sest} examines results from single-dish molecular gas
observations in NGC\,1377; Section~\ref{n4418} is a brief discussion
of some common properties of NGC\,4418 with the three galaxies studied
above; Section~\ref{scenarios} lists and discusses the various physical
scenarios which could account for the observed synchrotron deficiency.
These scenarios are related either to a very low cosmic ray density
stemming from a low supernova rate, or to a very low magnetic field
intensity; we conclude in favor of the former, as would be expected
in a nascent starburst. Section~\ref{stat_highq} then compares the
statistics of synchrotron-deficient galaxies to those of nascent
starbursts, and discusses possible triggering mechanisms.
Section~\ref{summary} contains a summary of our conclusions.

\section{Descriptive review of the galaxies}
\label{review}

\begin{figure}
\vspace*{-3cm}
\hspace*{-1.5cm}
\resizebox{20cm}{!}{\plotone{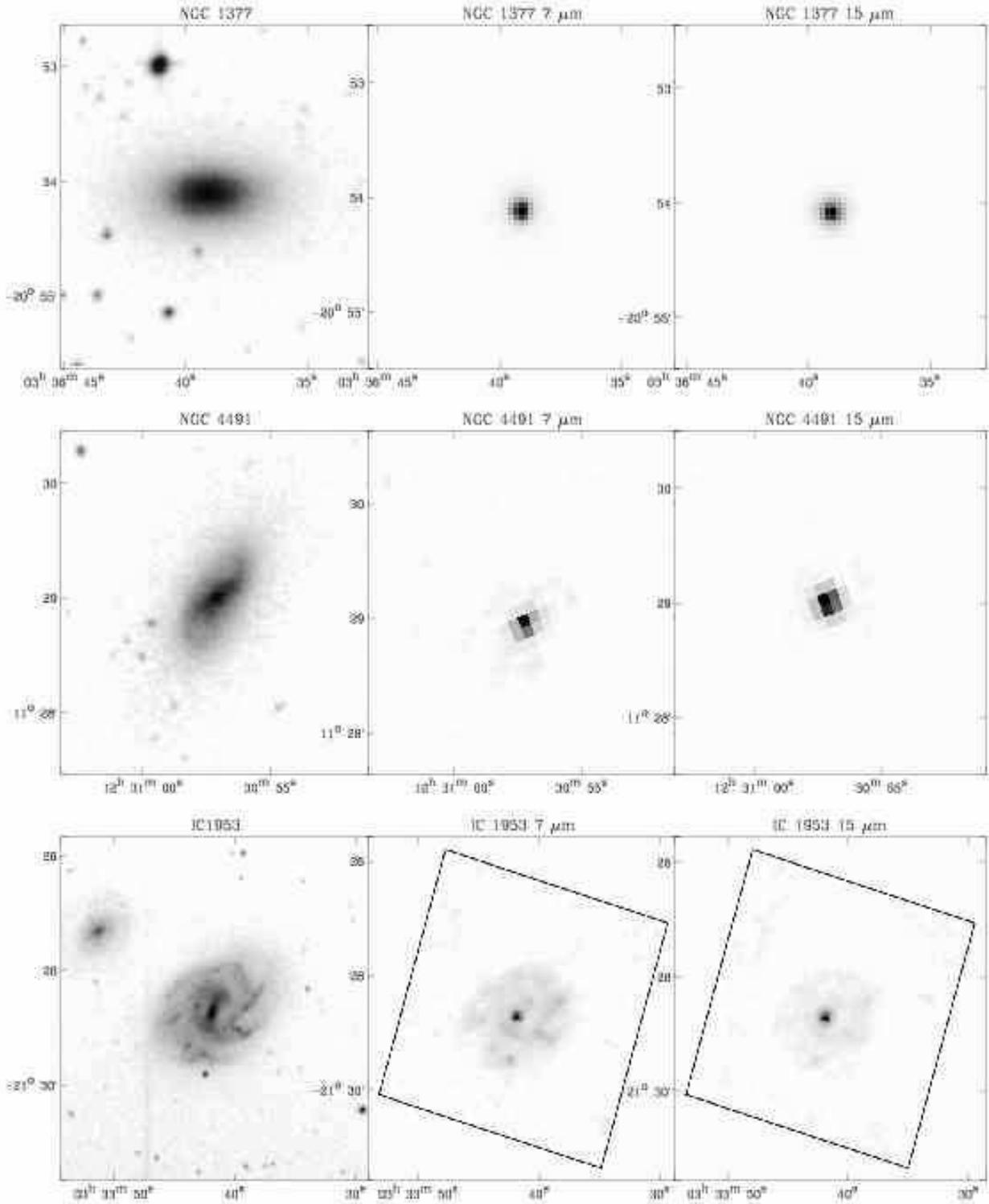}}
\vspace*{-3cm}
\caption{Red-band images from the DSS, and mid-infrared images
in the bandpasses 5--8.5\,$\mu$m and 12--18\,$\mu$m.
\label{images}}
\end{figure}

General information about the three galaxies under consideration can be
found in Table~\ref{tab_gen}. NGC\,1377 and IC\,1953 were identified
as radio synchrotron-deficient galaxies among the IRAS bright galaxy
sample mapped with the VLA \citep{Condon_atlas}. NGC\,4491, which is
fainter, was also found to be synchrotron-deficient in a search among
the small sample
of \citet{barres}, based on mid-infrared colors (see Sect.~\ref{irsed}).
All three galaxies are low-mass and quite unremarkable at first sight,
if it were not for their high dust temperatures. Being both faint and hot
in their dust emission, they are very peculiar with respect to the average
local $L_{\rm FIR}$--$F_{60}/F_{100}$ function, especially NGC\,1377
\citep{Chapman}. The average $F_{60}/F_{100}$ ratio at their infrared
luminosities is only $\approx 0.4$, and they deviate from this by about
$10 \sigma$ and $5 \sigma$. Although IC\,1953 seems to have a very close
companion (Fig.~\ref{images}), this galaxy is at much higher redshift
(see Section~\ref{envir}).

\subsection{Morphology}

NGC\,1377 is classified as a lenticular or amorphous galaxy, and does
not present any apparent disturbance. Its only peculiarity, noted by
\citet{Heisler_opt}, is a central dust lane along the southern part of
the minor axis, visible in the B-band isophotes.
NGC\,4491 is a barred dwarf Sa galaxy \citep{Binggeli_type} whose shape
is also regular, with a tenuous outer disk.
IC\,1953 is a strongly barred Sd galaxy (the bar is short but straight
and narrow, connecting at right angles to the spiral arms). It has an
unsually bright and well-defined core for its late type (very prominent
in the mid-infrared dust emission). One of the arms winds to join the
opposite end of the bar, and the other arm, winding for more than
360\degr, offers the aspect of three distinct branches.

The mid-infrared emission of NGC\,1377 and NGC\,4491, in the ISOCAM maps
at 7\,$\mu$m and 15\,$\mu$m, is dominated by the central regions, of
apparent sizes respectively about 2.3\,kpc and 800\,pc, and the dust
emission of IC\,1953 is also more concentrated (in a 1.5\,kpc region)
than in other galaxies of its late-type morphology (Fig.~\ref{images}).
Decomposing the surface brightness profile of NGC\,1377 at 7\,$\mu$m,
we find an unresolved source (at the resolution of ISOCAM, with
a point spread function of ${\rm FWHM} = 5.7$\arcsec) contributing
about 60\% of the total flux or more. For NGC\,4491, the central region
is also unresolved in the mid-infrared, and superimposed onto a very
diffuse component. The 7\,$\mu$m brightness profile decomposition
yields a point source accounting for about 65\% of the flux
of the infrared core, and 38\% of the total flux. An unresolved
source is also seen in the 7\,$\mu$m brightness profile of IC\,1953,
containing about 65\% of the flux of the circumnuclear regions.

\subsection{Environment}
\label{envir}

NGC\,1377 and IC\,1953 are both members of the Eridanus galaxy group.
According to \citet{Willmer}, this is a small group (54 members brighter
than ${\rm B} = 14.5$) in a young dynamical state, since it has an irregular
and subclustered aspect, is rich in spiral galaxies, and forms with the Fornax
cluster a bound system which is not yet virialized. In the soft X-ray maps
of \citet{Burrows} (0.1--0.9\,keV), Eridanus is not detected. NGC\,1377 and
IC\,1953 do not belong to any subcondensation, but to the same loose sub-group
lying in projection in the center, and are close to each other both
spatially and in radial velocity.

NGC\,4491 seems also to be part of a dynamically young system, the core
named group A in the Virgo cluster, containing M\,87 and 100 to 300 members,
and which is described by \citet{Binggeli_struct} as a ``turbulent core
with a long tail of low velocities''. NGC\,4491 belongs to this tail,
with a large velocity with respect to the systemic velocity of
$\approx -550$\,km\,s$^{-1}$.

None of the three galaxies shows any morphological disturbance, but they
have likely interacted with the ambiant medium, since they are deficient
in atomic gas. We did not find any published
HI mass measurement for NGC\,1377, but a spectrum from the HIPASS
survey, conducted with the Parkes telescope \citep{Barnes}.\footnote{
{\it http://www.atnf.csiro.au/research/multibeam/release/}.
The Parkes telescope is part of the Australia Telescope which is funded
by the Commonwealth of Australia for operation as a National Facility
managed by CSIRO.}
This spectrum has a sensitivity of about 10\,mJy per spectral channel,
and implies that M(HI$) < 2 \times 10^8$\,M$_{\sun}$.
Compared with the S0/a-Sab galaxies in \citet{Guiderdoni}, the
logarithmic HI deficiency of NGC\,1377, normalized by the
dispersion in the field reference sample, is higher than 1.0\,. This value
is $def = 1.34$ for IC\,1953 and $def > 1.73$ for NGC\,4491 \citep{barres},
and a galaxy can be considered HI-deficient when $def > 1.2$.

IC\,1953 has two small-mass neighbors, ESO\,548-G040, a very low-surface
brightness galaxy at about 2.6 arcmin in projected distance to the NW,
and ESO\,548-G036, at about 6 arcmin to the SE. ESO\,548-G040 has a much
higher receding velocity (4061\,km\,s$^{-1}$) than the members of the Eridanus
group ($\approx 1600$\,km\,s$^{-1}$), and is therefore most likely a background
galaxy. On the contrary, the velocity of ESO\,548-G036 (1480\,km\,s$^{-1}$)
identifies it as a true neighbor of IC\,1953, with a velocity difference of
400\,km\,s$^{-1}$ and a physical distance $\ge 36$\,kpc. It is thus tempting
to consider ESO\,548-G036 a candidate for tidal interaction with IC\,1953,
which could be responsible for the peculiar arm structure of IC\,1953.

The closest neighbor to NGC\,4491 is the blue compact dwarf IC\,3446 at
6.3 arcmin to the W, and the closest galaxy of comparable mass
is NGC\,4497, at about 12 arcmin to the NW. Their velocities
are 760 and 610\,km\,s$^{-1}$ larger than that of NGC\,4491.

\subsection{Far-infrared and radio properties}
\label{general}

In order to build a reference for far-infrared to radio flux ratios, obtained
with homogeneous IRAS and VLA data, we chose the sample of 69 nearby spiral
galaxies studied by \citet{barres}, with mid-infrared and other data
already at hand, and which comprises two of the three galaxies examined here,
NGC\,4491 and IC\,1953. We measured IRAS fluxes using the {\it Scanpi} tool
\citep{scanpi} with default parameters. Only for NGC\,4491 at 100\,$\mu$m was
the baseline fitted over a 30\arcmin\ range replaced by a baseline fitted over
the central 14\arcmin\ only, where it is reasonably flat. Given the brightness
of the three target galaxies, we used as a flux estimator for them the template
amplitude, except for IC\,1953 which is slightly extended at 25 and 60\,$\mu$m
(we used the flux integration of the scan between the first zero crossings).
Radio continuum fluxes or upper limits at 21\,cm were measured from NVSS maps
\citep{Condon_nvss}, with a beam FWHM of 45\arcsec\ and a detection limit of
the order of 1\,mJy. We obtain an upper limit for NGC\,1377 which is equal to
that given by the more sensitive survey of \citet{Condon_atlas} in a
21\arcsec-beam.

\begin{figure}
\resizebox{11cm}{!}{\rotatebox{90}{\plotone{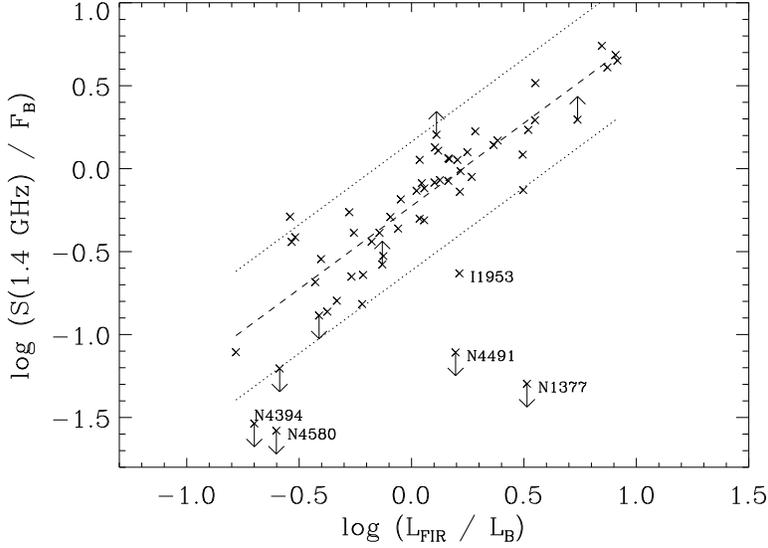}}}
\vspace*{-0.5cm}
\caption{Correlation of radio and far-infrared fluxes, both normalized by
the blue flux, in the sample of \citet{barres} with the additional NGC\,1377.
The dashed line represents the average linear relation, and the dotted lines
indicate the $2 \sigma$ interval of $q$. The galaxy with higher radio flux
than ${\bar q} - 2 \sigma$ is NGC\,4438, which has been disrupted by a collision,
and the radio lower limit close to ${\bar q} - 2 \sigma$ is NGC\,4388,
a Seyfert galaxy. The galaxy at ${\bar q} + 2 \sigma$ with a high
$L_{\rm FIR} / L_B$ ratio is NGC\,1022, an amorphous starburst.
\label{rad_fir_b}}
\end{figure}

\begin{figure}
\resizebox{11cm}{!}{\rotatebox{90}{\plotone{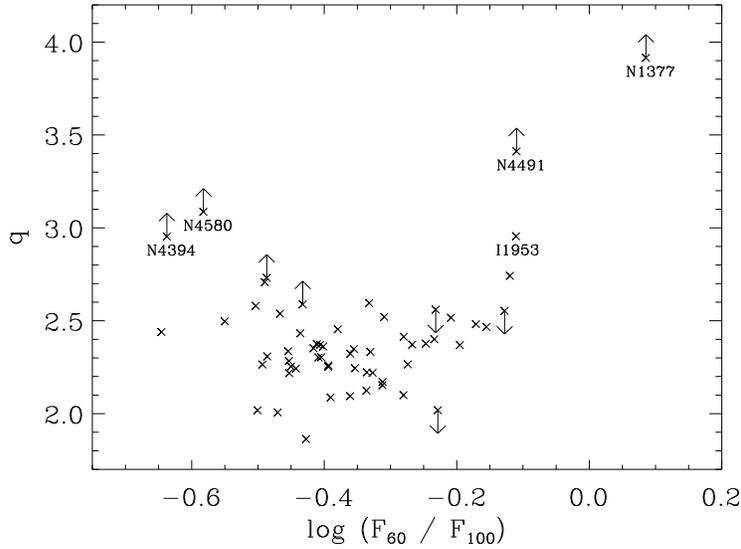}}}
\vspace*{-0.5cm}
\caption{Logarithmic far-infrared to radio flux ratio as a function of
$F_{60}/F_{100}$, indicating an average dust temperature.
\label{q_temp}}
\end{figure}

\citet{Helou_letter} defined a logarithmic ratio of far-infrared to radio flux
densities as: \\
$q = \log~[1.26 \times (2.58\,F_{60}+F_{100})\,/\,3.75\,{\rm (THz)}~/~S(21\,{\rm cm})]$,
where all flux densities, $F_{60}$, $F_{100}$ and $S$(21\,cm), are expressed
in the same unit.
In the reference sample, excluding radio upper and lower limits, we find an
average ratio ${\bar q} = 2.34 \pm 0.19$. The dispersion is the same as that
obtained by \citet{Condon_correl} for the IRAS bright galaxy sample. The
relationship between far-infrared and 21\,cm fluxes, both normalized by the
flux or the flux density in the blue band, is shown in Fig.~\ref{rad_fir_b}.
Five galaxies turn out to have far-infrared to radio flux ratios
$q > {\bar q} + 3 \sigma$: NGC\,1377 ($8.1 \sigma$), NGC\,4491 ($5.5 \sigma$)
and IC\,1953 ($3.2 \sigma$), with far-infrared to blue luminosity ratios higher
than 1.5; NGC\,4580 ($3.9 \sigma$) and NGC\,4394 ($3.2 \sigma$), two Virgo
galaxies having $L_{\rm FIR}/L_{\rm B} < 0.25$.
The latter two are most clearly separated from the first three by their
average dust temperature (Fig.~\ref{q_temp}): they have $F_{60}/F_{100}$
of respectively 0.26 and 0.23, and also low mid-infrared colors
$F_{15}/F_7$ of 0.9 to 1, whereas NGC\,4491 and IC\,1953 have
$F_{60}/F_{100} = 0.8$, and the flux density spectrum of NGC\,1377
peaks at 60\,$\mu$m.

The lack of radio synchrotron emission can be accounted for easily in the
cool-dust galaxies. If their star formation rate is presently very low,
as indicated by their dust emission properties (low colors and low surface
brightnesses), then the far-infrared-emitting dust can be expected to be
heated mainly by low-mass stars, well in excess of the emission produced
by the star formation activity. After the production of cosmic rays has
ceased, the timescale for synchrotron emission decay is less than
$10^8$\,years \citep{Condon_review}, so that the $q$ ratio should increase
on this timescale after an exhaustion of the star formation. The whole
disk of NGC\,4394 is extremely diffuse in H$\alpha$ \citep{Koopmann} and
in the mid-infrared as well \citep{atlas}. This is true also of NGC\,4580
outside its inner pseudo-ring. However, the high $F_{60}/F_{100}$ ratios
of NGC\,1377, NGC\,4491 and IC\,1953 exclude such an explanation for their
lack of synchrotron emission.

\begin{figure}
\resizebox{11cm}{!}{\rotatebox{90}{\plotone{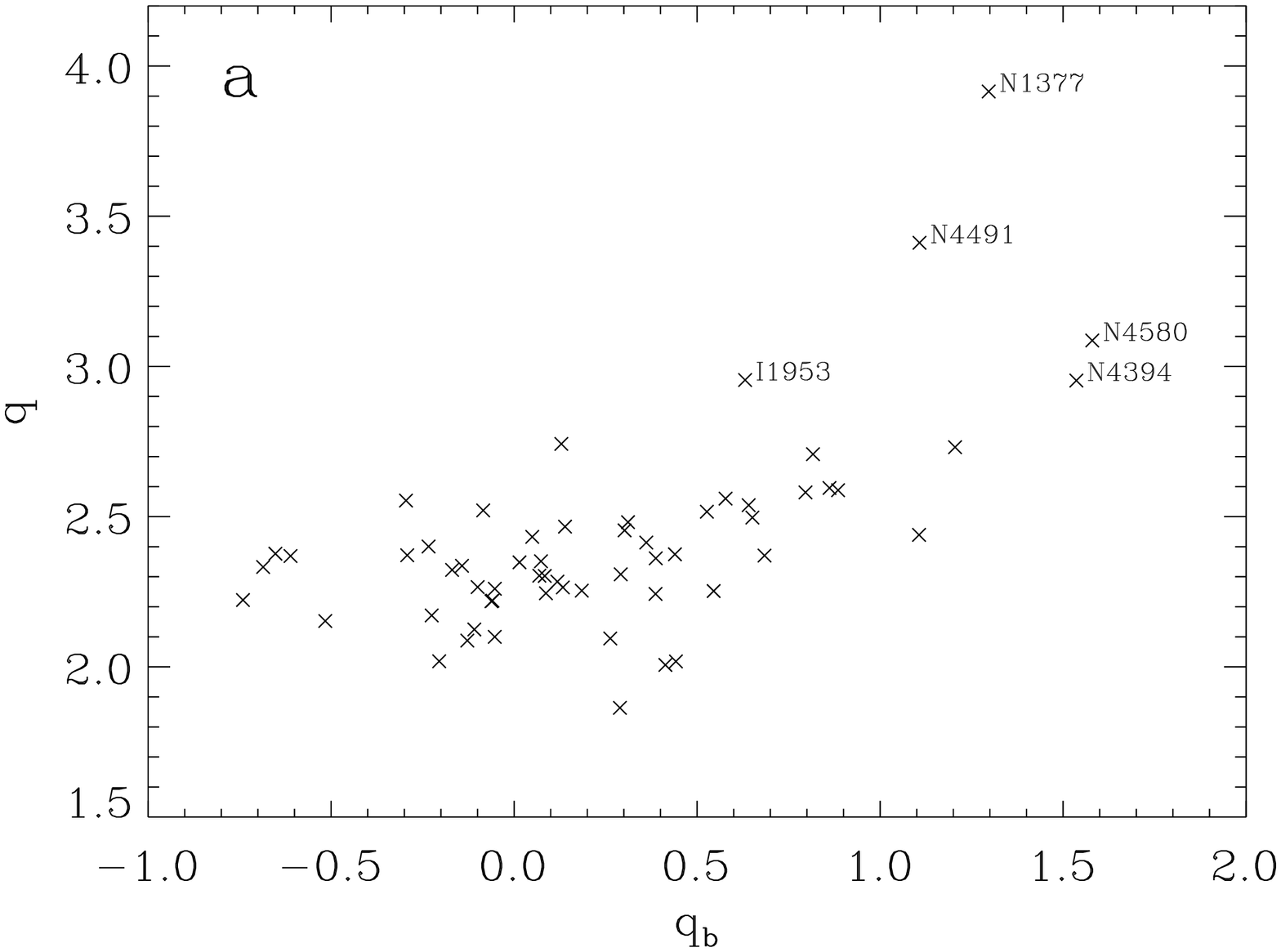}}} \\
\resizebox{11cm}{!}{\rotatebox{90}{\plotone{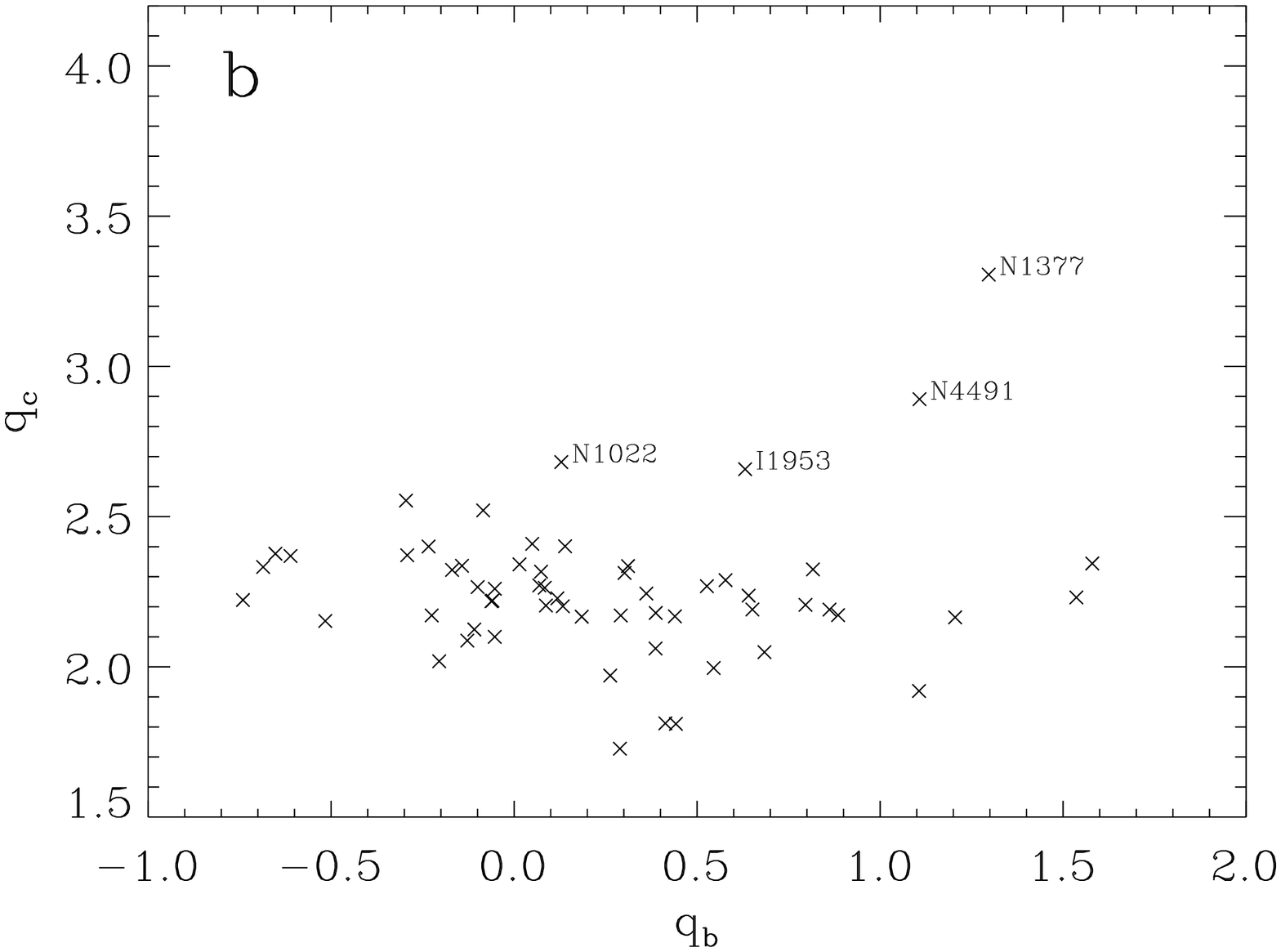}}}
\caption{{\bf a:} $q$ as a function of $q_{\rm b} = log(F_B/S$(20\,cm)), defined
by \citet{Condon_correl} to linearize and tighten the radio-infrared correlation.
{\bf b:} Corrected infrared to radio ratio $q_{\rm c}$ as a function of $q_{\rm b}$.
The galaxies more than $2 \sigma$ above the average $q_{\rm c}$ are indicated.
\label{q_qb}}
\end{figure}

The correlation between far-infrared and radio continuum luminosities
is known to be non-linear, the radio emission rising faster or decreasing
faster than the dust emission \citep{Devereux}. Faint galaxies such as those
studied here are thus expected to have higher $q$ ratios than brighter
galaxies. However, \citet{Hummel_correl} found evidence of a small decrease
of $F_{100}/S$(20\,cm) when the average dust temperature $F_{60}/F_{100}$
rises, along with an increase of the radio brightness temperature.
To linearize and tighten the relationship between infrared and radio fluxes,
\citet{Condon_correl} apply an empirical correction which assumes that the
radio emission is a good measure of the star formation rate, and decreases
the $q$ ratio in proportion to $q_{\rm b} = F_{\rm B}/S$(20\,cm), the blue
to radio flux ratio. Although they find that variations in $F_{60}/F_{100}$
cannot account for the dispersion in $q$, $F_{60}/F_{100}$ is correlated with
$q_{\rm b}$, and the trend of decreasing $q$ with increasing $F_{60}/F_{100}$
exists in our sample also (particularly illustrated in Fig.~\ref{q_temp}
by the behavior of NGC\,4580 and NGC\,4394). NGC\,1377, NGC\,4491 and IC\,1953
manifestly break this trend. If we apply the correction of \citet{Condon_correl}
to our data, as shown in Fig.~\ref{q_qb}, then the peculiarity of NGC\,1377,
NGC\,4491 and IC\,1953 with respect to the other galaxies is not destroyed:
they still lie at $5.9 \sigma$, $3.6 \sigma$ and $2.4 \sigma$ above the new
mean $q_{c} = 2.22 \pm 0.18$. The only other galaxy having
$q_{c} > {\bar q_{c}} + 2 \sigma$ is NGC\,1022
($q_{c} = {\bar q_{c}} + 2.5 \sigma$), an amorphous starburst with
$F_{60}/F_{100} = 0.76$. We note that the correction proposed by \citet{Devereux},
based on the blue to infrared luminosity ratio, cannot be appropriate, since
$q$ as a function of $L_B/L_{\rm FIR}$ shows no defined trend.

The high $F_{60}/F_{100}$  and low $F_{12}/F_{25}$ ratios signal the existence
of a dust component with high temperatures and thus high heating intensities,
which could be due to an extensive star formation burst or non-stellar activity,
or both \citep{Helou_colors}. In this respect, the spectral indices between
25 and 60\,$\mu$m of NGC\,1377 and NGC\,4491, -1.5 and -2.1, are ambiguous
\citep{Grijp}.

\subsection{Spectral energy distribution and power source}
\label{irsed}

\begin{figure}
\vspace*{-1cm}
\resizebox{9cm}{!}{\rotatebox{90}{\plotone{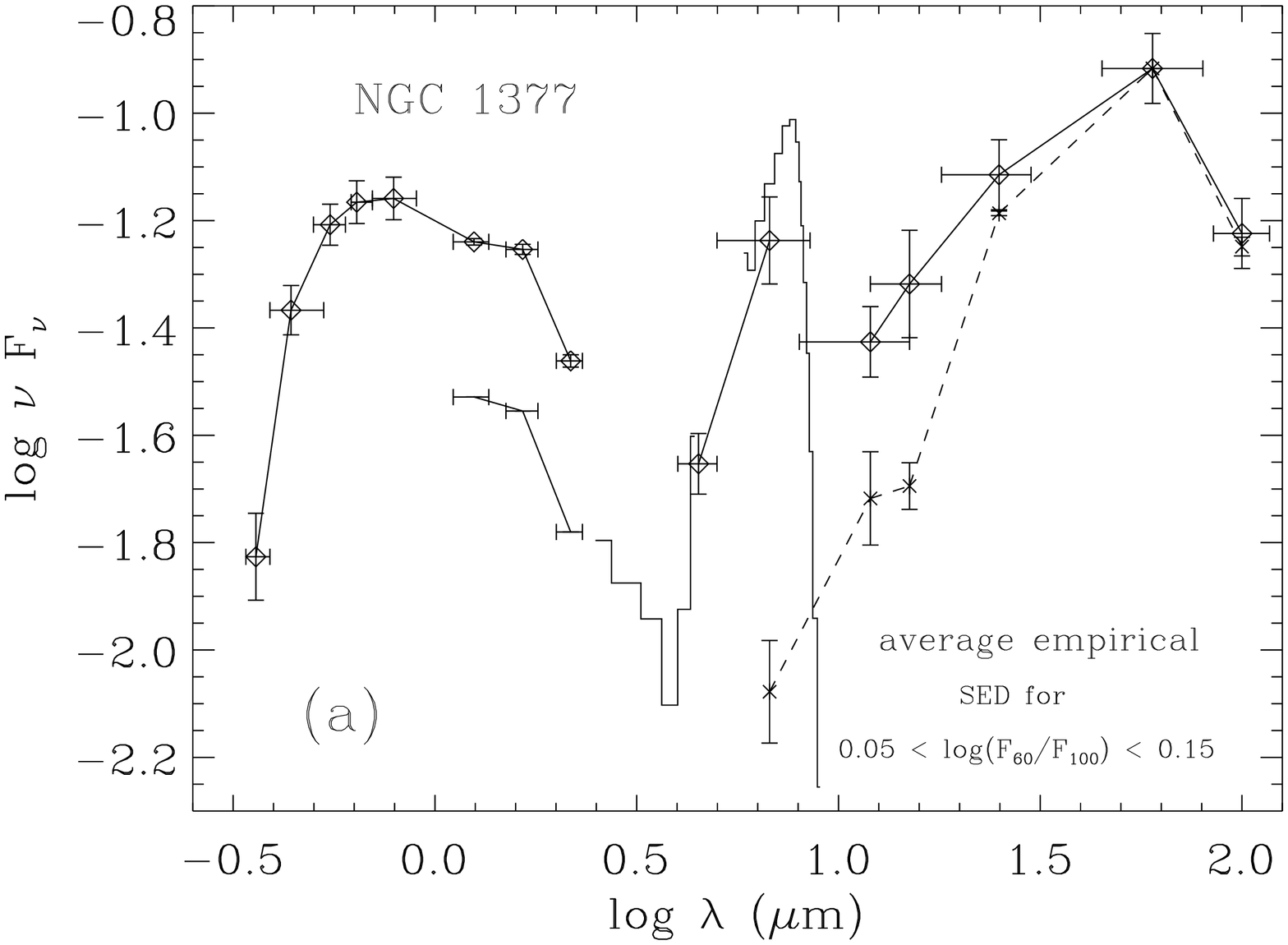}}}
\vspace*{-0.3cm} \\
\resizebox{9cm}{!}{\rotatebox{90}{\plotone{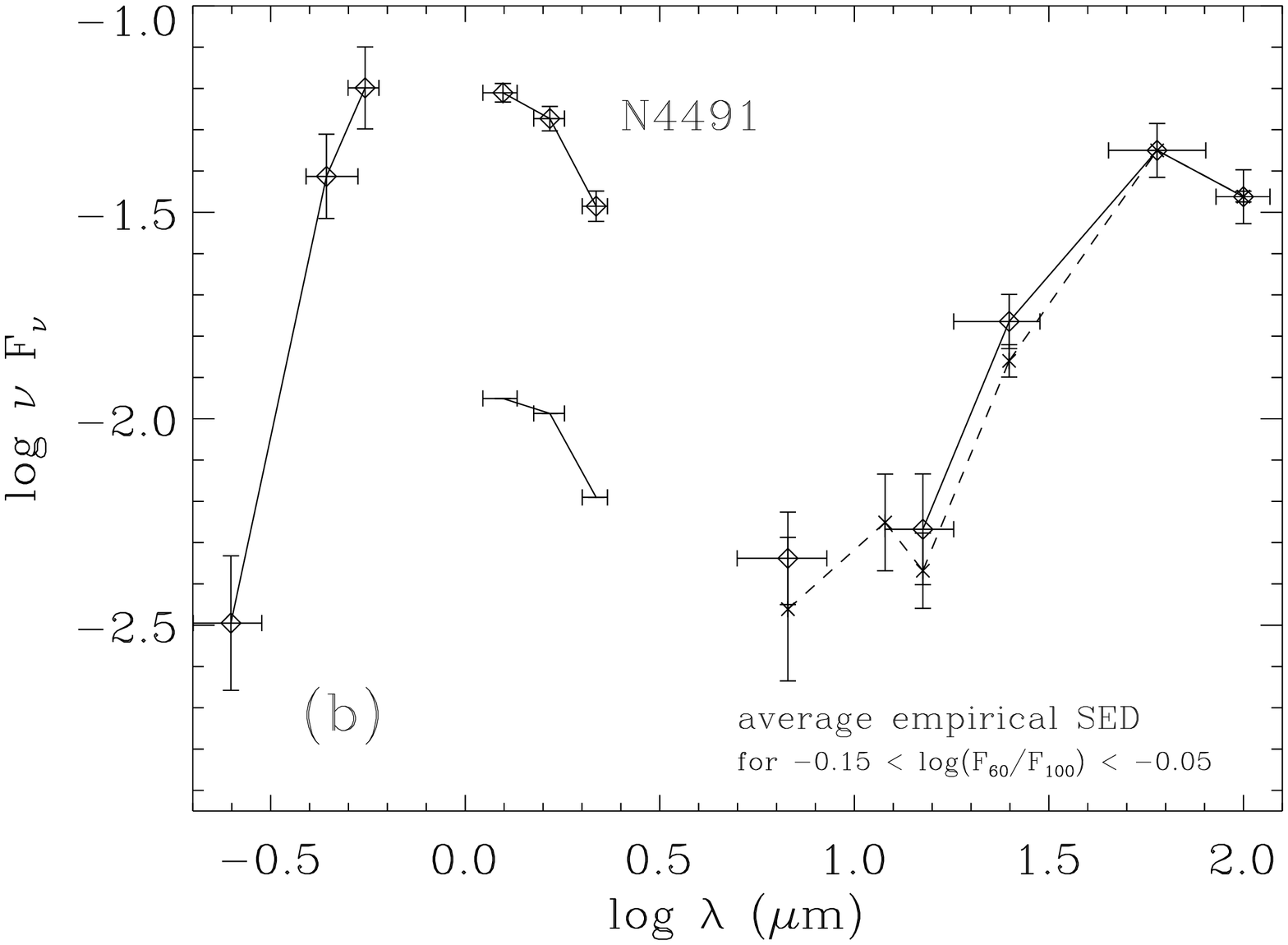}}}
\vspace*{-0.3cm}  \\
\resizebox{9cm}{!}{\rotatebox{90}{\plotone{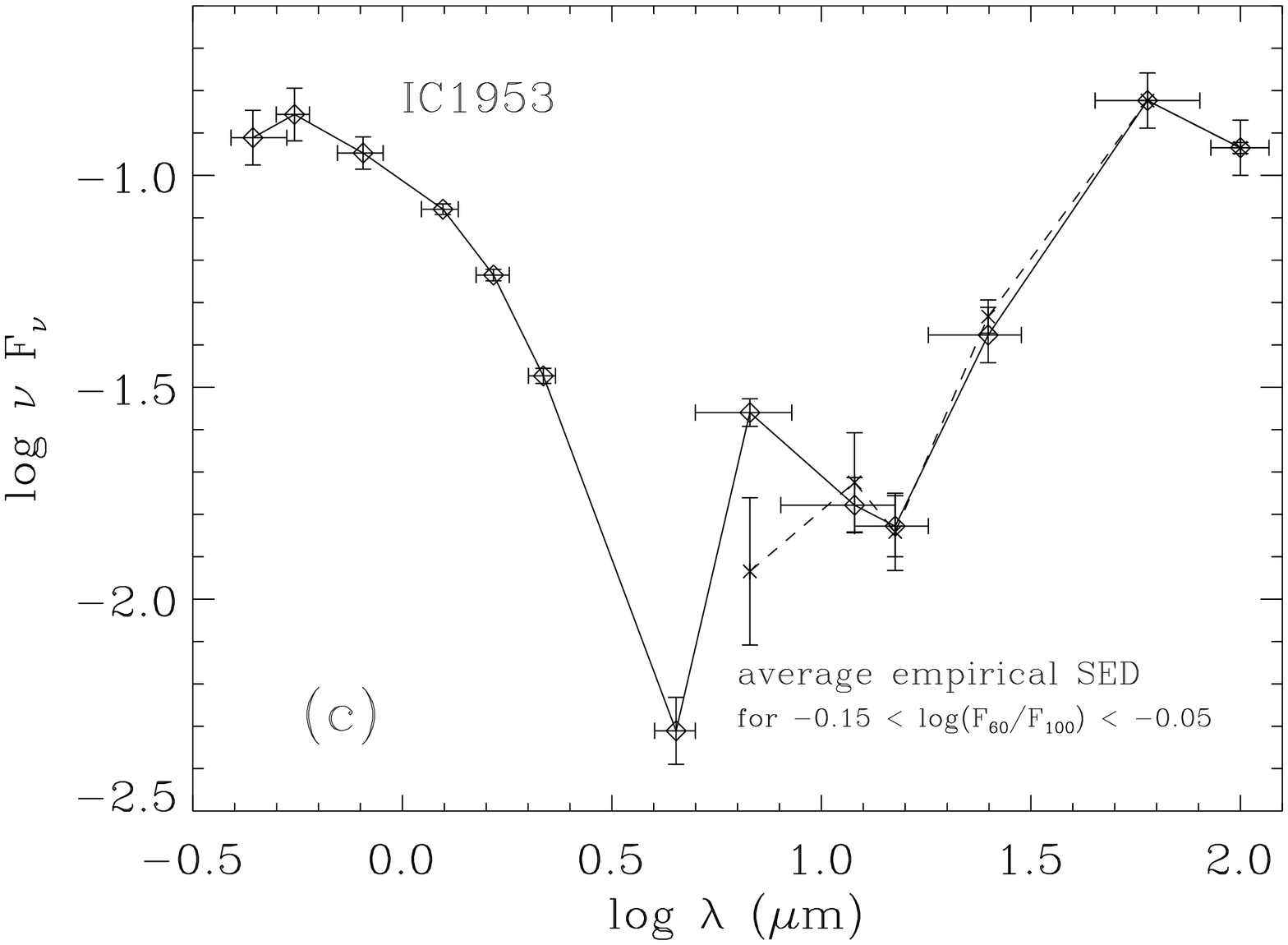}}}
\resizebox{9cm}{!}{\rotatebox{90}{\plotone{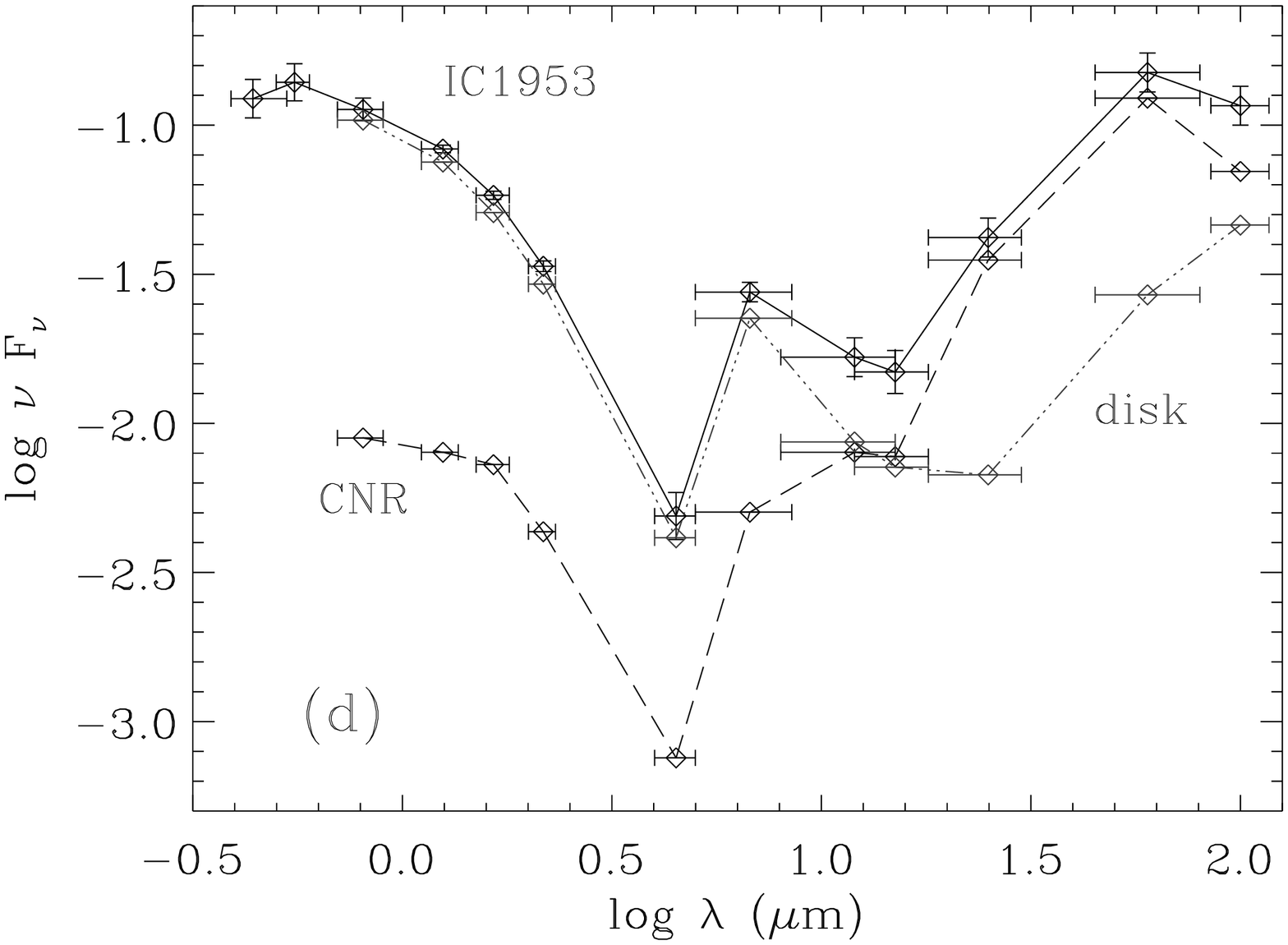}}}
\caption{Spectral energy distribution (in arbitrary unit), compared with
the average spectrum corresponding to the same $F_{60}/F_{100}$ ratio
from \citet{Dale} (dashed lines). The horizontal error bars represent
the approximate filter widths. {\bf (a)} The histogram superimposed on the SED
of NGC\,1377 reproduces the spectrum by \citet{Laureijs} in the
[2.5\,;\,4.8]\,$\mu$m and [5.8\,;\,9]\,$\mu$m ranges. {\bf (a) and (b)}
The lower lines in the range $\log \lambda \in [0.1\,;\,0.35]$ correspond to
the near-infrared spectra restricted to the mid-infrared emitting region.
{\bf (d)} The dot-dashed and dashed spectra are the decomposition of the
SED of IC\,1953 into the disk and the circumnuclear region, respectively.
\label{sed}}
\end{figure}

The broadband spectral energy distributions of the three galaxies are
shown in Fig.~\ref{sed}. J, H and Ks-band fluxes are derived from images
built for the 2MASS Large Galaxy Atlas \citep{Jarrett_atlas}, with a
preliminary photometric calibration valid for the CIT system. The U, V
and I-band fluxes of NGC\,1377 come from the colors given by
\citet{Heisler_opt}. Other optical data were found in the NED database.
We measured the 4--5\,$\mu$m fluxes of NGC\,1377 and IC\,1953
in ISOCAM maps.

\begin{figure}
\resizebox{12cm}{!}{\rotatebox{90}{\plottwo{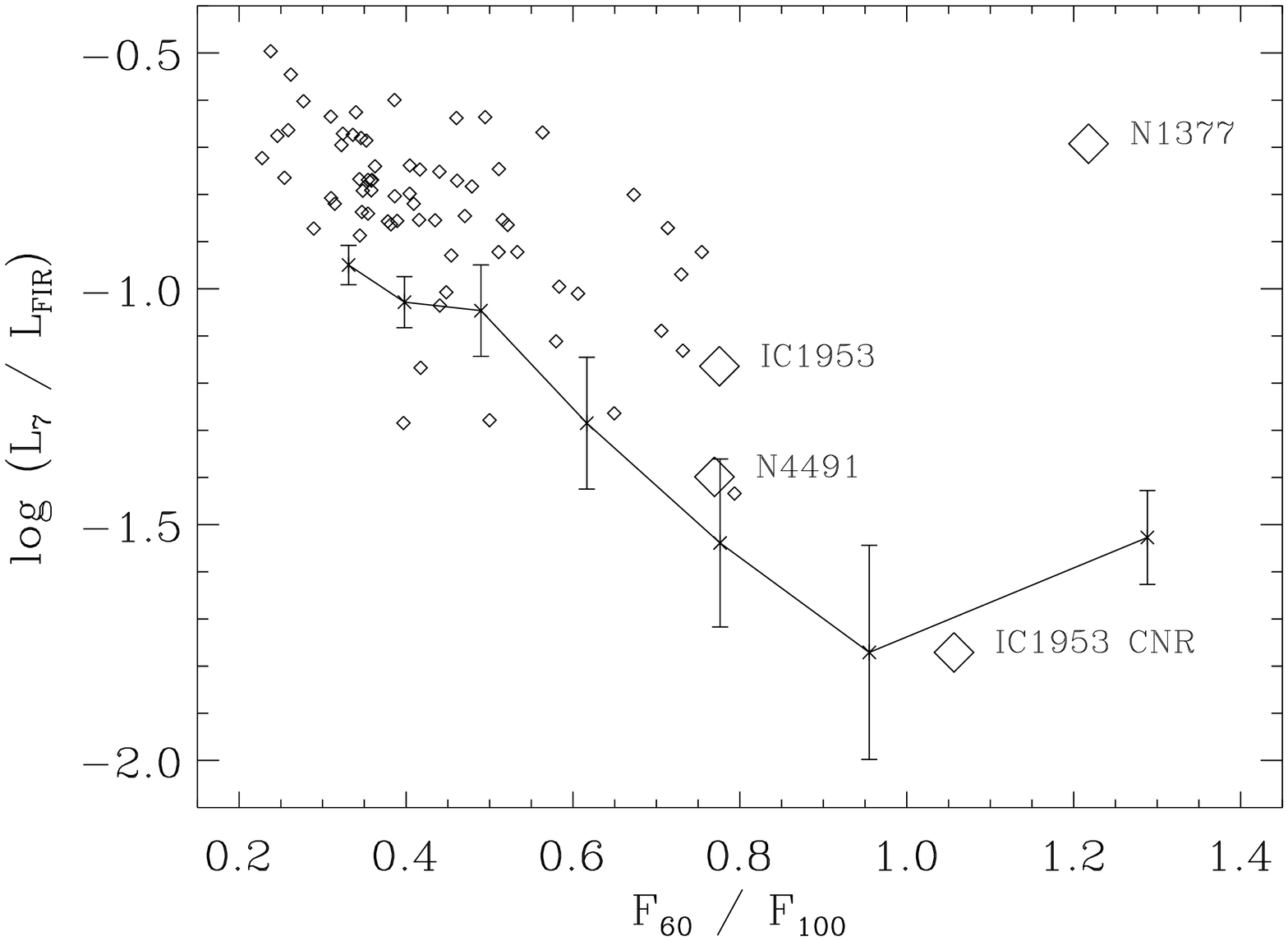}{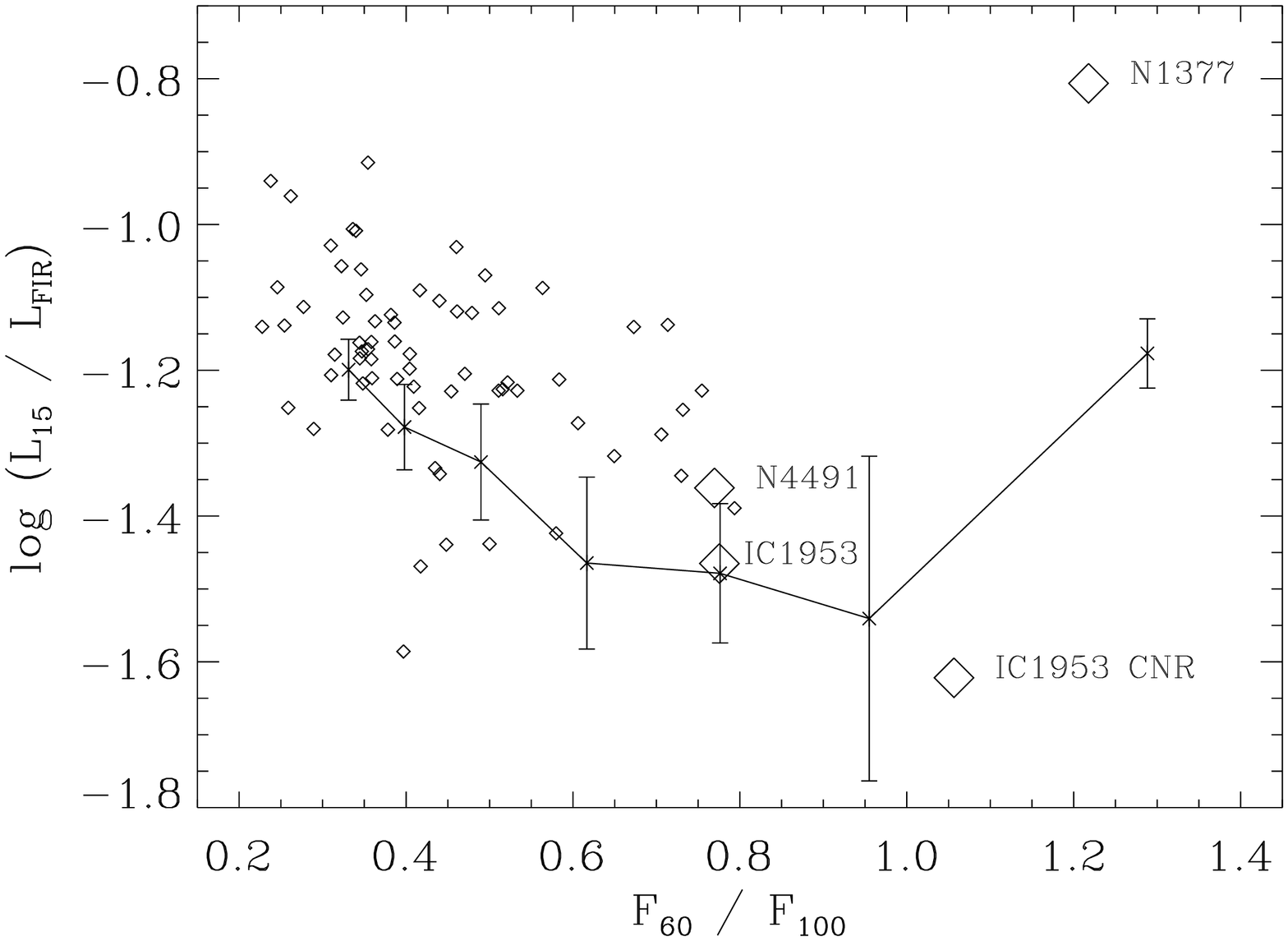}}}
\caption{Mid-infrared to far-infrared luminosity ratios as a function
of dust temperature in the sample of \citet{barres} (small lozenges),
average ratios of \citet{Dale} (dots connected with a line, with $1\sigma$
dispersions), and ratios of NGC\,1377, NGC\,4491 and IC\,1953 (big lozenges).
\label{mirsfir}}
\end{figure}

It appears that the dust emission spectrum of NGC\,1377 cannot be reproduced
by any empirical SED of \citet{Dale} (Fig.~\ref{sed}), ajusted to a sample
of normal galaxies, and parametrized by the $F_{60}/F_{100}$ ratio (note
however that the color bin of NGC\,1377 comprises only two galaxies). The
emission from transiently-heated dust particles, from 7 to 25\,$\mu$m, is
well in excess of what is expected for the large grain temperature (as
estimated by the $F_{60}/F_{100}$ ratio), by up to 8--$9\sigma$.
$L_7/L_{\rm FIR}$ is expected to decrease steadily with rising dust
temperature, and then starts to increase again at the highest
$F_{60}/F_{100}$ ratios, but NGC\,1377 overdevelops this trend
(Fig.~\ref{mirsfir}). For $F_{15}/F_{\rm FIR}$, which in normal galaxies
shows less dispersion than $F_{7}/F_{\rm FIR}$, the same remark applies.

The mid-infrared spectral shape between 2.5 and 11\,$\mu$m, observed with
{\small ISOPHOT} in $24\arcsec \times 24$\arcsec\ \citet{Laureijs}, is
quite unusual: instead of the typical spectrum of aromatic bands observed
in most star-forming galaxies \citep{Helou_spectres}, a broad asymmetric
feature between 6 and 8.5\,$\mu$m contributes the bulk of the emission.
This is reminiscent of some ultraluminous galaxies
\citep{Rigopoulou, Tran}. Such a broad feature, accompanied by very weak
6.2 and 11.3\,$\mu$m bands, is also observed in the peak E of the SMC
HII region N\,66 by \citet{Contursi}, in proto-planetary nebul\ae\
\citep{Kwok}, and bears some resemblance with the signature of
anthracite, used by \citet{Guillois} to reproduce proto-planetary nebula
spectra. All this suggests that the aromatic band carriers have not been
processed in the same way as in normal star-forming galaxies, and that the
dust responsible for the emission is subjected to hard radiation.
Since it contributes the same part in the infrared energy budget as
in quiescent galaxies, this dust must be more resilient in order not to
be destroyed. The abrupt fall-off of the spectrum from 9 to 11\,$\mu$m
further suggests that the absorption by silicates may be substantial,
but such an interpretation is very hazardous, because optically-thin
spectra of normal galaxies also have a minimum in this region.
Dust heating cannot be provided by a shock, in particular because
one would then observe either a prominent silicate emission feature,
or a negligible mid-infrared emission with respect to the far-infrared
emission, depending on the gas density \citep{Draine_shock, Dwek}.

\citet{Laureijs}, on the basis of this spectrum, propose that NGC\,1377
contains a buried Seyfert nucleus. However, several arguments suggest
otherwise: \\
{\bf 1)} The continuum emission at 3--4\,$\mu$m and 9--11\,$\mu$m is near
zero, contrary to what is observed in Seyfert nuclei. The short-wavelength
continuum, in particular, which is not affected as much by extinction,
is now routinely used as a diagnostic of the contribution of an active
nucleus to the energy budget \citep{Genzel, Laurent}. Among the sample
of \citet{Laureijs}, the spectrum of NGC\,1377 resembles much more that
of IRAS\,02530+0211, a galaxy of starburst type, than that of any of
the Seyfert galaxies. One could argue that the broad feature seen
between 6 and 8.5\,$\mu$m is similar to that seen in Centaurus\,A
\citep{Laurent}, and is produced by a continuum strongly absorbed
longward of 8\,$\mu$m by silicates; however, the similarity is
superficial, since the spectra of Centaurus\,A and other Seyfert nuclei
rise steeply longward of 10\,$\mu$m and are brighter at 11.5\,$\mu$m
than at 8\,$\mu$m, whereas the 11.5\,$\mu$m flux of NGC\,1377 is near
zero. This is a strong argument against pure silicate absorption,
because the red wing of the absorption band would otherwise be
exceptionally wide. Trying a quantitative fit of the 8\,$\mu$m to
100\,$\mu$m data, with the extinction law of \citet{Draine} (some
variants exist, but introduce differences essentially shortward of
8\,$\mu$m, which is why we exclude short wavelengths),
$A_{\rm V} \approx 70$ is required in order to reproduce the
8--10\,$\mu$m {\small ISOPHOT} spectrum with an absorbed dust continuum,
but no good fit can be performed: the solution is too high in the
10--11.5\,$\mu$m range; one can obtain a simultaneous fit of the
12\,$\mu$m, 25\,$\mu$m, 60\,$\mu$m and 100\,$\mu$m data with the sum
of two blackbodies of 370\,K and 100\,K, but the fit at 15\,$\mu$m is
then much higher than the observed flux, which is impossible to fix
by adding a new dust component. We thus conclude that the broad feature
between 6 and 8.5\,$\mu$m may contain a genuine emission band, and
may not be exclusively caused by absorption\footnote{Note that if
$A_{\rm V} = 70$ is adopted, then the emission of hot small grains up
to $\approx 25$\,$\mu$m would have to be boosted by a factor five to
obtain the intrinsic small grain luminosity. Since the complementary
power observed in the far-infrared is only 1.2 times the observed power
of small grains, at least $3 L_{\rm FIR}$ would have to be re-emitted
at longer wavelengths, in the submillimeter, which is inconceivable.
And this would only exacerbate the difficulty to account for the dust
luminosity of NGC\,1377 in the total absence of activity in the radio
window (this will be developed in Sect.~\ref{seyfert}).}. \\
{\bf 2)} The J, H and K-band surface brightness profiles computed by
\citet{Heisler_nir} are nearly flat inside the central 3--4\arcsec,
significantly below the best $r^{1/4}$ fit, whereas most
Seyfert galaxies with $F_{60} > F_{100}$ require an extra nuclear
point source. A high central absorption could flatten the profiles,
but this is unlikely, because 2MASS data indicate that the J-H and
H-K colors of the central 10\arcsec\ are not different from the
global colors within the error bars. H-K is relatively ``blue''
compared to the other galaxies in \citet{Heisler_nir}, and J-K
is very blue compared to the ultraluminous galaxies in \citet{Klaas}
(like the bluest starbursts and Liners), or to the 2MASS extended
sources brighter than ${\rm Ks} = 14$ \citep{Jarrett, Jarrett_atlas}.
The power emitted by dust between 4.5 and 100\,$\mu$m is only 1.8
times the total stellar power between 3600\AA\ and 4.5\,$\mu$m,
and about twice more considering the center alone. This means that
the luminosity-weighted extinction cannot be high enough to hide a
deep stellar potential well. \\
{\bf 3)} Seyfert galaxies either emit excess radio emission with respect
to their infrared emission, or follow the same correlation as normal galaxies
\citep{Sopp}. Although no starburst galaxy is known to be as radio-deficient
as NGC\,1377, no active nucleus either has been shown to be
synchrotron-deficient so far. X-ray data would be useful to rule out
Seyfert activity, but to our knowledge, none exist for NGC\,1377.

The 7 to 25\,$\mu$m emission of NGC\,4491 is somewhat in excess
of the corresponding SED of \citet{Dale}, but not significantly.
The blue-band surface brightness profile of \citet{Kodaira} is
exponential, with no central peak, and so are the near-infrared profiles.
The J-H and H-K colors are close to those of NGC\,1377, and the color
differences between the circumnuclear region and the whole galaxy are
small. The same conclusions thus apply.

As to IC\,1953, only the 7\,$\mu$m emission is 2\,$\sigma$ in excess of the
average empirical SED. Its global near-infrared colors are extremely blue,
and its circumnuclear colors are slightly bluer than those of NGC\,1377
and NGC\,4491. Since the disk infrared emission in IC\,1953 is significant,
and well resolved in ISOCAM maps, we decomposed the spectral energy
distribution into a disk component and a circumnuclear region (CNR) component,
which we identify to the core with a high $F_{15}/F_7$ ratio.
To do this, we estimated typical disk infrared colors using the ISOCAM
and IRAS fluxes of NGC\,4027 \citep{atlas}, which has high signal to noise
ratios and whose entire body can be assimilated to a disk:
$F_{15}/F_7 = 0.87$ (the small 7\,$\mu$m excess in the disk of IC\,1953
being ascribed to stellar emission), $F_{12}/F_7 = 0.85$, $F_{25}/F_7 = 1.37$,
$F_{60}/F_7 = 13.2$ and $F_{100}/F_7 = 37.7$\,. The result, shown in
Fig.~\ref{sed}d, outlines the fact that part of the peculiarity of IC\,1953
is hidden by normal disk properties: the center has hotter dust
and higher infrared to optical ratios than the galaxy as a whole.
The radio deficiency is likely more severe in the center; using
the radio measurement of \citet{Condon_atlas} inside 21\arcsec\ and the
above decomposition, we obtain $q_{\rm CNR} = 3.44 = {\bar q} + 5.7\sigma$.

The position of a galaxy in a mid-infrared surface brightness--color diagram
($\Sigma$(15\,$\mu$m) versus $F_{15}/F_7$) can give some sense of the relative
contribution of stellar populations of different ages, on a timescale of several
$10^8$\, years, in the dust heating \citep{barres}. The circumnuclear regions
of NGC\,4491 and IC\,1953 populate a part of the diagram otherwise
empty\footnote{The only other galaxies in the same part of the diagram
are NGC\,4388, a Seyfert galaxy, and NGC\,4519, a galaxy akin to IC\,1953
in many respects, but whose peculiar central regions may be diluted in the
emission of its bright disk.},
and have a central $F_{15}/F_7$ color more than twice the average value
corresponding to their surface brightness. This is fully consistent
with the hypothesis that they are undergoing a starburst in an early phase,
which is massive with respect to the mass of the underlying intermediate
stellar population. As for NGC\,1377, its $F_{15}/F_7$ excess is only
moderate. However, the emission in the 7\,$\mu$m bandpass being due to
an unusual dust species, the interpretation of its $F_{15}/F_7$ color
is not straightforward.

\section{Multi-wavelength radio continuum}
\label{effelsberg}

We mapped the galaxies with the Effelsberg 100\,m antenna\footnote{The
Effelsberg telescope is operated by the MPIfR Bonn on behalf of the
Max-Planck-Gesellschaft e.V.}, at 6.2\,cm and 3.6\,cm, during six observation
sessions between April and August 2002. These observations made use of
the 4850\,MHz HEMT dual-horn receiver (30\,K system temperature, 500\,MHz
bandwidth) and the 8350\,MHz HEMT single-horn receiver (25\,K system
temperature, 1100\,MHz bandwidth), both installed in the secondary focus
of the telescope. The beamwidths are 147\arcsec\ at 6.2\,cm and 82\arcsec\
at 3.6\,cm, thus comparable to the galactic disk sizes. Between 10 and 20
coverages per galaxy were obtained, scanned in azimuthal direction at
6.2\,cm and in right ascension and declination, alternatively, at 3.6\,cm.
3C138 and 3C286 were used as flux calibration sources. The data reduction
was performed using the NOD2 software package. The coverages were combined
by applying the ``basket-weaving'' technique \citep{Emerson}. The final
3.6\,cm maps were smoothed to a beamwidth of 90\arcsec\ to increase the
signal to noise ratio.

\begin{figure}
\vspace*{-1.5cm}
\hspace*{-2cm}
\resizebox{17cm}{!}{\plotone{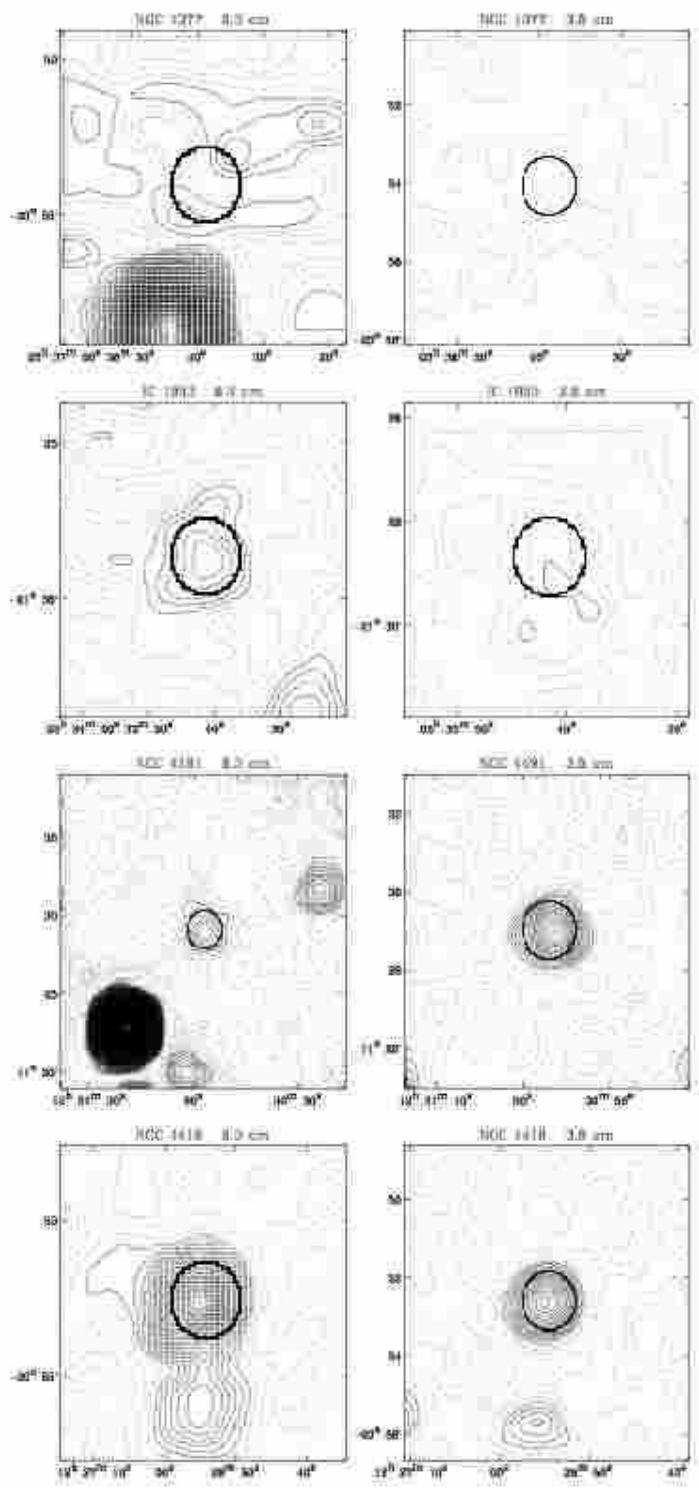}}
\vspace*{-1.5cm}
\caption{Contour maps at 6.2\,cm (left) and 3.6\,cm (right), spaced at
$1 \sigma$-intervals. The half-power beam width is indicated with a thick
circle at the position of the galaxy. The continuous contours outline
levels above $3 \sigma$; the grey dashed contours $1 \sigma$ and
$2 \sigma$ levels; the dotted contours the zero level; to visualize
any negative peaks which would indicate bad cleaning, the dot-dashed
contours ouline $-1 \sigma$ and $-2 \sigma$ levels (in grey), and
$-3 \sigma$ (in black).
\label{contours}}
\end{figure}

\begin{figure}
\hspace*{-2cm} \resizebox{10cm}{!}{\rotatebox{90}{\plotone{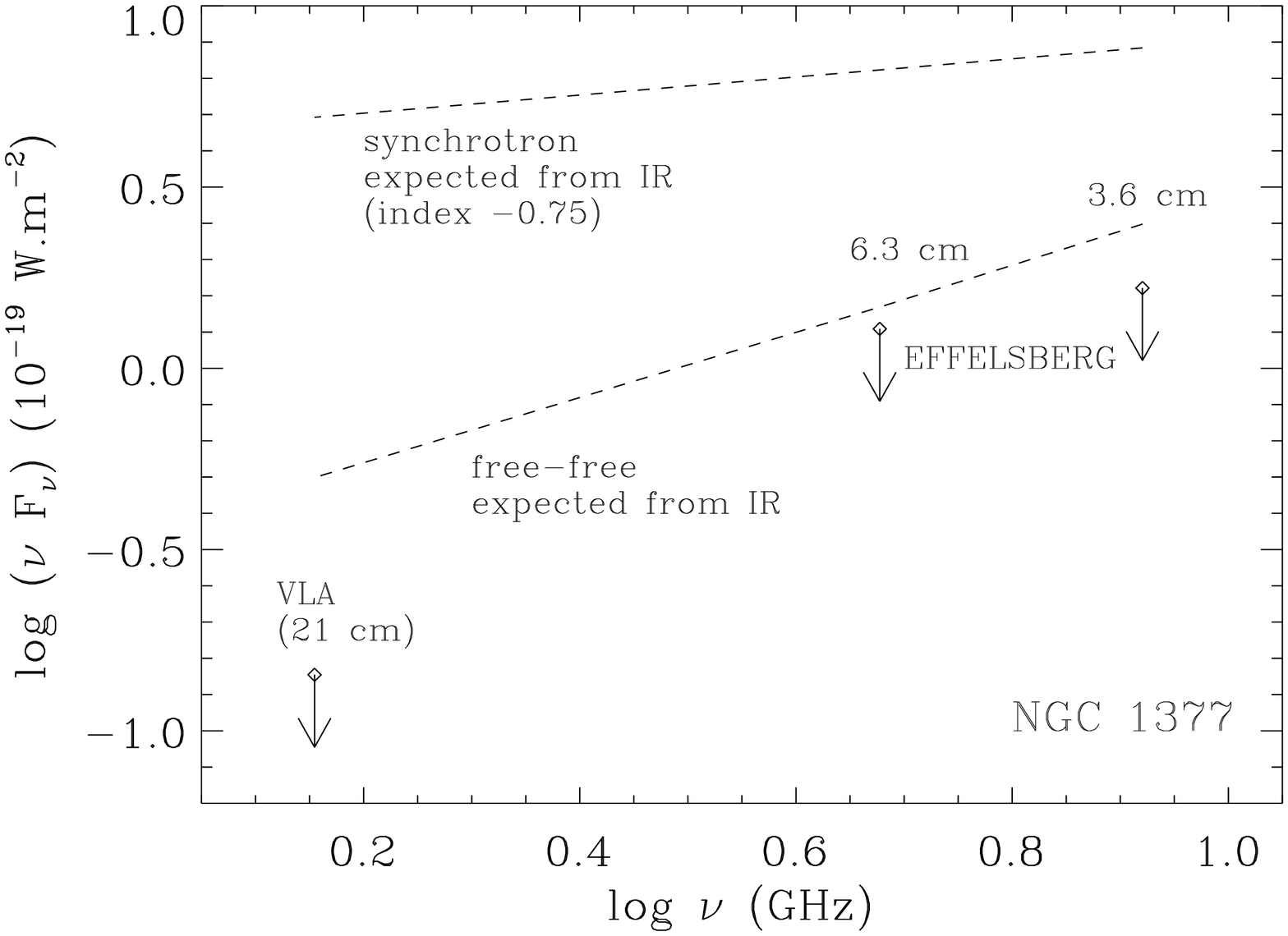}}}
\hspace*{-0.5cm} \resizebox{10cm}{!}{\rotatebox{90}{\plotone{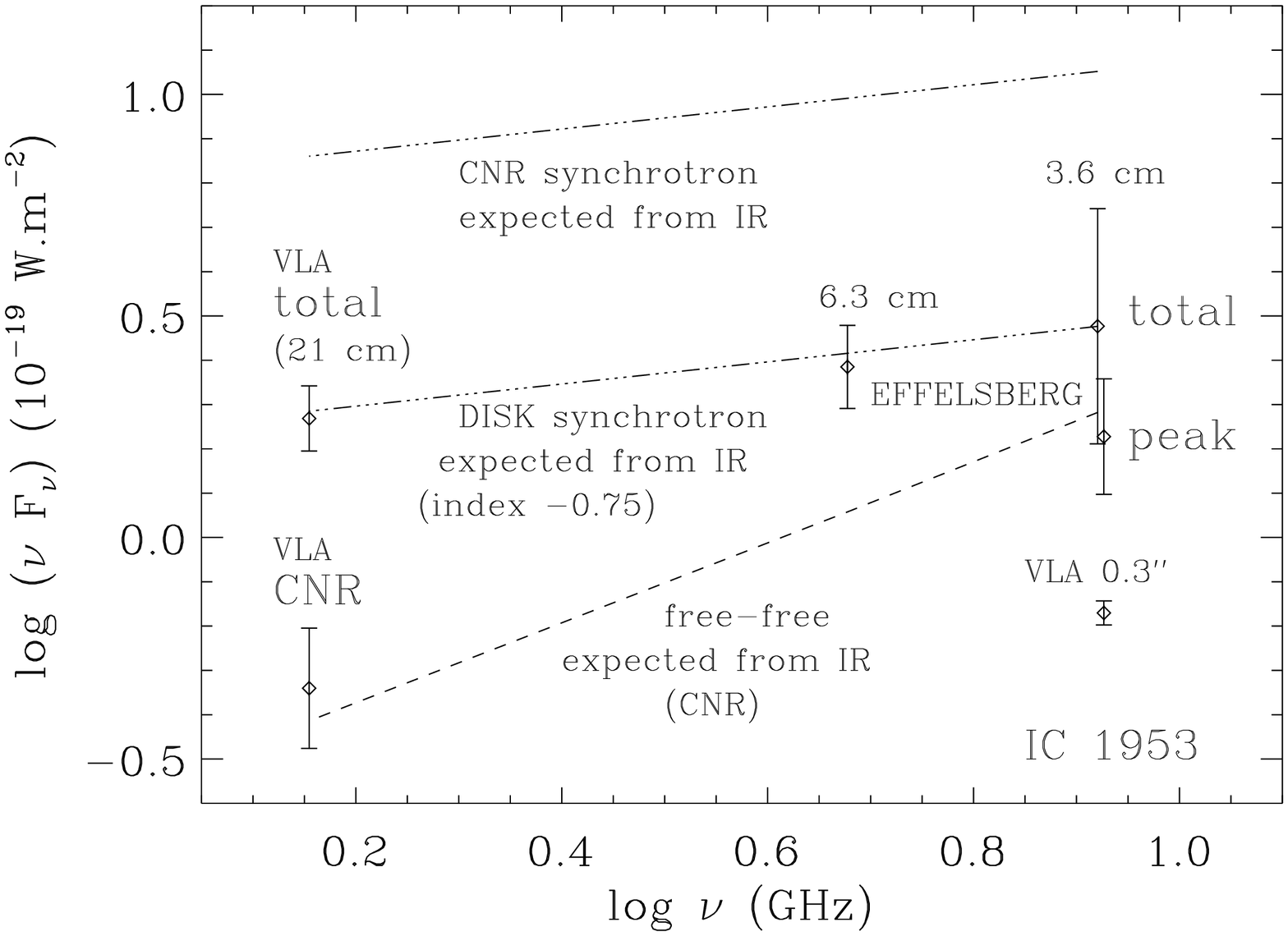}}}
\hspace*{-2cm} \resizebox{10cm}{!}{\rotatebox{90}{\plotone{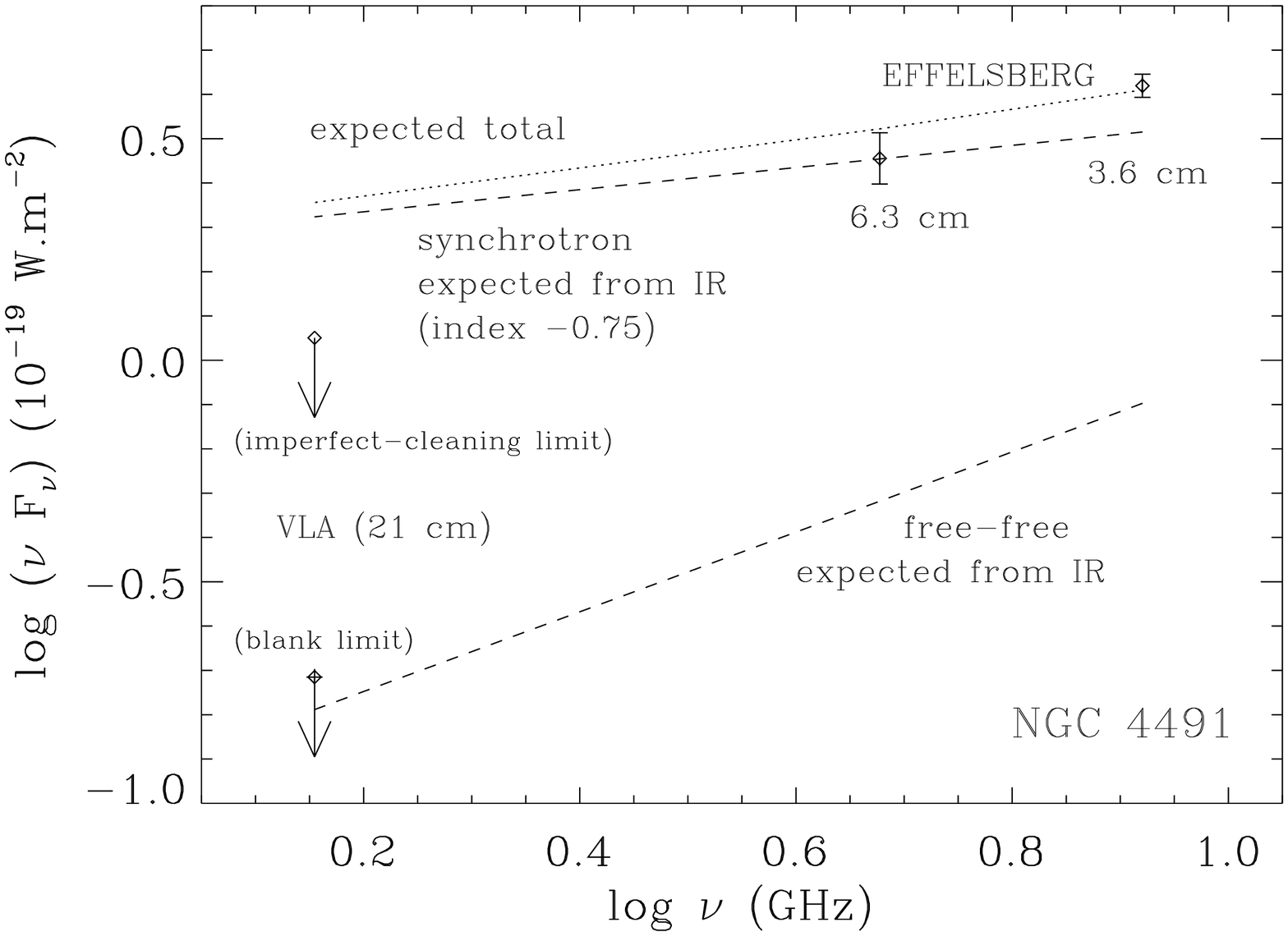}}}
\hspace*{-0.5cm} \resizebox{10cm}{!}{\rotatebox{90}{\plotone{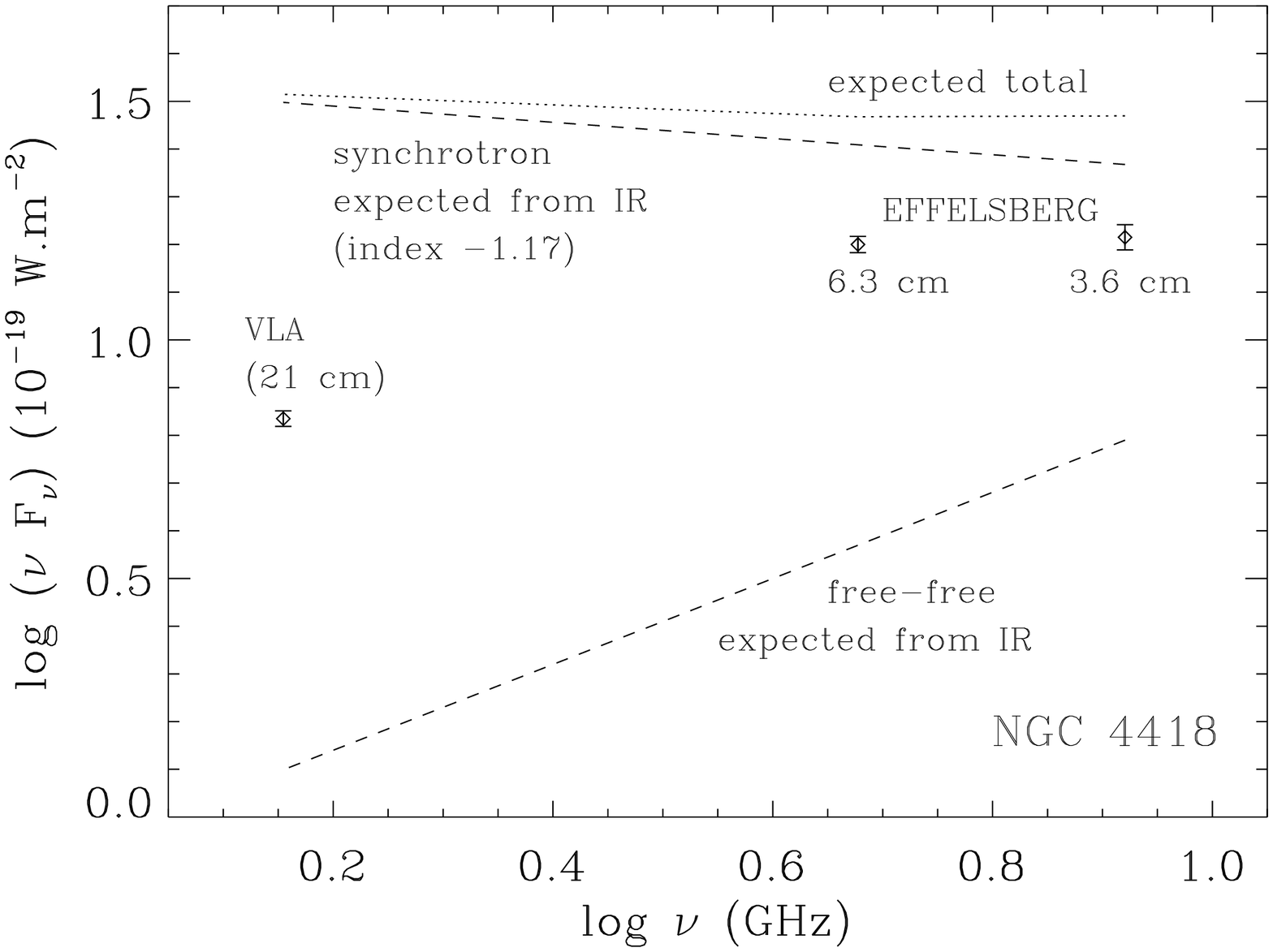}}}
\caption{Radio continuum measurements. The dashed lines indicate the
free-free and synchrotron emission expected from the far-infrared-derived
SFR and the infrared-radio correlation followed by normal galaxies (with
a thermal index $S \propto \nu^{-0.1}$ and non-thermal indices indicated
on the graphics). The radio spectrum of NGC\,4418 is added here (see
Section~\ref{n4418}).
\label{sed_radio}}
\end{figure}

The measured flux densities are contained in Table~\ref{tabradio}
and the contour maps shown in Fig.~\ref{contours}. Significantly
polarized emission was not detected in any galaxy.
Figure~\ref{sed_radio} compares the radio measurements with the free-free
and synchrotron spectra of normal galaxies matched to the far-infrared
fluxes of NGC\,1377, IC\,1953 and NGC\,4491. The total flux expected at
21\,cm was derived from the average $q$ ratio, the free-free component
from the ionizing photon flux estimate (assuming $T_e = 5000$\,K),
and we adopted the spectral indices $S_{\rm sync} \propto \nu^{-0.75}$
and $S_{\rm ff} \propto \nu^{-0.1}$, unless otherwise constrained
for the synchrotron component.

\subsection{NGC\,1377}
\label{n1377}

NGC\,1377 is not detected at any radio frequency, and the upper limits
even lie below fluxes expected from a mere free-free component
matched to the ionizing photon flux derived from the dust emission
(Fig.~\ref{sed_radio}). This fact suggests to us not only that NGC\,1377
is mostly devoid of synchrotron radiation, but also that the star
formation episode is so young that most massive stars are still
embedded in their parent molecular clouds, which would cause a large
fraction of the ionizing photons to be intercepted by dust before being
able to ionize the gas. NGC\,1377 was reobserved in 1990 at 3.55\,cm
with the VLA in the A configuration, yielding a beam of
$0.4\arcsec \times 0.25$\arcsec\ (J.J. Condon, unpublished). The galaxy
is still undetected at this high resolution, and the $3 \sigma$ upper
limit is 0.13\,mJy per beam, i.e. 15 times less than the 3.6\,cm upper
limit from Effelsberg. Interpreting this limit, we have to take into
account the possibility that the star-forming region be more extended
than 0.3\arcsec. Assuming a size of 100\,pc or $\approx 1$\arcsec\
(see the next to last paragraph in this section), the equivalent upper
limit on the total 3.55\,cm flux is of the order of 1.3\,mJy. 
\citet{Vader} derive from a ``snapshot'' observation with the VLA
$S$(6\,cm$) < 0.9$\,mJy for NGC\,1377, in a $4$\arcsec-beam, which
brings a stronger constraint. In such conditions, the data require a
reduction of ionizing photons by about 70\% or more (also by 70\% using
instead the 21\,cm upper limit). If we adopt the limit in a 0.3\arcsec-beam,
then the ionizing photons have to be suppressed by more than 95\%.
If indeed a larger fraction than usual
of the ionizing photons are absorbed by dust in NGC\,1377, then this
will have the simultaneous effect of decreasing the free-free emission
and enhancing the infrared emission, with respect to a normal galaxy
with the same star formation rate.

Thermal opacity is an alternative explanation for the non-detection of
the free-free emission. The upper limit at 3.55\,cm however places very
stringent a constraint in this case: with $S(3.55\,{\rm cm}) < 1.3$\,mJy,
the effective opacity is $\tau$(3.6\,cm$) > 0.83$, or $\tau$(21\,cm$) > 31$. 
Assuming a simple geometry and uniform electronic density, such a value
would imply that the scale of the emitting region be less than 28\,pc,
and the electronic density higher than 2700\,cm$^{-3}$. Since this is
unrealistic, the free-free opacity explanation is dismissed.

Assuming a normal initial mass function and no dust emission excess,
the starburst must have produced approximately $10^4$ O stars
(see Section~\ref{sfrate}).
Assuming that the dust emission at 60 and 100\,$\mu$m is due to a
pure blackbody of temperature 80\,K (which fits the 25 to 100\,$\mu$m
data with only a small observed excess of 25\% at 25\,$\mu$m),
its diameter is about 37\,pc. Using the model of \citet{Desert} for O5
stars, the size of the starburst region is rather of the order of 100\,pc
(1\arcsec) or larger. The emission region may thus be truly more
extended than the VLA A-configuration beam. To trigger star formation
quasi-instantaneously over such a scale perhaps needs a merger with
a satellite galaxy, which does not necessarily produce morphological
disturbances, but should be visible in the gas kinematics.

The dust mass required to produce the observed far-infrared emission
is $\approx 2.7 \times 10^5$\,M$_{\sun}$, according to the dust emissivity
derived by \citet{Bianchi} for the Galaxy, with $T_{\rm dust} = 54$\,K
(which produces an emission excess from small dust grains of 76\% at 25\,$\mu$m).
Using instead a blackbody fit, only $4.0 \times 10^4$\,M$_{\sun}$ of dust are needed.
Assuming that the gas to dust mass ratio is similar to the Galactic value of
$\approx 100$ \citep{Sodroski}, the gas mass associated with the starburst
in NGC\,1377 is then of the order of $4 \times 10^6$ to $3 \times 10^7$\,M$_{\sun}$,
which represents much less than 10\% of the observed stellar mass in the whole
central region. Since this mass ratio is reasonable, the hypothesis that
the dust is heated by a starburst is plausible, but still needs to be confirmed
by observations of the molecular gas: see Section~\ref{sest}. The upper limit
on the atomic gas mass, M(HI$) < 2 \times 10^8$\,M$_{\sun}$ (Section~\ref{envir}),
is not in contradiction with the starburst hypothesis.
We will discuss the possibility that a radio-quiet active nucleus
heats the dust in Section~\ref{seyfert}.

\subsection{IC\,1953}
\label{ic1953}

The entire radio spectrum lies well below a normal synchrotron spectrum,
but contrary to NGC\,1377, it is higher than the expected free-free
component alone. At 21\,cm, \citet{Condon_atlas} measure a flux density
of $\approx 2-3$\,mJy inside a 21\arcsec-beam, which is a bit larger
than the size of the circumnuclear regions seen in the mid-infrared.
This value is consistent with the free-free emission derived from the
infrared spectrum of the central regions. As explained in Section~\ref{irsed},
we performed a decomposition of the infrared spectral energy distribution
into disk and circumnuclear (CNR) components, based on an empirical color
model and on the true distribution observed at 7 and 15\,$\mu$m. This allows
us to estimate the free-free and synchrotron emission separately for
the disk and the galactic center, as shown in Fig.~\ref{sed_radio} (note
that no component has been tuned in any way to the radio data in this
graphic).

The central regions thus appear almost devoid of any synchrotron emission;
otherwise, the 21\arcsec\ measurement would be higher. This finds support
in the total flux densities at 21\,cm, 6.2\,cm and 3.6\,cm being in agreement
with the synchrotron radiation derived for the disk alone (with a reasonable
spectral index assumed for illustration purposes to be -0.75). Extremely
diffuse extended emission from the disk is detected at all three wavelengths.
The 3.6\,cm map contains extentions to the south, which do not correspond
to any structure at other wavelengths and are thus suspect; if the flux
measurement is limited to the beam area rather than the total extention,
however, then one obtains $2.4 \pm 1.0$\,mJy instead of $3.6 \pm 2.2$\,mJy.
The data therefore are consistent with free-free emission and disk synchrotron
emission, but not with the central synchrotron component expected from a
normal supernova rate.
One may question this result, since a minimum level of emission would
be produced by cosmic rays originating in the disk and diffusing into
the central region. In $10^7$~years,
cosmic rays may propagate to about 1\,kpc from their source, using a
diffusion coefficient of a few $10^{24}$\,m$^2$\,s$^{-1}$, as found by e.g.
\citet{Strong_diff} and \citet{Wallace} in the Galaxy. Scaling the
disk synchrotron brightness, which is very low, by a reasonably large
size of 5\arcsec\ for the starburst, and by an increase of the magnetic
field strength from 10\,$\mu$G to 50\,$\mu$G (see Section~\ref{magfield}
for a justification of these numbers), we estimate the synchrotron
emission from disk cosmic rays to be at most of the same order as the
free-free emission within the central region, even at 20\,cm. Furthermore,
if the central magnetic field is much larger than the disk magnetic field,
then it will reflect cosmic rays coming from the disk \citep{Fermi}.

IC\,1953 was reobserved at 3.55\,cm with the VLA in the A configuration
($0.4\arcsec \times 0.25$\arcsec\ beam). The detected flux is
$0.80 \pm 0.05$\,mJy, and the size of the emitting region of the order
of $0.17\arcsec \times 0.09$\arcsec, with a position angle of 11\degr\
(J.J. Condon, unpublished). This compact component accounts for less
than half the peak flux density measured at 3.6\,cm with the Effelsberg
antenna, which is $2.0 \pm 0.6$\,mJy. It is consistent with normal
free-free emission provided that about 65\% of the flux comes from
a region larger than 0.3\arcsec, i.e. 30\,pc, which is very likely,
since the lower limit on the size of the starburst, obtained for a
blackbody of 70\,K matched to the CNR far-infrared emission, is 46\,pc.

\subsection{NGC\,4491}
\label{n4491}

Only for NGC\,4491 does a flux measurement at another radio wavelength exist.
\citet{Niklas_data} measured $S$(2.8\,cm$) = 5 \pm 1$\,mJy, with the
Effelsberg telescope, in a 69\arcsec-beam. Since the observed flux density
at 3.6\,cm in NGC\,4491 ($5.0 \pm 0.3$\,mJy), confirming the measurement
of \citet{Niklas_data} at 2.8\,cm, is in good agreement with what is expected
from the spectra of normal galaxies (4.9\,mJy for a non-thermal index of
-0.75 between 3.6\,cm and 21\,cm), and the free-free emission can contribute
only of the order of 20\% of the total emission at this wavelength, we have
evidence that NGC\,4491 produces synchrotron emission. However, this
emission seems to vanish at 21\,cm: the NVSS map indicates that
$S(20\,{\rm cm}) < 1.35$\,mJy. NGC\,4491 was not detected either in
the (much less sensitive) 12.6\,cm survey of \citet{Dressel}.
Source confusion could be responsible for this inconsistency.
Using the 8.4\,GHz source counts of \citet{Fomalont}, which indicate
$9 \times 10^{-4} < N(S > 4\,{\rm mJy}) ({\rm arcmin}^{-2}) < 2 \times 10^{-3}$,
and a beam of 80\arcsec, the probability of observing a confusing source
at 3.6\,cm lies between 1/360 and 1/800.
Given that the radio positions at 3.6\,cm and 6.2\,cm are in good agreement
with the optical center, as well as the position at 6.2\,cm of a bright quasar
seen to the SW in the NVSS map, and that the observed flux coincides
with the flux expected from the infrared emission, confusion is unlikely.

As NGC\,4491 lies only 54.5\arcmin\ away from M\,87, which emits
about 220\,Jy at 20\,cm, it is possible that the NVSS map does not
retrieve the correct flux density of NGC\,4491, due to imperfect
cleaning of the beam sidelobes which can generate negative sources.
To evaluate this effect, we selected a random sample of 50 NVSS fields
(centered on NED extragalactic sources), at distances to M\,87 between
52.5\arcmin\ and 56.5\arcmin, and looked in each one for the minimum
brightness. The histogram of the minima forms a smooth distribution
down to -6.52\,mJy per beam, and two outliers at $\approx -12$ and
-16\,mJy per beam. The latter are characteristically associated with
a chain of other deep negative points, of amplitude scaled by about
one third. Since this pattern is not present in the NGC\,4491 map, we
are confident that such a deep negative component cannot be superimposed
onto NGC\,4491 (furthermore, it would have to coincide accurately with
the galaxy's position, the probability of which is low). We thus adopt
as an absolute upper limit to any missing flux density the value
6.52\,mJy, and we derive $S(20\,{\rm cm}) < 7.9$\,mJy. We will use
this limit in the following discussion, but we insist that deeper
observations at 20\,cm are needed to set any more significant limit.

The new infrared to radio flux ratio is only constrained to be
$q > 2.65 = {\bar q} + 1.6 \sigma$. However, the spectral indices
are very flat: $S$([3.6--6.2\,cm]$) \propto \nu^{-0.34}$ and
$S$([6.2--20\,cm]$) \propto \nu^{\alpha}$ with $\alpha > -0.22$.
The second value is very uncertain, but the [3.6--6.2\,cm] index
is more reliable.
From radio continuum studies carried out at high angular resolution
\citep{Tarchi, SBeck, Turner, Kobulnicky}, several starburst galaxies,
e.g.  NGC\,2146, NGC\,4214, NGC5253, He\,2-10 -- for the most part with
Wolf-Rayet signatures and being either irregular, amorphous, blue compact
or merging dwarf galaxies -- have been found to contain sources with flat
or inverted radio spectra, revealing optically-thick free-free emission
regions. However, we find that considering {\it total} infrared and radio
fluxes, all of these galaxies have normal $q$ ratios (between 2.23 and
2.67). Moreover, NGC\,4491 emitting synchrotron emission at 3.6\,cm,
it is much more difficult to explain the shape of its radio spectrum
by free-free absorption, for the thermal electrons would have to fill
a very compact distribution, and at the same time encompass the
relativistic electrons. Assuming nevertheless that this is the case,
the effective opacity at 20\,cm would have to be larger than
0.7\,. \citet{Condon_ulig} found three ultraluminous galaxies with
$q > 2.9$, which are argued to be compact starbursts in merging systems,
and whose radio-quiescence is best explained by high free-free opacity.
However, these very luminous objects are not comparable at all with the
galaxies studied here. For two of them do stellar continuum observations
exist, and their 60\,$\mu$m to near-infrared power ratios are more than
20 to 40 times greater than in NGC\,1377, NGC\,4491 and IC\,1953.
Furthermore, the required free-free opacities are only 0.5--0.7 at 20\,cm.
Thus, NGC\,4491 is more likely intrinsically deficient in synchrotron
emission rather than opaque at 20\,cm.

The hypothesis of non-stellar activity powering the infrared and radio
emission will be discussed in Section~\ref{seyfert}.
Another possibility is that the population of relativistic electrons
is very young, so that no significant energy losses occurred yet.
We address here the likelihood of such a scenario.
The median spectral index of radio supernova remnants, which are the
likely sites of cosmic ray acceleration, is about $-0.5$ (with a total
distribution between $-0.2$ and $-0.8$), independent of age
\citep{Clark, Gordon}. This is significantly flatter than the spectral
index of diffuse synchrotron emission in galactic disks, $-0.8$
on average \citep{Niklas_nth}. Very flat indices of supernova remnants
can be reproduced in diffusive shock acceleration models, when
second-order Fermi acceleration is also included \citep{Droge}
or when backreaction on the shock due to the pressure of the
relativistic particles is taken into account \citep{Pelletier}.
Since the lifetime of radio supernova remnants is short,
$\approx 10^5$~years \citep{Ulvestad}, and they normally account for
a small fraction of the total synchrotron emission in a galaxy, it is
unrealistic to assume that the radio continuum emission of NGC\,4491
be due predominantly to flat-spectrum supernova remnants.
Steepening of the energy spectrum is a natural effect of synchrotron and
inverse Compton losses, once the relativistic electrons have escaped
their acceleration sites, but it gradually takes place over their whole
lifetime of several $10^7$~years. Cosmic-ray diffusion also contributes
to steepen the radio spectrum: with a diffusion length scaling as
$E^{-0.6}$ for $E > 1.5$\,GeV \citep{Engelmann}, the spectral index
is decreased by $-0.3$.

The flat radio spectrum of NGC\,4491 can therefore be explained if any
old population of cosmic rays has completely decayed, and a fresh
population has started building up within the last few Myr. This is
consistent with the observations presented in Section~\ref{n4491_pifs}.
Assuming constant spectral indices between each measurement pair, the total
radio power of NGC\,4491 between 3.6\,cm and 20\,cm is of the order of
74\% of the radio power expected of a normal spectrum, or less, owing
to the 20\,cm upper limit and to the fact that unobserved frequencies
lower than 1.4\,GHz were not included in the computed power. It is thus
conceivable that not all cosmic rays were yet released by supernova remnants
and that the interstellar medium of NGC\,4491 is out of equilibrium
(in which case the assumption of equipartition between cosmic ray and
magnetic energy densities is invalidated).

\section{Star formation indicators}
\label{mid_ir}

The H$\alpha$ recombination line, classically used as tracer of massive
stars, is not suitable to estimate the star formation rates of the three
galaxies, because nebular extinction in the circumnuclear core can be
expected to be high.
In a 2\arcsec-wide slit placed on the nucleus of NGC\,1377,
\citet{Kim} detect [NII] (6583\,\AA) and [SII] (6716 and
6731\,\AA) in emission, but not H$\alpha$, and furthermore a high
[SII]$_{\,6716}$ to [SII]$_{\,6731}$ ratio, indicating that the
detected nebular emission arises in low-excitation and low electronic
density regions, presumably in shocks. Modelling the starburst nucleus
of NGC\,4569, \citet{Gabel} found that the [SII] and [NII]
emission arises predominantly from low-density gas at large distances
from the starburst core; the same may apply to NGC\,1377.

IC\,1953 was observed in (H$\alpha$ + [NII]) by \citet{GarciaB},
who kindly provided their map to us. In the H$\alpha$ light, the
circumnuclear region is fainter than several star-formation
complexes in the disk, whereas it is of much higher surface brightness
than any other part of the galaxy in the 7 and 15\,$\mu$m dust emission.
The bar is not detected in H$\alpha$, which is not customary in late-type
galaxies, but more characteristic of strong bars.

The three galaxies do not feature in the optical range any indication of
a starburst, but have starburst-like far-infrared and mid-infrared colors.
In order to better constrain the star formation and supernova rates, we
undertook an observation program of near-infrared emission lines.

\subsection{Near-infrared spectroscopy}
\label{nirspec}

We observed NGC\,4491 on February 1 and 2, 2002, at the 5-m Hale telescope at
the Palomar observatory, with the Palomar Integral Field Spectrograph, PIFS
\citep{Murphy}. The hydrogen recombination lines Pa$\beta$ and Br$\gamma$
were chosen to estimate the star formation rate and the extinction in the
circumnuclear region, and the [FeII]\,1.644 line was used to set a constraint
on the supernova rate. We also observed NGC\,4102, as a bright comparison
galaxy with a very young circumnuclear starburst \citep{Jogee, barres}, but
normal infrared and radio properties; its $q$ ratio is 2.37, and
its central mid-infrared color, $F_{15}/F_7 = 3.1$, is comparable to those
of the three galaxies studied here.
On October 17-19, 2002, we pursued this program by observing NGC\,1377
in Pa$\beta$, Br$\gamma$, [FeII]\,1.644 and the H$_2$\,(1-0)\,S(1)
line at 2.12\,$\mu$m.
We chose as a second comparison galaxy NGC\,1022, a bright starburst
galaxy with a moderately high $q$ ratio (Sect.~\ref{general}).
In view of the raw data, we decided to observe also H$_2$\,(2-1)\,S(1)
at 2.25\,$\mu$m and H$_2$\,(1-0)\,S(0) at 2.22\,$\mu$m.
The observations and data reduction are described in Appendix~\ref{app_pifs}.
The spectra are shown in Fig.~\ref{spec} and the images and velocity maps
in Fig.~\ref{pifsima}. In both NGC\,4491 and NGC\,1377, the velocity
gradient derived from the line centroids is much smaller than the velocity
resolution.

\subsubsection{NGC\,4491}
\label{n4491_pifs}

Using the Pa$\beta$/Br$\gamma$ decrement as an extinction estimator,
we obtain $A$(Pa$\beta) = 0.8$ in the central $2\arcsec \times 2\arcsec$
of NGC\,4491, and $A$(Pa$\beta) = 2.4$ in the central $4.83\arcsec \times 4.83\arcsec$
of NGC\,4102. This is assuming an intrinsic Pa$\beta$/Br$\gamma$ ratio
of 5.58, valid for $T_e = 5000$\,K and $n_e = 300$\,cm$^{-3}$
\citep{Storey}, and adopting the extinction law of \citet{Cardelli}.
These extinctions are equivalent, in the H$\alpha$ line,
to respectively 2.4 and 7.3.

Although the absolute central star formation rate of NGC\,4491,
$(0.04 \pm 0.02)$\,M$_{\sun}$.yr$^{-1}$ using the calibration of \citet{Kennicutt},
is much less than that of NGC\,4102, $(7.6 \pm 0.3)$\,M$_{\sun}$.yr$^{-1}$,
the Pa$\beta$ equivalent widths of the two galaxies are more comparable:
we measure respectively 6.65\,\AA\ in $2\arcsec \times 2\arcsec$ (2.24\,\AA\
in $5.7\arcsec \times 5.7\arcsec$) and 9.16\,\AA.
NGC\,4102 is experiencing a powerful outflow (accounting for about one third
of the total H recombination line fluxes), at a velocity of
$\approx 800$\,km.s$^{-1}$, which confirms the youth of its starburst.

The [FeII]\,1.644 emission is thought to be due to shock excitation in
young supernova remnants \citep{Vanzi}. As supernov\ae\ appear only
a few Myr after the starburst onset, the ratio of the [FeII] line to
the hydrogen recombination lines, which are produced almost instantaneously,
rises as a function of time elapsed since the beginning of star formation.
The [FeII]/Br$\gamma$ flux ratio in the center of NGC\,4102, corrected for
extinction, amounts to about 1.3, which is in the low range of values
found in spiral starbursts, between 0.9 and 3 \citep{Moorwood}.
In the central $2\arcsec \times 2\arcsec$ of NGC\,4491, [FeII]/Br$\gamma$ is
constrained to be less than 0.5 at the $3\sigma$ level. Comparing this
to what is expected from the synthesis model of \citet{Leitherer}
for continuous star formation over a short period, with default parameters
(solar metallicity and Salpeter IMF between 1 and 100\,M$_{\sun}$),
the last star formation episode is less than 7.1\,Myr old.
In the same model, the starburst of NGC\,4102 is at least twice older
($\approx 16.1$\,Myr), but has not reached steady-state either.
NGC\,4491 is very faint, so our observations are not sensitive enough to produce
more stringent a constraint. However, the data are consistent with the idea that
NGC\,4491 is observed very shortly after the onset of a new episode of star
formation, even more recent than in NGC\,4102.

For the central $2\arcsec \times 2\arcsec$ of NGC\,1022, the galaxy used
for comparison with NGC\,1377, we derive $A$(H$\alpha) = 4.9$,
$EW$(Pa$\beta) = 13.88$\,\AA, and [FeII]/Br$\gamma \approx 1.6$. This galaxy
has not reached steady-state either, due to its moderately high infrared to
radio flux ratio (Section~\ref{general}), and hosts an extranuclear star-formation
complex about 2\arcsec\ to the north of the nuclear complex.

\subsubsection{NGC\,1377}
\label{n1377_pifs}

If most ionizing photons are intercepted by dust as suggested by the
radio spectrum of NGC\,1377, then the hydrogen recombination lines
severely underestimate the star formation rate.
In agreement with the lack of free-free emission, we detect none of the
hydrogen lines. In an area of $2\arcsec \times 2\arcsec$
centered on the nucleus, the $3 \sigma$ upper limits are
$F$(Pa$\beta) < 2.24 \times 10^{-19}$\,W.m$^{-2}$ and
$F$(Br$\gamma) < 4.70 \times 10^{-19}$\,W.m$^{-2}$.

We do not detect either the shock tracer [FeII], but a strong H$_2$\,(1-0)\,S(1)
emission line is observed. The H$_2$\,(2-1)\,S(1) to H$_2$\,(1-0)\,S(1) ratio
is less than 0.03 and the H$_2$\,(1-0)\,S(0) to H$_2$\,(1-0)\,S(1) ratio is
less than 0.07, which points to low-velocity shocks as the main mechanism
exciting the molecular gas, rather than fluorescence, unless the gas density
is high enough to de-excite the molecules by collisions \citep{Hollenbach}
(but the hydrogen nucleus density would have to be higher than
$10^6$\,cm$^{-3}$). The emission arises in a region smaller than 2\arcsec,
or 200\,pc. In the hypothesis of very high nebular extinction, the H$_2$
emission is probably foreground, but it could also originate from the molecular
material embedding the newly-formed stars. The young starburst
NGC\,1022 emits much less flux in H$_2$ than in any other line (Table~\ref{pifs})
and has a H$_2$\,(2-1)\,S(1) to H$_2$\,(1-0)\,S(1) ratio of the order of 0.2,
so that both shocks and ultraviolet photons can contribute to the molecule
excitation.

The lack of hydrogen emission in NGC\,1377 may be due partly to lack of
ionizing photons and partly to nebular extinction. In the assumption that dust
absorbs as much as $f = 95$\% of the ionizing photons, as suggested by the
lowest radio continuum upper limit (Sect.~\ref{n1377}), then the dust emission
could easily overestimate the star formation rate by a factor of two.
With $SFR_{\rm true} = SFR_{\rm dust} / (1 + f)$, our most stringent limit
on the recombination rate still implies a nebular absorption
$A({\rm Pa}\beta) > 3.7$, or equivalently $A({\rm H}\alpha) > 11$.
If only 70\% of the ionizing photons are absorbed by dust (Sect.~\ref{n1377}),
then $A({\rm Br}\gamma) > 2.4$, or $A({\rm H}\alpha) > 21$. Either the nebular
extinction is very high, or the quasi-totality of the ionizing photons are
absorbed by dust, i.e. no H{\small II} regions have yet appeared.

To prevent the development of H{\small II} regions, massive stars
have to still undergo an active accretion-outflow phase, which lasts less
than $10^5$\,years \citep{Walmsley}. This phase could be responsible for
shocking the molecular gas in NGC\,1377. \citet{Garay} discuss the properties
of a Galactic IRAS source which is a massive star-forming region, still
accreting material, and with no free-free emission, hence in a pre-H{\small II}
region stage. \citet{Osorio} modelled the dust emission from an accreting
OB star, and obtained spectral energy distributions strikingly similar to
the 3--100\,$\mu$m spectrum of NGC\,1377 in its qualitative features.
We thus propose that NGC\,1377 may host a proto-cluster of this type of
object, accounting for an unknown fraction of its infrared luminosity.
In this case, a large part of dust heating might not be provided
by the direct radiation from massive stars, but by the accretion luminosity
\citep{Osorio}, and in view of the luminosities computed for a range of
model parameters, of the order of $5 \times 10^4$ O stars would be needed to
generate the total infrared luminosity of NGC\,1377, i.e. a few times
more than for main-sequence stars.

The continuum emission in the PIFS field of view has the same morphology
and orientation as the large-scale stellar emission (Fig.~\ref{images})
and shows no disturbance.  The velocity gradient in the H$_2$\,(1-0)\,S(1)
line is much less than the velocity resolution, and the continuum emission
does not indicate any stellar density cusp. However, the continuum map
ratios reveal an emission excess in the Ks band, in the central 1--2\arcsec\
(slightly elongated parallel to the major axis). The color excesses with
respect to the surrounding bluer area, $\Delta (H_{1.655}-K_{2.142}) \approx 0.5$
and $\Delta (J_{1.29}-H_{1.655}) \approx 0.2$, are not compatible with pure
extinction ($A(H_{1.655}-K_{2.142}) / A(J_{1.29}-H_{1.655}) = 0.7$ using
the extinction law of \citet{Cardelli}), but consistent with the presence
of hot dust contributing to the Ks-band emission.
Assuming that the hot dust contributes only to the Ks-band flux, and not
to the J and H fluxes, we find that $F_{\nu}(7.7) / F_{\nu}({\rm K_{dust}})$,
the flux density ratio of the mid-infrared peak at 7.7\,$\mu$m to the Ks-band
excess, is of the order of 500. As a comparison, \citet{Lu} found an average
ratio of 150 in a sample of normal star-forming galaxies (see their Figure~6a).
The 6--8\,$\mu$m mid-infrared emission in NGC\,1377 is thus abnormally high
not only with respect to the far-infrared emission, but also with respect
to the hot dust emission in the Ks band, although its (H-K) color excess
is more extreme than in all but one of the galaxies studied by \citet{Hunt},
drawn from the same sample as the galaxies in \citet{Lu}.

\subsection{Star formation rate estimates}
\label{sfrate}

Table~\ref{sfr} lists the {\it total} star formation rates based on
an extrapolation of far-infrared fluxes to the 8--1000\,$\mu$m range
\citep{Kennicutt}, using the observed spectral energy distribution to
compute the 8--42\,$\mu$m flux and the model of \citet{Dale} to
estimate the contribution from 122 to 1000\,$\mu$m, as a function of
$F_{60}/F_{100}$.
Star formation rates in the circumnuclear regions can be derived
from the spatially-decomposed mid-infrared fluxes, applying the
calibration established in \citet{sfr}. IC\,1953 and NGC\,4491 are depleted
in 7\,$\mu$m emission with respect to the far-infrared, which could mean
that the aromatic band carriers are destroyed in these high dust-temperature
galaxies; the 7\,$\mu$m calibration is therefore not applicable to them
without correcting for the grain destruction effect. Considering that
galaxies from which the calibration was derived have $L_7 / L_{\rm FIR}$
ratios between 0.12 and 0.25, and that this quantity is about 0.02 in
the center of IC\,1953 and 0.04 in NGC\,4491, we estimate star formation
rates of respectively $1.65 \pm 0.60$\,M$_{\sun}$.yr$^{-1}$ and
$0.3 \pm 0.1$\,M$_{\sun}$.yr$^{-1}$. These numbers are not far below
the {\it total} star formation rates.
NGC\,1377, on the contrary, has a 7\,$\mu$m to far-infrared flux ratio
similar to normal and disk-dominated galaxies (Fig.~\ref{mirsfir}).
Although the 6--11\,$\mu$m spectrum of NGC\,1377 reveals that the dust
responsible for the emission in this range is peculiar, the star formation
rate derived from $F_7$ agrees to within 3\% with that derived from the
total dust emission for NGC\,1377 (Table~\ref{sfr}).

To further characterize the star formation activities in the central regions
of IC\,1953 and NGC\,4491, a comparison of their mid-infrared data with
those of circumnuclear starbursts and merging galaxies, for which
ionizing photon flux densities derived independently are available,
can be instructive. Using the circumnuclear regions of the galaxy sample
of \citet{atlas}, the starburst galaxies M\,82, NGC\,253 and NGC\,1808,
and the merging systems Arp\,220 and NGC\,6240 to span a range of more
than five orders of magnitude in star formation surface density, we found
a well-defined relationship between the mid-infrared color and the ionizing
photon flux density \citep{sfr_stb}.
7\,$\mu$m fluxes show no appearance of saturation and remain
valid star formation tracers up to ionizing photon flux densities as high as
$5 \times 10^{48}$\,s$^{-1}$\,pc$^{-2}$, but decline steadily above that
threshold \citep{sfr_stb}.
The colors of IC\,1953 and NGC\,4491 suggest that their star formation
is more compact than in the M\,82 core. If this assumption
is correct, this corresponds to ionizing photon flux densities above
$10^{16}$\,s$^{-1}$\,m$^{-2}$. Combined with star formation
rates derived from the total infrared emission, upper limits on the sizes
of the star-forming cores can be inferred: diameters of less than 125\,pc
and 70\,pc are respectively obtained. In the blackbody assumption,
we obtain lower limits of 46\,pc and 35\,pc.

The derived star formation rates are only of the order of
0.3--2\,M$_{\sun}$.yr$^{-1}$. However, the three galaxies have low masses.
Comparing these rates with the underlying stellar mass, as traced by the
Ks-band light, we find that the timescale to build up an equivalent mass
is only of the order of 0.5--2\,Gyr in the central $\approx 5$\arcsec\ of
NGC\,1377, IC\,1953 and NGC\,4491
(using $M/L_{\rm Ks} = 0.7\,M_{\sun}/L_{\sun \rm Ks}$,
as found by \citet{Grosbol}, with Ks$_{\sun} = 3.33$\,mag, from \citet{Colina}).
As a comparison, this timescale is similar for the starburst in NGC\,4102
(1.7\,Gyr) and much longer in the bright star-forming center of the normal
spiral NGC\,4535 (of the order of 10\,Gyr).
The star formation activity in the three galaxies can thus be truly
identified with starbursts.

Using the star formation rate range derived from dust emission, and comparing it
with the extinguished star formation rate derived from the H$\alpha$ line, we 
derive an absorption in H$\alpha$ between 4.3 and 4.9 for the circumnuclear
region of IC\,1953. If the [NII] to H$\alpha$ ratio were unusually high
such as in NGC\,1377, then the nebular absorption could be even higher.

For NGC\,4491, we have measured the star formation rate from near-infrared
hydrogen recombination lines in the PIFS field of view. The 7\,$\mu$m-emitting
region may be larger than this, but the strong discrepancy between the
recombination estimate and the dust estimate (Table~\ref{sfr}) points to
either a major fraction of the nebular emission being completely screened,
or an overestimation of the star formation rate by the infrared emission,
unless photospheric absorption from a young stellar population strongly
affects the hydrogen line strengths.

\subsection{Excitation of the interstellar medium}

Far-infrared lines due to collisional excitation in
photodissociation regions and ionized regions were observed by ISO-LWS
with an effective beam of 78\arcsec\ (LWS handbook). This beam is much
larger than the size of the infrared emission of NGC\,1377 and
NGC\,4491, but encompasses only 60\% of the total 7\,$\mu$m emission
of IC\,1953. Fluxes are collected in Table~\ref{lws}.

The [CII]\,157.7 line is the major coolant of photodissociation regions.
Since aromatic band carriers dominating the emission in the 7\,$\mu$m
bandpass are of planar structure, they are thought to be more efficient
sources of photoelectric heating than three-dimensional dust grains
\citep{Hendecourt}. This finds a confirmation in the constancy of the
[CII] to 7\,$\mu$m flux ratio found by \citet{Helou_cii} in star-forming
spiral galaxies. NGC\,4491 and IC\,1953 behave exactly as this sample:
their $F_{\rm [CII]}/F_7$ ratios are within $1.5 \sigma$ of the average value.
Their $F_{\rm [CII]}/F_{\rm CO}$ ratio is of the order of 2000--4000,
depending on the assumed CO spatial distribution, or less, owing to the
smaller CO beams in the observations of \citet{Boselli} and \citet{Combes}
(respectively 33\arcsec\ and 44\arcsec). These values are typical of normal
star-forming galaxies, and argue for near-solar metallicities, since
metal-deficient galaxies will have much higher ratios \citep{Madden}.
However, NGC\,1377 appears deficient in [CII] by at least an order of magnitude
and lies more than $6.5\sigma$ (in dex units) below the average $F_{\rm [CII]}/F_7$.
Interestingly, the 7\,$\mu$m flux in NGC\,1377, which is not produced by
classical aromatic particles, is also 0.85\,dex higher than expected from
its 60 and 100\,$\mu$m fluxes (Fig.~\ref{sed}).
The species responsible for the 7\,$\mu$m emission in NGC\,1377 is thus
either not easily photoionized, because already highly ionized in its
equilibrium state, or is heated mainly by low-energy photons, which is
unlikely owing to the observed high dust temperatures in this galaxy,
or is not an efficient purveyor of photoelectrons if its structure
is not planar.

C$^+$ is collisionally de-excited above relatively low densities, such that
the [CII] deficiency of NGC\,1377 could be due to increased density; but
[OI]\,63 remains an efficient cooling agent at densities 100 times higher.
Since the $F_{\rm [OI]}/F_{\rm FIR}$ ratio of NGC\,1377 is also low compared
to the values it takes in \citet{Malhotra}, this explanation can be ruled out.
\citet{Malhotra} also discard high optical depths in C$^+$ as the cause
of the [CII] deficiency (relative to the far-infrared flux) in their
high-$F_{60}/F_{100}$ galaxies, because they would expect [OI] to become
optically thick before [CII], which they do not observe. A plausible
origin of the [CII] deficiency in NGC\,1377 is thus a radiation field
intense enough to highly ionize the small dust particles emitting in the
mid-infrared, which then cannot heat the interstellar medium, and
different structural properties of these particles than in normal galaxies.
Alternatively, since H{\small II} regions seem to be absent in NGC\,1377
(Sect.~\ref{n1377} and \ref{n1377_pifs}), C$^+$ ions may not exist and
thus be unable to act as a coolant. In any case, the [CII] deficiency
is unusual and puzzling.

The upper limits on $F_{\rm [NII]}$ are within the range of detections
at lower dust temperatures of \citet{Malhotra}, when normalized by the
far-infrared flux.

\section{Molecular gas}
\label{sest}

NGC\,1377 was observed in the CO(1-0) and CO(2-1) lines at the SEST
in January 2003, during three sessions of 7\,h each, as part of the
ESO time share. Important details on the observations can be found
in Appendix~\ref{app_sest}. The measurements and results are given
in Table~\ref{tabco}.

We measure $\int {\rm T}_A^* {\rm dV} = (1.75 \pm 0.12)$\,K\,km\,s$^{-1}$
for the CO(1-0) line and a CO(2-1) to CO(1-0) brightness temperature
ratio of $0.53 \pm 0.14$. The spectra are shown in Figure~\ref{co_maps}.
The averages of the spectra of the five centermost
pointings are shown in Figure~\ref{co_tot}. The CO(2-1) line profile is
perfectly fitted by a gaussian of FWHM 66\,km\,s$^{-1}$; the CO(1-0)
line profile can be fitted by a gaussian of the same width and centered
at the same velocity, but shows a wing toward high velocities. This
excess is only at the $2 \sigma$ level. For lack of adequate constraints
on the H$_2$ mass to CO luminosity ratio, we adopt a standard value of
$2.3 \times 10^{24}$~molecules\,m$^{-2}$\,(K\,km\,s$^{-1}$)$^{-1}$
\citep{Strong}, which is uncertain by a factor of at least a few.
For infrared-luminous starbursts in which CO is subthermally excited,
assuming virialized clouds, \citet{Radford} derive factors between
about 2 and $7 \times 10^{24}$, in the same unit. Due to its low
CO(2-1) to CO(1-0) brightness ratio, collisional excitation of the
molecular gas in NGC\,1377 seems to be also subthermal.
\citet{Garay_co} similarly found very low CO(2-1) to CO(1-0) brightness
ratios in starburst galaxies, that suggest that most of the emission
comes from low-density and very clumpy gas \citep{Park}, while the
star-forming cores are probably hidden by their envelopes.
We derive a hydrogen mass of the order of $2.0 \times 10^8$\,M$_{\sun}$.
This is at least ten times more than the minimum mass required
by the starburst hypothesis (Sect.~\ref{n1377}).
The far-infrared to CO flux ratio is about seven times higher than
the average for normal galaxies, but within the broad range observed
for starbursts \citep{Sanders}, which is not surprising owing to the
high dust temperature and emissivity in NGC\,1377.
The total gas mass to blue luminosity ratio lies at the upper bound
of the range measured by \citet{Welch} for a volumed-limited sample
of lenticular galaxies. The timescale to consume all the molecular
gas, including the mass of helium and metals, at the current star
formation rate derived from the far-infrared emission, can be estimated
to only $\approx 170$\,Myr.

The CO(1-0) line was also observed in IC\,1953, NGC\,4491 and NGC\,4418
(see below a discussion of this galaxy), with smaller signal to noise
ratios \citep{Combes, Boselli, Sanders_n4418}. In all cases, the velocity
profile is very narrow (approximate FWHM of 120, 80 and 120\,km\,s$^{-1}$),
and the direct effect of rotation invisible, which argues for a
concentration of the molecular gas in a small central area.
The size of the molecular disk, even if it cannot be directly estimated,
can be coarsely constrained by indirect arguments. If one assumes that
NGC\,1377 must follow the same Schmidt law as characterized by
\citet{Kennicutt} from normal spirals to starbursts, within the observed
scatter, then the diameter of the gas disk should be ($150 \pm 300$)\,pc.
Dynamics can be used to set an independent
constraint, but one has to make the unverified assumption that the
CO disk is coplanar with the stellar disk, which could be false if
the gas was accreted from another galaxy. The maximum velocity, corrected
for inclination, is about 90\,km\,s$^{-1}$. In case the gas is decoupled
from stars (i.e. the stars are distributed in a much more extended
structure with a significant plane-perpendicular scale height), the
diameter is about 330\,pc. If the gas is confined within
this region, then it accounts for a large fraction of the total enclosed
mass: at least a quarter, estimating the stellar mass from the Ks image.

We now compute the Toomre parameter for local gravitational stability,
$Q_g = v_s \kappa / (\pi G \mu_g)$ \citep{Toomre}, in the same manner
as usually done for galactic centers, i.e. assuming a gas density
$\mu_g$ equal to the average, and the epicyclic angular velocity
$\kappa = 2 V / R$ in the linear part of the rotation curve. The
sound speed $v_s$ is chosen of the order of 10\,km\,s$^{-1}$, and the
diameter 4\arcsec\ (400\,pc). In these conditions, $Q_g \simeq 0.3$, which
means that the gas is locally very unstable. We neglected stars, although
they influence the gas stability, but they are expected to only decrease it
\citep{Jog}. The parameters entering $Q_g$ are all ill-constrained,
and there is no guarantee that the gas is distributed in a thin disk,
but if it is the case, since values close to unity are routinely found
in the literature (indicating gravitational stability allowing the mild
growth of perturbations), it is at least likely that the
molecular gas disk of NGC\,1377 be overcritical. This would explain
the instantaneous trigger of a starburst throughout $\approx 100$\,pc.
Sound evidence clearly requires more detailed data, difficult to
obtain owing to the low brightness of NGC\,1377.

Finally, the CO flux can be coupled with the H$_2$ line measurements
described in Section~\ref{n1377_pifs}, to estimate the mass fraction
of hot molecular gas. The H$_2$\,(1-0)\,S(0) upper limit does not
bring any useful constraint, but the lower limit on the
H$_2$\,(1-0)\,S(1) / H$_2$\,(2-1)\,S(1) flux ratio implies temperatures
below 1500\,K. With standard assumptions (due to lack of constraints
by the observations), the hot H$_2$ mass is then above 250\,M$_{\sun}$,
and the fraction of hot gas is above $\approx 10^{-6}$. Such a fraction
is in the normal range found for bright starburst galaxies and
cannot be further constrained.

\section{NGC\,4418: similarities and differences}
\label{n4418}

NGC\,4418, mentioned by \citet{Yun} as a high-$q$ galaxy, is an interesting
case for which very valuable data exist. Its far-infrared luminosity
is $\approx 5 \times 10^{10}$\,L$_{\sun}$, with $L_{\rm FIR}/L_{\rm B} \approx 35$,
$F_{60}/F_{100} = 1.44$ and $F_{25}/F_{60} = 0.21$. The NVSS map shows a
second source without any optical counterpart about 1.25\arcmin\ to the
east, which is certainly included in the IRAS fluxes; depending if this
source is included or not in the 21\,cm flux, the $q$ ratio of NGC\,4418
is 3.02 or 3.09, between 3.5 and $3.9\sigma$ above the average ratio.

\citet{Spoon} show that its mid-infrared spectrum presents a broad
feature similar to that observed in NGC\,1377, but with strong ice absorption
bands superposed, which are almost never observed. Other peculiarities
of this galaxy are the detection of the H$_2$ line at 2.12\,$\mu$m
in emission \citep{Ridgway} and of an OH megamaser \citep{Martin}.
\citet{Dudley} suggested that, like in Arp\,220, the infrared emission
arises from an extremely compact energy source.

NGC\,4418 is also deficient in [CII], but $F_{\rm [CII]}/F_7$ is only
constrained to be more than $2\sigma$ below the average ratio in
\citet{Helou_cii}. There are two other major differences with respect
to NGC\,1377: NGC\,4418 has much lower mid-infrared to far-infrared flux
ratios, in agreement with the predictions from its high dust temperature,
and a very compact radio source is detected at its nucleus, with a
brightness temperature of $10^{4.9}$--$10^{5.0}$\,K both at 2.3\,GHz
inside 0.1\arcsec\ \citep{Kewley} and at 1.49\,GHz inside
$0.5\arcsec \times 0.3\arcsec$ \citep{Condon_atlas}. NGC\,4418 is not
detected in X-rays \citep{Cagnoni}. Although the compactness and extreme
extinction in NGC\,4418 suggest the presence of a Seyfert nucleus, a
highly compact starburst is also plausible, and the nuclear 2.2\,$\mu$m
emission is shown to be more extended than the PSF by \citet{Eales}.
The 3.55\,cm radio continuum, observed with the VLA in a
$0.4\arcsec \times 0.25$\arcsec\ (J.J. Condon, unpublished) confirms
that the radio source is extended, with a size of
$0.20\arcsec \times 0.16$\arcsec.

During our radio continuum observations of NGC\,1377, NGC\,4491 and
IC\,1953 at Effelsberg, we also mapped NGC\,4418 (Fig.~\ref{sed_radio}).
We measured ($33.3 \pm 1.3$)\,mJy at 6.2\,cm and ($19.7 \pm 1.2$)\,mJy
at 3.6\,cm (the beam including the source located 1.25\arcmin\ eastward).
Combined with the NVSS flux at 21\,cm, ($47.9 \pm 1.8)$\,mJy (with the
contaminating eastward source added to NGC\,4418), this
indicates a pronounced spectral bend from a steep index at high frequencies
($S_{\rm [3-6\,cm]} \propto \nu^{-0.94}$) to a flat index at low frequencies
($S_{\rm [6-21\,cm]} \propto \nu^{-0.30}$). The implied high free-free opacity
is consistent with the high extinction and compactness of NGC\,4418.
We find that the combined effect of both thermal opacity and a reduction
of the intrinsic synchrotron emission (with respect to the level expected
from the far-infrared emission) provides the best fit to the spectrum.
The lowest possible opacity and most reasonable non-thermal spectral
index, $\tau$(21\,cm$) = 0.9$ and $\alpha_{\rm nth} = -1.17$, are
obtained for an intrinsic deficit of synchrotron emission of about 50\%.
Diminishing the synchrotron deficit in the model causes the opacity
to rise and the spectral index to become steeper; assuming no deficit,
$\tau$(21\,cm) would have to be as high as 1.6 and $\alpha_{\rm nth}$
as low as -1.57\,.

J.J. Condon obtained with the VLA-A a higher flux at 3.6\,cm: 27.4\,mJy,
including the contaminating eastward source. Using this measurement
instead of the single-dish flux, we obtain a maximum opacity
$\tau$(21\,cm$) = 0.5$, the steepest possible non-thermal spectral
index is $\alpha_{\rm nth} = -0.7$, and in such conditions, the intrinsic
synchrotron emission is reduced by 70\% with respect to a galaxy with normal
$q$ ratio.

Based on this new result, and on the other existing evidence, we argue
that the infrared emission of NGC\,4418 originates from a very young and
compact starburst. Seyfert activity is not necessary to explain the data.
The emission line properties of NGC\,4418 are also strikingly similar to
those of NGC\,1377. It was observed with the PIFS spectrograph by
\citet{Dale_pifs}, who detected neither Pa$\beta$, Br$\gamma$ nor
[FeII]\,1.644, but a clear H$_2$\,(1-0)\,S(1) line. This fact strengthens
the similarity between the two galaxies, and suggests a common excitation
mechanism; since NGC\,4418 is a synchrotron source, though, it must
have reached a later evolutionary stage than NGC\,1377, and is possibly
a composite source containing both a nascent starburst and a more evolved
component.

\section{Scenarios for the synchrotron deficiency}
\label{scenarios}

We now discuss plausible physical scenarios to account for the high
infrared to radio ratios of the three galaxies we have studied. If they
are due to a deficit of radio emission, then the primary possibility
is a lack of cosmic rays, which would follow either from a nascent
starburst or from an initial mass function biased towards low-mass
stars; the alternative is a very weak magnetic field.
If the radio emission is not depressed, then it might be absorbed,
pointing to optically-thick 20\,cm emission. Finally, we discuss
the possibility of non-stellar activity providing dust heating
but remaining radio-quiet.

\subsection{Nascent starbursts}

The hypothesis that we favor to explain the infrared activity and the
radio deficiency of NGC\,1377, NGC\,4491 and IC\,1953 is the onset
of a star formation burst in their centers. In this view, the synchrotron
weakness derives from a lack of relativistic electrons, which can happen
if the previous major star formation episode occurred more than $\approx 10^8$
years ago, and the current episode started less than a few (1 to 7) Myr before
the epoch of observation. In NGC\,1377 and the central region of IC\,1953,
no synchrotron component is detected, which implies in this scenario that no
or very few supernov\ae\ have yet been produced. In NGC\,4491, the synchrotron
component does exist but has a much flatter than usual spectral index.
We interpret this as due to an energy distribution of cosmic rays inflated
with high energies, rather than the effect of free-free opacity; the number
of relativistic electrons released by supernov\ae\ is significant, but may
still be rising, and the near-infrared spectroscopic results indicate that
shocks are weaker than expected from a steady-state situation.

Other observations consistent with this scenario are: the high dust
temperatures, indicating starburst-like radiation intensities; the extreme
$F_{15}/F_7$ flux ratios of NGC\,4491 and IC\,1953, suggesting that very
young stellar populations are dominating the dust heating; the mid-infrared
spectrum of NGC\,1377, which suggests that dust nanoparticles have not
been processed into classical aromatic components radiating in normal
star-forming galaxies.

\subsection{Initial mass function}
\label{imf}

Since the mass spectrum of stars able to heat the dust extends to much lower
values than that of stars producing supernov\ae, high infrared to radio flux
ratios could be achieved if the initial mass function had a very steep slope,
making high-mass stars rarer than in normal galaxies. In this hypothesis, the
extreme dust temperatures would however require extreme stellar densities.
To be compatible with the 21\,cm upper limit of NGC\,1377, the cutoff of a
truncated Salpeter IMF would have to be less than $8\,M_{\sun}$, since
supernov\ae\ have to be totally suppressed. However, according to the model
of \citet{Desert}, the $F_{60}/F_{100}$ ratio of NGC\,1377 implies an average
radiation field about 400 times higher than the solar neighborhood field
throughout the dust emission region. Reaching this radiation field with only
$M < 8\,M_{\sun}$ stars is impossible to reconcile with the data, as we will
show with a simplistic model assuming a uniform stellar distribution. The Ks-band
image gives us a constraint on the stellar mass density. The H$\alpha+$[NII] map
of \citet{Heisler_ha}, of better resolution than the infrared maps, delineates a
central region of radius $\approx 5$\arcsec. Using the model of \citet{Leitherer},
we computed the bolometric power per unit of stellar mass produced by a star
formation episode lasting 10\,Myr (IMF-truncated) and by an older population
of 1\,Gyr, as well as their mass to light ratios in the Ks band (close to
1\,M$_{\sun}/{\rm L}_{\sun \rm Ks}$), and then the resulting radiation field
in the center as a function of the mass fraction of the younger population.
Using the value of the solar neighborhood radiation field given by \citet{Mathis},
we find that the radiation field is high enough only if more than 80\% of the
mass is locked in the IMF-truncated starburst, hence the mass of any old
stellar population is negligible. Since this is unrealistic, an abnormal
IMF is unlikely to cause the observed synchrotron deficiency in NGC\,1377.
In IC\,1953, the average radiation field intensity implied by the dust
temperature is only 100 times the solar neighborhood value (in fact more
than 200 times in the circumnuclear region, according to the decomposition
shown in Fig.~\ref{sed}), but the required stellar density is still much higher
than that observed, unless more than 65\% of the central mass had been generated
by a recent starburst.

\subsection{Magnetic field and inverse Compton losses}
\label{magfield}

The hypothesis of a weak magnetic field remains difficult to rule out.
However, if the magnetic energy density of some galaxies happened to be
10--40 times lower than in normal galaxies, it would be difficult to explain
why this is so rare and why this deficiency is connected with high dust
temperatures, which are observed in all high-$q$ galaxies, unless it is causally
linked with the occurrence of a starburst. If the magnetic pressure is unusually
low, then cloud collapse is more efficient and a starburst can be triggered
at relatively low gas densities; however, the question remains of what caused
the low magnetic pressure, and why the starburst was not triggered before.
It is also possible, if magnetic fields are usually amplified by a starburst
due to increased turbulence, that there is a time delay for the adjustment
of the magnetic field; this would become a likely scenario if we detected
significant shock excitation by supernov\ae\ by means of the [FeII] emission
line. Note however that the radio spectra of NGC\,4491, NGC\,4418 and the
disk of IC\,1953 bring evidence of a normal magnetic field strength,
because they show substantial synchrotron emission, and that no signs
of supernova shocks exist yet for NGC\,1377.

Random magnetic fields of a few $\mu$G, about half the typical strength
of total magnetic fields in spiral galaxies, are expected from turbulence
alone, even in the absence of large-scale dynamo action. Field strengths
of several $\mu$G are easily obtained in elliptical galaxies although
they are devoid of rotation \citep{Lesch, Moss}, comparable to strengths
in spirals. Unbiased statistical information about synchrotron emission
of ellipticals is uneasy to find, but they are much weaker radio emitters
than spirals, and their activity is mostly due to Seyfert nuclei
\citep{Condon_ell}; their star formation rates are too low to produce
enough cosmic rays.
Magnetic field amplification by large-scale dynamo processes occurs
on timescales as large as $5 \times 10^8$~years \citep{Beck_magnetic}.
However, the growth of random fields requires much less time.
Depending on the turbulence characteristics, in particular the
cut-off scale, the fastest-growing modes in a predominantly neutral
medium may develop on timescales as small as $10^4$~years, according
to \citet{Subramanian}.

As pointed out by \citet{Lisenfeld}, the fact that quiescent galaxies
and starbursts follow the same infrared-radio correlation, despite
the large expected differences in inverse Compton losses, suggests
that the total magnetic field grows subsequent to a starburst.
If inverse Compton losses are dominant over synchrotron losses,
the lifetime of relativistic electrons is inversely proportional
to the radiation energy density. Assuming a size of 100\,pc for the
infrared-emitting region of NGC\,1377 (see Sect.~\ref{n1377}), the
radiation energy density integrated between 3 and 120\,$\mu$m, at the
surface of the region, is of the order of $5 \times 10^{-10}$\,J\,m$^{-3}$.
In the ultraluminous compact starbursts studied by \citet{Condon_ulig},
the typical radiation energy density is estimated to be six times higher
than this, and yet they are strong radio continuum emitters, so that
synchrotron losses must at least balance inverse Compton losses: 
mechanisms at work to amplify magnetic fields seem to conspire
to produce a constant ratio of magnetic energy density to radiation
energy density. This must occur on timescales comparable to or shorter
than the growth time of the supernova rate, otherwise significant
deviations from the correlation would be observed.

The issue of magnetic field strength is unsettled for NGC\,1377,
but we would like to stress that a weak field is not required to
explain the synchrotron deficiency, since it is easily achieved by
lack of cosmic rays, consistent with the fact that no supernova
activity is observed. Furthermore, a separate cause has to be sought
to explain the absence of free-free emission, which is unaffected by
the magnetic field strength.

Alternatively, if the magnetic field in a galaxy were amplified
beyond normal values, it would cause more intense radio emission for the
same injected population of cosmic rays, since the synchrotron emissivity
scales as a large power of the field strength (index between 2 and 4).
As a result, the infrared-to-radio ratio would decrease, but the radio
emission would fade faster after an episode of star formation.
If the magnetic field became so strong that the cosmic rays decayed
very fast, the synchrotron emission would appear in intense bursts
followed by periods of radio quiescence.
In a steady-state situation, allowing some uncertainty range for the
IMF slope, one would expect from the infrared luminosity of NGC\,1377
a type-II supernova explosion, on average, at least every 100 years.
Thus, cosmic rays should be injected into the interstellar medium,
after a time delay, at this average rate.
For the lifetime of cosmic-ray electrons to be reduced to 100 years or less
at GHz frequencies, a magnetic field of the order of 50\,mG would be
required over large portions of the interstellar medium
\citep{Condon_review}, that is to say $5 \times 10^3$ times
higher than the typical strength in spiral galaxies \citep{Beck_review}.
No amplification mechanism is known to achieve such a result on large
spatial scales, and such a strong field would clearly not be confined
\citep{Parker}. Invoking such absurd values of the magnetic field
is much less credible than the nascent starburst hypothesis.
As a comparison, \citet{Carilli} estimate fields
strengths up to 350\,$\mu$G in the radio-loud galaxy Cygnus~A, using
the assumption of minimum energy density for the sum of cosmic rays
and magnetic fields. Starburst galaxies may have magnetic fields of
a few 10\,$\mu$G, generally calculated under the same mimimum-energy
assumption; for instance, \citet{Klein_m82} derived an equipartition
total field of 50\,$\mu$G in M\,82.

\subsection{Electron opacity}

We can discard high opacity at 21\,cm as the reason for radio weakness,
because it would require NGC\,1377 and IC\,1953 to have respectively
$\tau$(21\,cm$) > 2.8$ and $\tau$(21\,cm$) > 0.5$, only in order to have
an intrinsic $q$ of ${\bar q} + 2\sigma$, and NGC\,4491 would have to
feature an unrealistic geometry. High opacities are very unlikely,
since the high-$q$ ultraluminous compact starbursts studied by
\citet{Condon_ulig} (with $q > {\bar q} + 3\sigma$), much more active
and dense, have estimated opacities of only 0.5--0.7.

\subsection{Seyfert activity}
\label{seyfert}

The infrared-radio properties of NGC\,1377 and NGC\,4491 may be
suggestive of the presence of active nuclei in these galaxies.
However, spectral diagnostics do not support this view, since
emission lines are either absent or very weak and narrow.

Two indirect arguments can be put forward to support the presence of a
Seyfert nucleus: warm infrared colors, and the compactness of the infrared
emission. We consider the argument of deep silicate absorption in the
9--11\,$\mu$m range in NGC\,1377 speculative, especially since the continuum
emission longward of 11\,$\mu$m has not been observed, and because it
would be difficult to reconcile with the moderate dust to stellar emission
power ratio; diagnostics based on optical spectroscopy cannot be applied.
However, the above two characteristics support equally well the hypothesis
of a young starburst. The mid-infrared spectrum is an extremely ambiguous
diagnostic, since it does not show the aromatic bands associated with
star formation, but does show an unusual emission feature, in place of
the continuum expected in a pure Seyfert nucleus. Concerning the infrared
broadband colors, \citet{Yun}, for instance, have shown that they are not
valid discriminants between starburst and Seyfert activities -- even if a
larger percentage of energetically-dominating Seyfert galaxies is found in any
sample pre-selected to have warm colors, than in a sample with cooler colors.
The concentration factor is not a strong argument either, since a young
starburst will be compact (to trigger a star formation burst instantaneously
over a large area requires uncommon conditions); the data for NGC\,1377
suggest that the size could be as small as $\approx 100$\,pc, i.e. 1\arcsec\
(Sect.~\ref{n1377}), and is smaller than $\approx 2$\arcsec, i.e. 200\,pc
(Sect.~\ref{n1377_pifs}). In addition, Seyfert
nuclei powerful enough to dominate dust heating tend to occur in massive
galaxies with a deep potential well, whereas NGC\,1377 is a low-mass galaxy
and with a shallow potential.  But perhaps the most problematic fact
for Seyfert activity is the radio deficiency.

Many active nuclei are known to be radio-quiet. However, their $q$
ratios are the same as in normal galaxies. For instance, the
infrared-loud and radio-quiet quasar studied by \citet{Beichman} has
an observed $q = 2.48 = {\bar q} + 0.75 \sigma$, even if it is strongly
absorbed at 20\,cm. In the sample of \citet{Ho}, containing a large
number of radio-quiet active nuclei, no nucleus has a
$q$ ratio significantly higher than the average for normal galaxies;
all the Seyfert-type galaxies which are not detected at 20\,cm are not
detected by IRAS either. We have further verified this fact by selecting
in the NED database all galaxies in the quasar category with
$F_{60} > 0.5$\,Jy and no mention of a radio association; on the 53
objects thus selected, 7 were excluded for lack of any radio observations,
and all the others have either normal or low $q$ ratios.
The infrared brightness of NGC\,1377 thus presents serious difficulties
for the hypothesis of a radio-quiet active nucleus.

An examination of the energetics provides another argument against
the production of the infrared power of NGC\,1377 by an active nucleus.
Assuming that the radiated power is equal to the Eddington limit,
$L_{\rm E} {\rm (W)} = 1.3 \times 10^{31} M_{\rm BH}$ (M$_{\sun}$),
and that it goes entirely into dust heating, an upper limit on
the black hole mass provides an upper limit on the infrared power.
For a given black hole mass, the radio loudness can vary considerably
(e.g. \citet{Woo}); but for radio-quiet active nuclei exclusively,
estimating black hole masses from reverberation mapping techniques,
\citet{Nelson} found a correlation between the black hole mass and the
radio power, with a lower envelope of all their data of the form 
$P{\rm (20\,cm) (W\,Hz}^{-1}) = 10^{15.2} (M_{\rm BH}/M_{\sun})^{0.84}$.
Using the 20\,cm upper limit for NGC\,1377, the alleged
black hole cannot be more massive than $2.4 \times 10^5$\,M$_{\sun}$.
Using alternatively the kinematic information derived from the
H$_2$ emission line (Sect.~\ref{n1377_pifs}), another limit can
be set. With a velocity gradient below 40\,km\,s$^{-1}$,
the inclination-corrected rotation velocity is
less than 25\,km\,s$^{-1}$ (with a signal to noise ratio of
$\approx 5$ throughout the central 2\arcsec). Assuming virial
equilibrium, and provided the H$_2$ lies in the plane of the galactic
disk, we then do not expect more than $1.5 \times 10^5$\,M$_{\sun}$
to be enclosed in the central parsec. Using the first upper limit,
we thus find that the maximum infrared power (in extremely unrealistic
conditions) radiated by an active nucleus is $3 \times 10^{36}$\,W.
In fact, it is expected to be a factor of several below this value,
because the dust heating efficiency cannot be 100\%. The infrared
luminosity of NGC\,1377 being $3.9 \times 10^{36}$\,W, an active nucleus
cannot easily account for it.

The absence of synchrotron emission might be attributed to lack of
magnetic field (Sect.~\ref{magfield}), or extreme inverse Compton losses,
due to the high radiation field intensity in NGC\,1377. Nevertheless,
an active nucleus must also produce some free-free emission, which is not
detected either. Inside the region where dust grains are sublimated,
the Lyman continuum photons are free to ionize the gas, contrary to the
situation in an embedded starburst. \citet{Elvis} have compiled spectral
energy distributions of quasars and estimated, in particular, the ionizing
photon flux and the total power between 1000\AA\ and 100\,$\mu$m. The
ratio between these two quantities is
$N_{\rm ion} / L_{UV-IR} = (8.1 \pm 2.9) \times 10^{16}$\,s$^{-1}$\,W$^{-1}$
(or $L_{\rm ion} / L_{UV-IR} = 0.5 \pm 0.2$).
Taking the infrared luminosity of NGC\,1377 as a lower limit to $L_{UV-IR}$
and assuming that the ionizing photon flux available in the dust-free region
is equal to $N_{\rm ion} / 2$, the free-free emission at 3.6\,cm, where
thermal opacity is negligible, should amount to $\approx (2.8 \pm 1.0)$\,mJy
(with $T_{\rm e} = 5000$\,K). It would thus have been detected, especially
by the sensitive observation with the VLA in the A configuration. The
single-dish upper limit is 2\,mJy, and the VLA upper limit is 0.13\,mJy.
Although we cannot rule out completely the presence of an active nucleus,
because the properties of NGC\,1377 remain challenging,
the above arguments rather support our hypothesis of a nascent starburst.

The properties of NGC\,4491 may in turn be indicative of non-stellar
activity, because some active nuclei are known to have flat radio spectra.
Dust heating by an active nucleus is however very unlikely, owing to the
following facts: NGC\,4491 has no bulge; the velocity gradient inferred
from the Pa$\beta$ line (see Section~\ref{nirspec}) is much less than
100\,km\,s$^{-1}$, which confirms that there is no central gravitational
cusp (the inclination on the line of sight is about 62\degr); optical and
near-infrared emission lines do not give any indication of an active nucleus;
the dust spectral energy distribution is fully consistent with the properties
of star-forming galaxies, whereas one would expect a high mid-infrared to
far-infrared power ratio if dust were heated by an active nucleus.

\section{Frequency of occurrence}
\label{stat_highq}

Other high-$q$ galaxies, deviating by more than $3\sigma$ from the average
infrared to radio flux ratio, are known, though NGC\,1377 is the most
extreme case. Among the sample studied by \citet{Vader} in a series of
papers, comprised of galaxies whose flux density is higher at 60\,$\mu$m
than at 100\,$\mu$m, 31 lie in the NVSS sky coverage (including NGC\,1377);
of these 31, we find that five galaxies satisfy the above high-$q$ criterion.

Of the 40 very luminous galaxies belonging to the IRAS bright galaxy
sample ($F_{60} > 5.24$\,Jy and log($L_{\rm FIR}/L_{\sun}) > 11.25$),
\citet{Condon_ulig} have found three more high-$q$ galaxies, which are
distant (200 to 300\,Mpc) and whose flux densities peak at 60\,$\mu$m;
they classify them as compact starbursts. The activity of two of them
is obviously driven by on-going mergers, with double nuclei separated
by less than 5\arcsec\ \citep{Scoville}, and the third galaxy presents
a strong warp \citep{Lehnert}. However, \citet{Condon_ulig} claim that
their radio deficiency is best explained by high free-free opacity at
20\,cm, rather than a lack of cosmic rays.

Nine galaxies with $q > 3$ are mentioned by \citet{Yun} in a complete
sample of 1809 galaxies with $F_{60} > 2$\,Jy (i.e. with an occurrence
rate of 0.5\%), two of which in common with \citet{Condon_ulig}.
However, reexamining these galaxies, we find that five of them are false
high-$q$ detections, due partly to confusion affecting IRAS fluxes,
and partly to NVSS catalog fluxes missing low-surface brightness
or extended structures. On the other hand, \citet{Yun} have missed
NGC\,1377, NGC\,4491 and two galaxies of the sample of \citet{Vader},
which, according to the selection criteria of \citet{Yun}, must be
part of their sample and have $q > 3$. Because of inevitable errors
due to automatic handling of very large samples, the census of high-$q$
galaxies they provide is therefore uncertain. The two new high-$q$
galaxies that they found have $F_{60}/F_{100}$ ratios of $\approx 0.9$
and 1.44\,.

In view of the above numbers, it seems that the most efficient way of
finding synchrotron-deficient galaxies is to select high dust
temperatures ($F_{60}/F_{100} \ge 0.8$ for all known high-$q$ galaxies),
and that the infrared luminosity matters much less.
The occurrence rate in the color-selected sample of \citet{Vader} is
as high as 16\%. 60\,$\mu$m-peakers being scarce objects, however,
the implied probability for a random local galaxy to be synchrotron-deficient
must be very low. The flux-limited sample of \citet{Condon_atlas}
and \citet{Condon_atlas2} (500 galaxies with $F_{60} > 5.24$\,Jy)
contains at least 6 high-$q$ galaxies identified in the above various
samples, which implies a rate of the order of 1\%.

The incidence of high dust temperatures is an additional argument in
favor of the nascent starburst hypothesis that we have developed for
NGC\,1377, IC\,1953 and NGC\,4491. In this scenario, the scarcity
of high-$q$ galaxies derives naturally from the timescales involved.
The likelihood for a galaxy to have a high $q$ ratio is the product
of the probabilities of two occurrences: it has to have been quiescent
for at least $10^8$~years, and a starburst has to have started less
than 3--4\,Myr before the epoch of observation.
Less than $\approx 15$\% of optically-selected galaxies have
infrared color excesses indicative of on-going intense star formation,
as apparent from the analysis of \citet{Huang} (of their ``HP sample'');
from this we may infer that galaxies spend, on average, less than
15\% of the time in starburst events. Given a typical starburst
duration of $f \times 4$\,Myr, with $f \geq 1$, the second probability
listed above is necessarily lower than $0.15 / f$. While the probability
of the first occurrence (quiescence for $10^8$~years) is difficult to
assess, it is sufficiently unlikely that high-$q$ systems should
be rare, consistent with the statistics.
However, at high redshifts, before the epoch of maximum star
formation density, the probability of observing radio-deficient galaxies
may be higher. It is unknown beyond what redshift metallicity
effects, reducing the amount of dust and thus of infrared emission, will
start to compete with the nascent starburst effect. \citet{Garrett}
recently estimated $q$ ratios of distant galaxies ($z$ up to 1.3) and
found no significant deviation of the average from the local value.

\subsection{Triggering mechanisms}

The data shown in this paper are consistent with the high infrared
to radio flux ratios being caused by a delay in either the magnetic field
strength or the cosmic ray density, in either case a consequence of a
nascent starburst.

In contrast to the above infrared-luminous galaxies, NGC\,1377,
NGC\,4491 and IC\,1953 have a shallow potential well, such that the
triggering mechanism of a central starburst is not obvious. In IC\,1953,
tidal interaction, likely at work, could be responsible for
strengthening the bar perturbation (which seems unusually strong for
the morphological type of IC\,1953) and causing massive gas inflow
due to the bar potential. In this view, we would be witnessing the
first burst caused by the bar, and IC\,1953 will likely have an
earlier-type aspect afterwards. NGC\,4491 is also strongly barred,
and this bar may also have grown due to tidal forces with other
Virgo galaxies. However, no evidence of a bar nor of an intrusion
exists in the most extreme galaxy of the three, NGC\,1377. 
Are tidal interactions experienced by galaxies in clusters and groups
efficient at producing gas inflows inside them, even in the absence
of any visible potential perturbation~? Other HI-deficient and
non-Seyfert Virgo galaxies, besides NGC\,4491, indeed have surprisingly
high mid-infrared colors in their centers: N4293, N4569, N4579
\citep{barres}. Note that all these galaxies either lie close to M\,87
in projection and have radial velocities more than 600\,km\,s$^{-1}$
away from the velocity centroid \citep{Cayatte}, or lie at the periphery
of the cluster in projection, which could mean that they are part
of a very turbulent subsystem. \citet{Helou_bcd} found such an
environmental effect in the Virgo unrelaxed subsystem called W,
where blue compact dwarf galaxies are more numerous and have hotter
dust than in the rest of the cluster.

Hidden perturbations are best revealed
by observing the HI morphology and velocity structure, but NGC\,1377
seems to have too little HI gas for such a diagnostic to be applied.
An assessment of the nature and cause of the activity seen in the three
galaxies will have to wait for complementary data.

\section{Summary and final remarks}
\label{summary}

In order to understand the cause of the synchrotron deficiency
observed in the three closest high-$q$ galaxies, we have presented
and analysed extensive data in several windows: broadband dust
emission, radio continuum at several frequencies, emission line
diagnostics of photon ionization, supernova shocks and hot molecular
gas excitation, and emission of the cold molecular gas.

We argued that the galaxies are seen at the very onset of a
starburst episode, following a long period of quiescence, based
on these data, and consistent with the statistics of high-$q$
galaxies inferred from starburst timescales. In this scenario,
cosmic rays generated by past star formation had the time to
decay, and massive stars produced by the current star formation
episode have not evolved into supernov\ae\ yet (NGC\,1377, central
region of IC\,1953) or, in the case of NGC\,4491, the supernova rate
is still rising and the cosmic ray population is in its formative
stages.

All three galaxies host unusually hot dust, requiring an intense
radiation field consistent with that produced by a young starburst.
NGC\,1377, whose properties are the most peculiar, is not only
devoid of synchrotron radiation, but also of free-free emission.
We propose that the starburst is so young that massive
stars are still accreting material, preventing the development of
H{\small II} regions. Along the same line, dust emission in the
mid-infrared has an unusual spectral signature, suggesting that
grains were processed differently from grains found in almost
every other star formation environment; the molecular gas disk
may be overcritical, which would explain the quasi-instantaneous
trigger of a starburst. While this is the most likely explanation,
the properties of NGC\,1377 remain very intriguing.

We can rule out abnormal initial mass function, electron opacity and
fast synchrotron losses as causes of the depressed radio continuum.
Other scenarios are not formally ruled out, but create serious
difficulties: radio-quiet Seyfert activity is unlikely owing to
the energetics arguments discussed in Section~\ref{seyfert}; a weak
magnetic field (relevant only for NGC\,1377) is possible, but
difficult to explain in light of what is known of turbulent field
amplification, and still requires a young starburst after a period
of quiescence.

We would like to stress that the galaxies studied in this paper
are not akin to irregular or blue compact dwarf galaxies,
which would at first sight be easily susceptible of synchrotron
deficiency for various reasons: fast cosmic ray escape and
interstellar medium disruption and loss; evidence of
optically-thick free-free emission in some of them.
Although these galaxies seem to have much more dispersed
infrared to radio flux ratios than spirals, they follow the
same average relationship and none of them is known to deviate
from it significantly.

Among galaxies of the IRAS bright galaxy sample, selected by their
60\,$\mu$m flux, the occurrence rate of high-$q$ galaxies (in our
interpretation nascent starbursts) is of the order of 1\% or
slightly more. The best way to select these galaxies is to search
for high dust temperatures; no other simple criterion is evident.
Although very few examples are known, from this phenomenon arises
a potential problem for the identification of distant infrared
galaxies with high-resolution radio observations, especially
with possibly changing star formation timescales at high
redshifts.

The galaxies studied here offer a unique glimpse at starbursts
during a critical period of their evolution, and will allow us
to address questions concerning the pre-starburst accumulation
of molecular material, the triggering mechanisms, the spatial
development of a starburst, and the start, build-up and
evolution of a cosmic ray population.

\acknowledgments

We are grateful to Antonio Garc\'{\i}a-Barreto for providing us his
(H$\alpha$+[NII]) and I-band maps of IC\,1953, to Sangeeta Malhotra
for mentioning the peculiar infrared properties of NGC\,4418, thereby
arousing our curiousity, to the persons who assisted us in the
observations at Effelsberg, Palomar and La Silla, and to James Brauher
for communicating us some fluxes in advance of publication.
This publication makes use of data products from the Two Micron All Sky Survey,
which is a joint project of the University of Massachusetts and IPAC/Caltech,
funded by NASA and the National Science Foundation.
The ISOCAM data presented in this paper were analyzed using and adapting
the CIA package, a joint development by the ESA Astrophysics Division and
the ISOCAM Consortium.
This research also benefited from the NASA/IPAC Extragalactic Database
(NED), which is operated by the Jet Propulsion Laboratory of Caltech.

\appendix

\section{PIFS observations}
\label{app_pifs}

In February, exposures times were 300\,s for all lines but [FeII] in
NGC\,4491, for which it was increased to 400\,s. Except for the
beginning of Feb. 1, the sky was very clear, and the observations were
photometric. Star number 9148 in \citet{Persson} was used as a photometric
standard. In October, exposure times for NGC\,1377 and NGC\,1022 were
respectively 260 and 220\,s for all lines; the flux calibration was
performed with 2MASS images.

The field of view of PIFS is sliced into eight slits, 0.67\arcsec-wide and
9.7\arcsec-long. We dithered to reconstruct cross-slit spatial information,
and obtained maps of $5.8\arcsec \times 10$\arcsec, with a pixel size of
0.167\arcsec. The high-resolution grating yields resolutions between 950 and
1500 at the wavelengths of interest.

Observations were carried out as outlined in the PIFS manual (which can be
found in {\it http://www.its.caltech.edu/~btsoifer/pifs/}). The flat-field and
atmospheric transmission function was calibrated by sweeping G0V--G8V stars
across the array with the chopping secondary, using a fast triangle wave,
at the same airmass as the galaxy to within 10\%.
The same stars were chopped in the cross-slit direction
in order to establish a flux reference and to calibrate spatial curvature.
Noble gas spectra served to correct for spectral distorsion and fit the
wavelength function. The data reduction consists of the following steps:
(1) define a mask for bad pixels and cosmic rays; (2) subtract the bias
for each quadrant of the array; (3) subtract the sky, tying the OH line
intensities in the off-source spectra to their intensities in the on-source
spectra (in the edges of the field, where the contribution from the galaxy
is minimal); (4) divide spectra by the flat-field and atmospheric function;
(5) construct a wavevelength map using the noble gas spectra; (6) correct for
the spectral and spatial distortions, and combine the rectified spectra in
a cube. Table~\ref{pifs} offers a summary of observing parameters and
measured fluxes. In order to minimize residuals from OH sky lines in the
faint spectra of NGC\,4491, we subtracted the spectral variations of an
off-center region around its average flux density.

\section{SEST observations}
\label{app_sest}

We used dual beam switching, and connected the
{\small IRAM}\,115/230 receivers to the low-resolution acousto-optical
spectrometers. The initial velocity
resolution was 3.6\,km\,s$^{-1}$ and 1.8\,km\,s$^{-1}$, respectively.
System temperatures ranged between 200\,K and 400\,K at 230\,GHz and
between 230\,K and 460\,K at 115\,GHz. The beam width is 45\arcsec\
at 115\,GHz and 23\arcsec\ at 230\,GHz. Baseline fitting and removal
in each separate spectrum, and averaging of spectra obtained at the
same position were performed with the {\small CLASS}
software, which is developed by the Observatoire
de Grenoble and IRAM. We used main beam efficiencies of 0.7 at
115\,GHz and 0.5 at 230\,GHz, and a forward efficiency of 0.9\,.

Pointing corrections were determined on the closest bright SiO masers,
o~Ceti and Orion~SiO, twice per observing session, at the beginning and
after sunset. These sources are respectively about 15\degr\ and 30\degr\
away from NGC\,1377 in azimuth, and about 15\degr\ away to the north.
Due to these large distances, the pointing accuracy is
likely degraded with respect to the quoted value of 3\arcsec\ rms
in azimuth and elevation. Furthermore, a hardware limit obliged us to
rotate the antenna by 360\degr\ in azimuth once per night; because the
CO(2-1) spectrum obtained on the center of NGC\,1377 before the rotation
was not reproduced within the error bar after the rotation, we
suspect that this introduced additional pointing errors, and we
discarded the redundant data obtained after rotating the antenna.

In order to know whether the CO source is extended, we performed
minimal mapping, integrating on the galaxy center and eight adjacent
positions stepped by half the CO(2-1) beam width. The spectra are shown
in Figure~\ref{co_maps}. The resulting eighteen total flux measurements
were least-squares-fitted with a gaussian source of adjustable width,
allowing for some random pointing errors of up to 3\arcsec\ at each
position, plus some systematic larger pointing errors after each large
rotation of the antenna. If no systematic errors are allowed, then the
fit is not acceptable (normalized $\chi^2 > 8$).
We can reproduce the data provided these systematic pointing errors
are of the order of 5\arcsec\ or slightly more. The best fit is then
always obtained for the smallest source size of the explored grid.
Due to the large beams, this parameter is not well constrained,
but the derived total fluxes are robust results.

\clearpage   

\begin{deluxetable}{lcccclc}
\tablecaption{General data. \label{tab_gen}}
\tablehead{
galaxy & J2000 & group & $h_{75}~D$ & $\Delta$(c$z$) & type & $D_{25}$ \\
~ & ~ & membership & (Mpc) & (km.s$^{-1}$) & ~ & (arcmin)
}
\startdata
NGC\,1377 & 03 36 39.1 $-$20 54 07 & Eridanus & 21 & $\approx +200$ & amorphous\tablenotemark{a} & 1.78 \\
IC\,1953  & 03 33 41.8 $-$21 28 43 & Eridanus & 21 & $\approx +280$ & SBd & 2.75 \\
NGC\,4491 & 12 30 57.2 $+$11 28 59 & Virgo\,A & 17 & $\approx -550$ & SBa dwarf\tablenotemark{b} & 1.70 \\
\enddata
\tablenotetext{a}{from \citet{Heisler_opt}, who note the presence of a dust lane in the center.}
\tablenotetext{b}{from \citet{Binggeli_type}.}
\end{deluxetable}

\begin{deluxetable}{lcccccccc}
\tablecaption{Infrared and radio flux densities and derived quantities. \label{tab_flux}}
\tablehead{
galaxy & $F_{100}$ & $F_{60}$ & $F_{25}$ & $S_{1.4 {\rm GHz}}$ &
$L_{\rm FIR}$ & $L_{\rm FIR}/L_{\rm B}$ & $F_{60}/F_{100}$ & $q$ \\
~ & \multicolumn{3}{c}{(Jy)} & (mJy) & ($10^9\,L_{\sun \rm bol}$) & ~ & ~ & ~
}
\startdata
NGC\,1377 & 5.97  & 7.27 & 1.92 & $ < 1.01$\tablenotemark{a} & 4.3 & 3.3 & 1.22 & $>3.92$ ($8.1 \sigma$) \\
IC\,1953  & 11.62 & 9.01 & 1.05 &     13.00\tablenotemark{b} & 6.1 & 1.6 & 0.78 &   2.95  ($3.2 \sigma$) \\
NGC\,4491 & 3.45  & 2.68 & 0.43 & $ < 1.35$\tablenotemark{c} & 1.2 & 1.6 & 0.78 & $>3.41$ ($5.5 \sigma$) \\
\enddata
\tablenotetext{a}{This value is in agreement with the limit
$S_{1.49 {\rm GHz}} < 1.0$\,mJy of \citet{Condon_atlas} measured in a 21\arcsec-beam with the VLA.}
\tablenotetext{b}{Our measurement on the NVSS map coincides with \citet{Condon_nvss}.}
\tablenotetext{c}{See Sect.~\ref{n4491} for an important discussion of this limit.}
\end{deluxetable}

\clearpage

\begin{deluxetable}{lccccccc}
\tablecaption{Mid-infrared data. \label{tab_mir}}
\tablehead{
galaxy & $F_{\rm 15\,tot}$ & $F_{\rm 7\,tot}$ & $D_{\rm CNR}/D_{25}$ &
  $D_{5\,\mu{\rm Jy}}/D_{25}$\tablenotemark{a} &
  $F_{\rm 15\,CNR}$ & $F_{\rm 7\,CNR}$ & $(F_{15}/F_7)_{\rm CNR}$ \\
~ & \multicolumn{2}{c}{(mJy)} & ~ & ~ & \multicolumn{2}{c}{(mJy)} & ~
}
\startdata
NGC\,1377\tablenotemark{b} & $721 \pm 162$ & $391 \pm 68$ & 0.22 & 0.88 & $705$ & 343 & $\approx 2.1$ \\
IC\,1953  & $223 \pm 37$ & $186 \pm 14$ & 0.09 & 0.79 & 116 & 34 & 3.4 \\
NGC\,4491 & $81 \pm 25$ & $31 \pm 8$ & 0.10 & 0.46 & 73 & 18 & 4.0 \\
\enddata
\tablenotetext{a}{Relative size of the isophote $5\,\mu$Jy\,arcsec$^{-2}$ at 7\,$\mu$m. }
\tablenotetext{b}{The nucleus of NGC\,1377 is slightly saturated both at 15 and 7\,$\mu$m.
The fluxes were measured on maps constructed by projecting together only the images which
are not affected by saturation (since there is a shift of a non-integer number of pixels
between each image, the brighter pixel receives a variable fraction of the flux of the
central region). The total fluxes obtained without discarding the saturated images are
$F_{\rm 15\,sat} = 705$\,mJy and $F_{\rm 7\,sat} = 366$\,mJy.}
\end{deluxetable}

\begin{deluxetable}{lllll}
\tablecaption{Single-dish radio continuum fluxes. \label{tabradio}}
\tablehead{
~ & $S$(3.6\,cm) & $S$(6.2\,cm) & $S$(21\,cm) & $\alpha$(3--6\,cm)\tablenotemark{a} \\
~ & \multicolumn{3}{c}{(mJy)} & ~
}
\startdata
beam size & 82\arcsec & 147\arcsec & 45\arcsec & ~ \\
\noalign{\smallskip}
\hline
\noalign{\smallskip}
NGC\,1377 & $< 2.0 (3 \sigma)$ & $< 2.7 (3 \sigma)$ & $< 1.0$ & ~ \\
IC\,1953 & $3.6 \pm 2.2$\tablenotemark{b} & $5.1 \pm 1.1$\tablenotemark{c} & 13.0 & $-0.64 \pm 1.19$ \\
NGC\,4491 & $5.0 \pm 0.3$ & $6.0 \pm 0.8$ & $< 7.9$\tablenotemark{d} & $-0.34 \pm 0.27$ \\
NGC\,4418 & $19.7 \pm 1.2$ & $33.3 \pm 1.3$ & $47.9 \pm 1.8$ & $-0.97 \pm 0.13$ \\
\enddata
\tablenotetext{a}{Spectral index between 3\,cm and 6\,cm ($S \propto \nu^{\alpha}$).}
\tablenotetext{b}{Including the southern extensions (the noise in the map is 0.6\,mJy/beam).}
\tablenotetext{c}{The emission is extended, and the noise in the map is 0.5\,mJy/beam.}
\tablenotetext{d}{See text (Sect.~\ref{n4491}).}
\end{deluxetable}

\clearpage

\begin{deluxetable}{rlccccc}
\tablecaption{PIFS spectroscopy. \label{pifs}}
\tablehead{
~~~~~~~~line & date & $T_{\rm int}$\tablenotemark{a} & airmass &
  line flux & continuum & area \\
~ & ~ & (min) & ~ & ($10^{-18}$\,W\,m$^{-2}$) &
  ($10^{-15}$\,W\,m$^{-2}\,\mu$m$^{-1}$) & ~
}
\startdata
\multicolumn{7}{l}{NGC\,4491:} \\
Pa$\beta$   & Feb. 1  & 80  & $\approx 1.1$ & $4.25 \pm 1.27$ & 18.9 & $5.67\arcsec \times 5.67\arcsec$ \\
Pa$\beta$   & ~       & ~   & ~ 	    & $2.05 \pm 0.21$ & 3.09 & $2\arcsec \times 2\arcsec$ \\
$[$FeII$]$  & Feb. 2  & 133 & 1.1--1.3      & $< 0.22~(3\sigma)$ & 1.66 & $2\arcsec \times 2\arcsec$ \\
Br$\gamma$  & Feb. 2  & 40  & 1.4--2.0      & $0.56 \pm 0.24$ & 0.95 & $2\arcsec \times 2\arcsec$ \\
\multicolumn{7}{l}{NGC\,4102:} \\
Pa$\beta$   & Feb. 1  & 20  & 1.1--1.2      & $120.8 \pm 4.8$ & 131.8 & $4.83\arcsec \times 4.83\arcsec$ \\
$[$FeII$]$      & Feb. 1  & 20  & $\approx 1.1$ & $76.7 \pm 3.2$ & 123.6 & $4.83\arcsec \times 4.83\arcsec$ \\
Br$\gamma$  & Feb. 2  & 20  & 1.6--1.8      & $60.1 \pm 6.5$ & 117.2 & $4.83\arcsec \times 4.83\arcsec$ \\
\multicolumn{7}{l}{NGC\,1377:} \\
Pa$\beta$   & Oct. 17 & 26  & 1.7--1.8      & $< 0.22~(3\sigma)$ & 4.51 & $2\arcsec \times 2\arcsec$ \\
$[$FeII$]$      & Oct. 17 & 26  & 1.7--1.8      & $< 0.29~(3\sigma)$ & 3.96 & $2\arcsec \times 2\arcsec$ \\
Br$\gamma$  & Oct. 18 & 39  & 1.7--1.9      & $< 0.47~(3\sigma)$ & 3.05 & $2\arcsec \times 2\arcsec$ \\
H$_2$\,2.12 & Oct. 18 & 35  & 1.7--1.8      & $4.59 \pm 0.78$ & 7.34 & $5.17\arcsec \times 5.17\arcsec$ \\
H$_2$\,2.25 & Oct. 19 & 26  & 1.7--1.8      & $< 0.14~(3\sigma)$ & ~ & $2\arcsec \times 2\arcsec$ \\
H$_2$\,2.22 & Oct. 19 & 30  & 1.7--1.8      & $< 0.33~(3\sigma)$ & ~ & $2\arcsec \times 2\arcsec$ \\
\multicolumn{7}{l}{NGC\,1022:} \\
Pa$\beta$   & Oct. 17 & 18  & 1.4--1.5      & $19.3 \pm 0.4$ & 13.9 & $2\arcsec \times 2\arcsec$ \\
$[$FeII$]$      & Oct. 17 & 15  & 1.3--1.4      & $8.96 \pm 0.69$ & 12.1 & $2\arcsec \times 2\arcsec$ \\
Br$\gamma$  & Oct. 18 & 15  & 1.7--1.9      & $8.14 \pm 0.35$ & 8.46 & $2\arcsec \times 2\arcsec$ \\
H$_2$\,2.12 & Oct. 18 & 40  & 1.4--1.7      & $3.77 \pm 0.17$ & ~ & $2\arcsec \times 2\arcsec$ \\
H$_2$\,2.25 & Oct. 19 & 15  & $\approx 1.3$ & $0.72 \pm 0.17$ & ~ & $2\arcsec \times 2\arcsec$ \\
\enddata
\tablenotetext{a}{Total on-source integration time (an equal time was spent
on the sky).}
\end{deluxetable}

\clearpage

\begin{deluxetable}{lccc}
\tablecaption{SFR estimates. \label{sfr}}
\tablehead{
galaxy & SFR$_{\rm CNR}$(7\,$\mu$m) & SFR$_{\rm tot}$(FIR) &
  SFR$_{\rm CNR}$(H) \\
~ & \multicolumn{3}{c}{(M$_{\sun}$.yr$^{-1}$)}
}
\startdata
NGC\,1377\tablenotemark{a} & 1.84 & 1.79 & not applicable \\
IC\,1953  & $1.65 \pm 0.60$ & 1.83 & ext. H$\alpha$: 0.02\tablenotemark{b} \\
NGC\,4491 & $0.3 \pm 0.1$   & 0.38 & Pa$\beta$-Br$\gamma$: 0.04\tablenotemark{c} \\
\enddata
\tablenotetext{a}{The star formation rates derived from the dust emission
may be overestimates, due to various reasons (see text).}
\tablenotetext{b}{From the (H$\alpha$ + [NII]) flux measured on the map
given by Antonio Garc\'{\i}a-Barreto (uncorrected for extinction), with an
approximate calibration based on the I-band measurement of \citet{Giovanelli},
and after correcting for a slight over-subtraction of the continuum with the
help of the I-band image, assuming the light follows the same distribution
in I and in R.}
\tablenotetext{c}{In $5.67\arcsec \times 5.67\arcsec$.}
\end{deluxetable}

\begin{deluxetable}{lllll}
\tablecaption{Far-infrared spectroscopy. \label{lws}}
\tablehead{
galaxy & [CII]\,157.7 & [OI]\,63.2 & [NII]\,121.9 & [OIII]\,88.4 \\
~ & \multicolumn{4}{c}{($10^{-16}$\,W\,m$^{-2}$)}
}
\startdata
NGC\,1377\tablenotemark{a} & $< 1.95$        & $< 2.82$ & $< 1.16$ & $< 2.17$ \\
IC\,1953\tablenotemark{a}  & $5.82 \pm 0.40$ & $< 2.82$ & $< 1.44$ & $< 4.31$ \\
NGC\,4491\tablenotemark{b} & 1.3 & ~ & ~ & ~ \\
\enddata
\tablenotetext{a}{from \citet{Brauher}.}
\tablenotetext{b}{from \citet{Leech}.}
\end{deluxetable}

\clearpage

\begin{deluxetable}{llllllllll}
\tablecaption{Molecular gas in NGC\,1377. \label{tabco}}
\tablehead{
~ & \multicolumn{9}{l}{individual pointings ($\Delta$N, $\Delta$E)\tablenotemark{a}} \\
~ & (0, 0) & (0, +1) & (0, +2) & (0, -1) \\
~ & (-1, 0) & (+1, 0) & (-1, +1) & (-2, 0) & (-1, -1)
}
\startdata
CO(1-0) & $1.62 \pm 0.11$ & $1.64 \pm 0.09$ & $< 0.9$ & $1.48 \pm 0.12$ \\
~ & $1.67 \pm 0.12$ & $1.42 \pm 0.12$ & $1.10 \pm 0.15$ & $< 0.11$ & $< 0.73$ \\
CO(2-1) & $1.91 \pm 0.06$ & $1.85 \pm 0.08$ & $< 0.6$ & $0.98 \pm 0.09$ \\
~ & $2.32 \pm 0.08$ & $1.25 \pm 0.09$ & $< 0.26$ & $< 0.22$ & $< 0.10$ \\
\noalign{\smallskip}
\hline
\noalign{\smallskip}
\multicolumn{10}{c}{fitted model parameters:} \\
\noalign{\smallskip}
\multicolumn{10}{l}{total CO(1-0) flux: ($1.75 \pm 0.12$)\,K\,km\,s$^{-1}$} \\
\multicolumn{10}{l}{total CO(2-1) flux: ($2.55 \pm 0.49$)\,K\,km\,s$^{-1}$} \\
\multicolumn{10}{l}{source FWHM: ($0 \pm 8$)\, arcsec} \\
\multicolumn{10}{l}{systematic pointing errors to obtain $\chi^2 = 1$:
   (+4, -2), (-5, 0), (-6, 0)\,arcsec\tablenotemark{b}} \\
\enddata
\tablenotetext{a}{The offsets with respect to the center are given in
units of 11.5\arcsec, half the CO(2-1) beam width. All flux values,
$\int {\rm T}_A^* {\rm dV}$, are in K\,km\,s$^{-1}$.}
\tablenotetext{b}{One offset after each large rotation of the antenna (see
Appendix~\ref{app_sest}).}
\end{deluxetable}

\clearpage   

\clearpage

\clearpage

\clearpage

\clearpage

\clearpage

\clearpage

\clearpage

\begin{figure}
\hspace*{-1.8cm}
\resizebox{6cm}{!}{\rotatebox{90}{\plotone{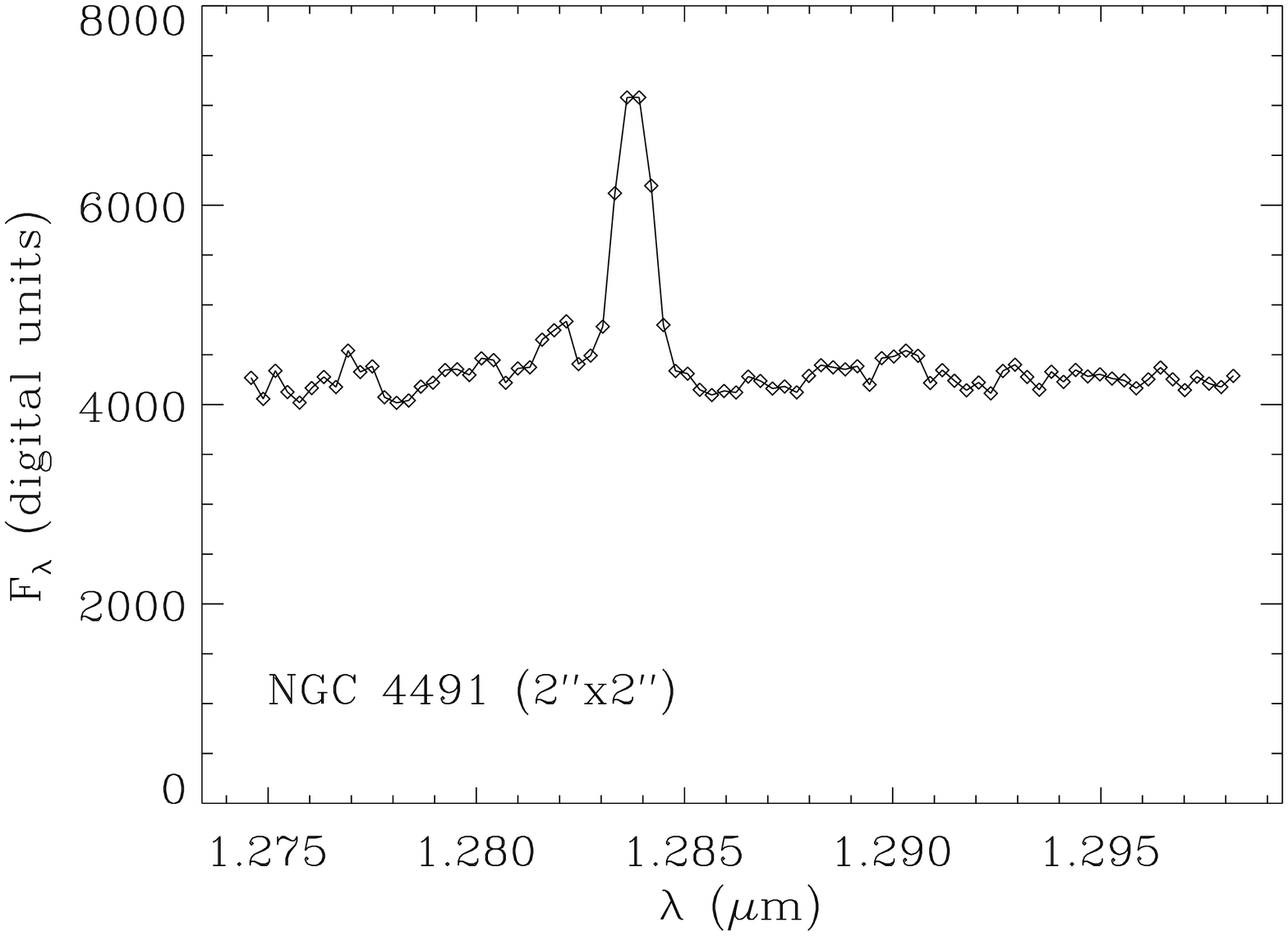}}}
\resizebox{6cm}{!}{\rotatebox{90}{\plotone{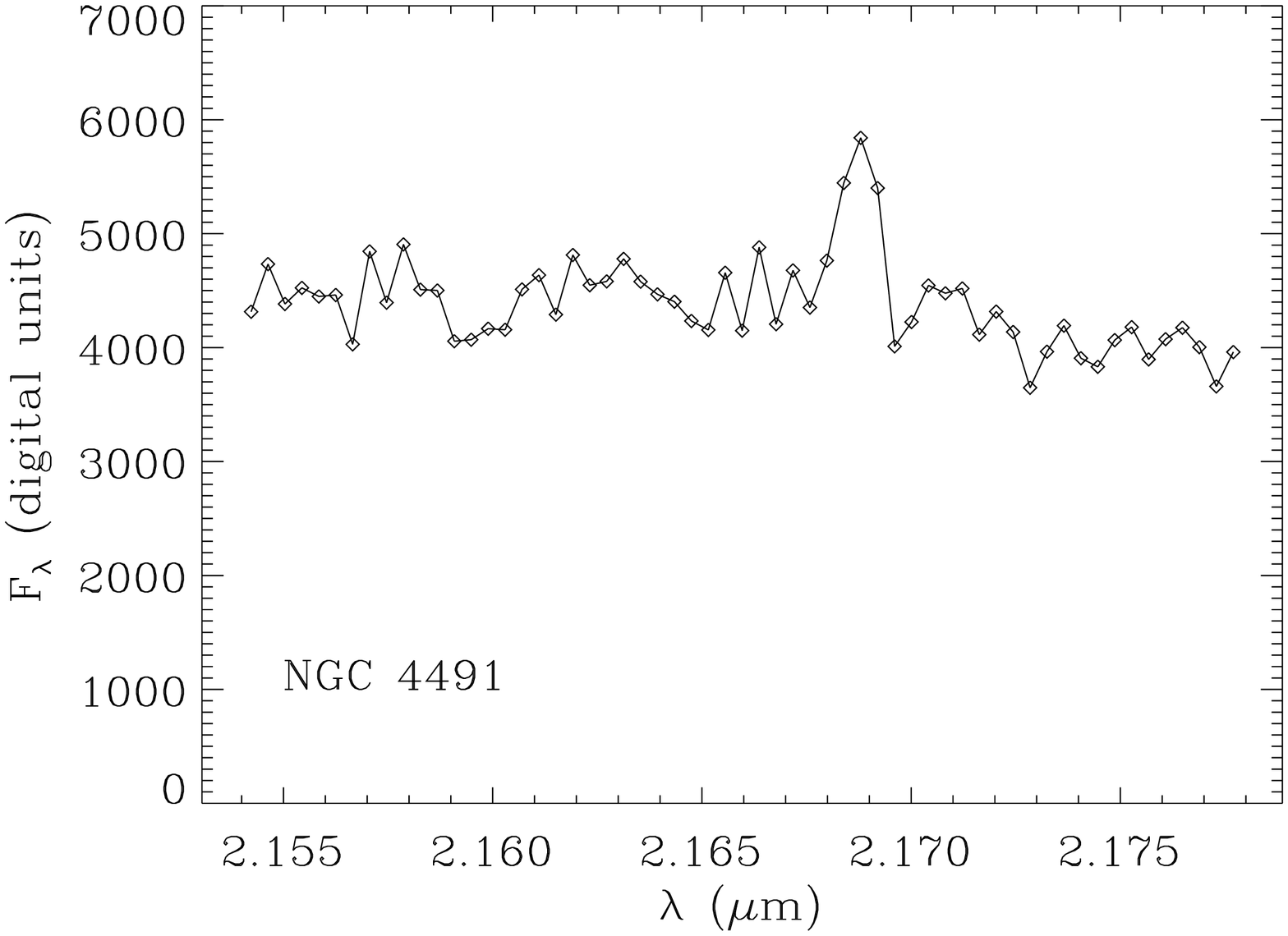}}}
\resizebox{6cm}{!}{\rotatebox{90}{\plotone{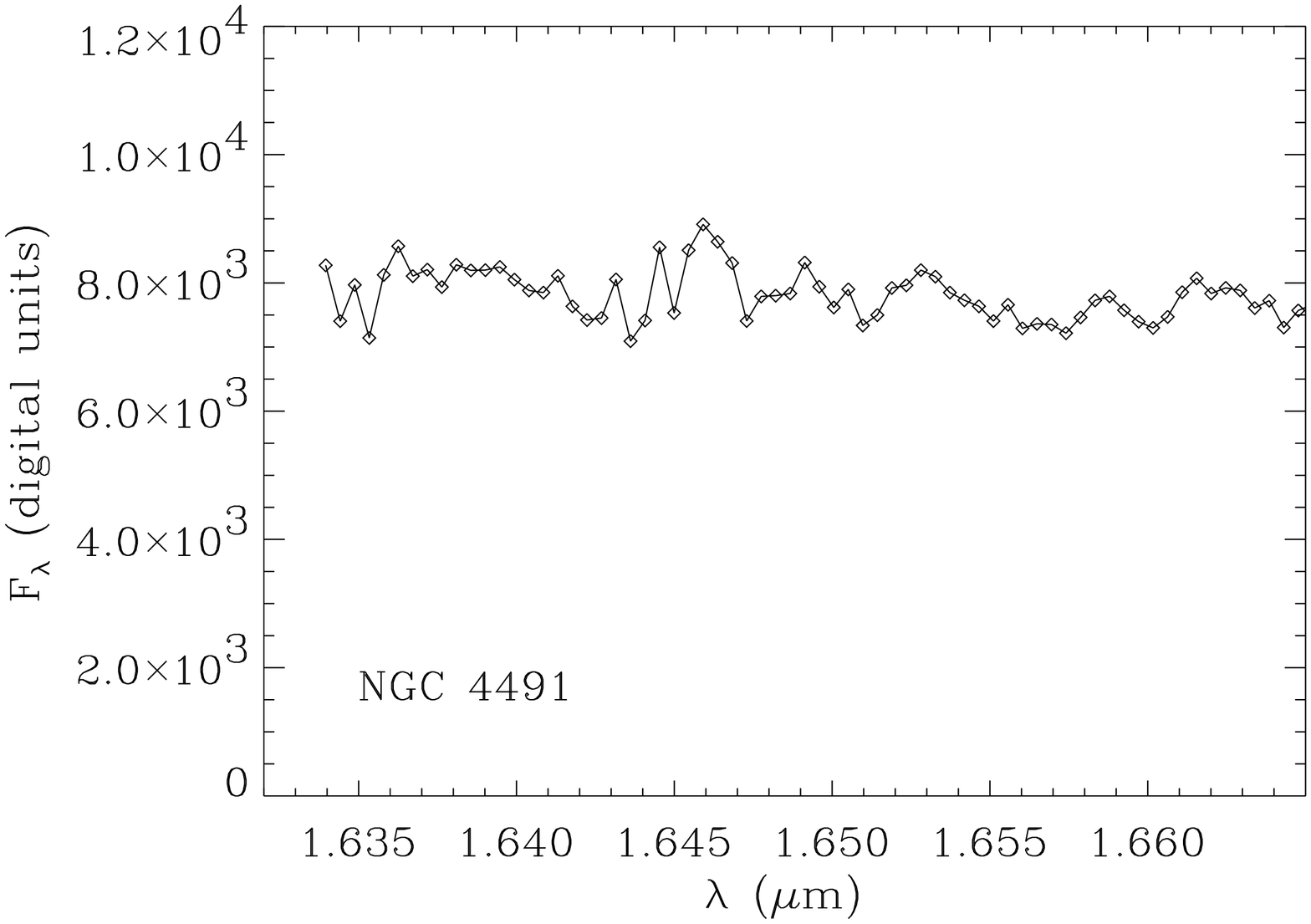}}} \\
\hspace*{-1.8cm}
\resizebox{6cm}{!}{\rotatebox{90}{\plotone{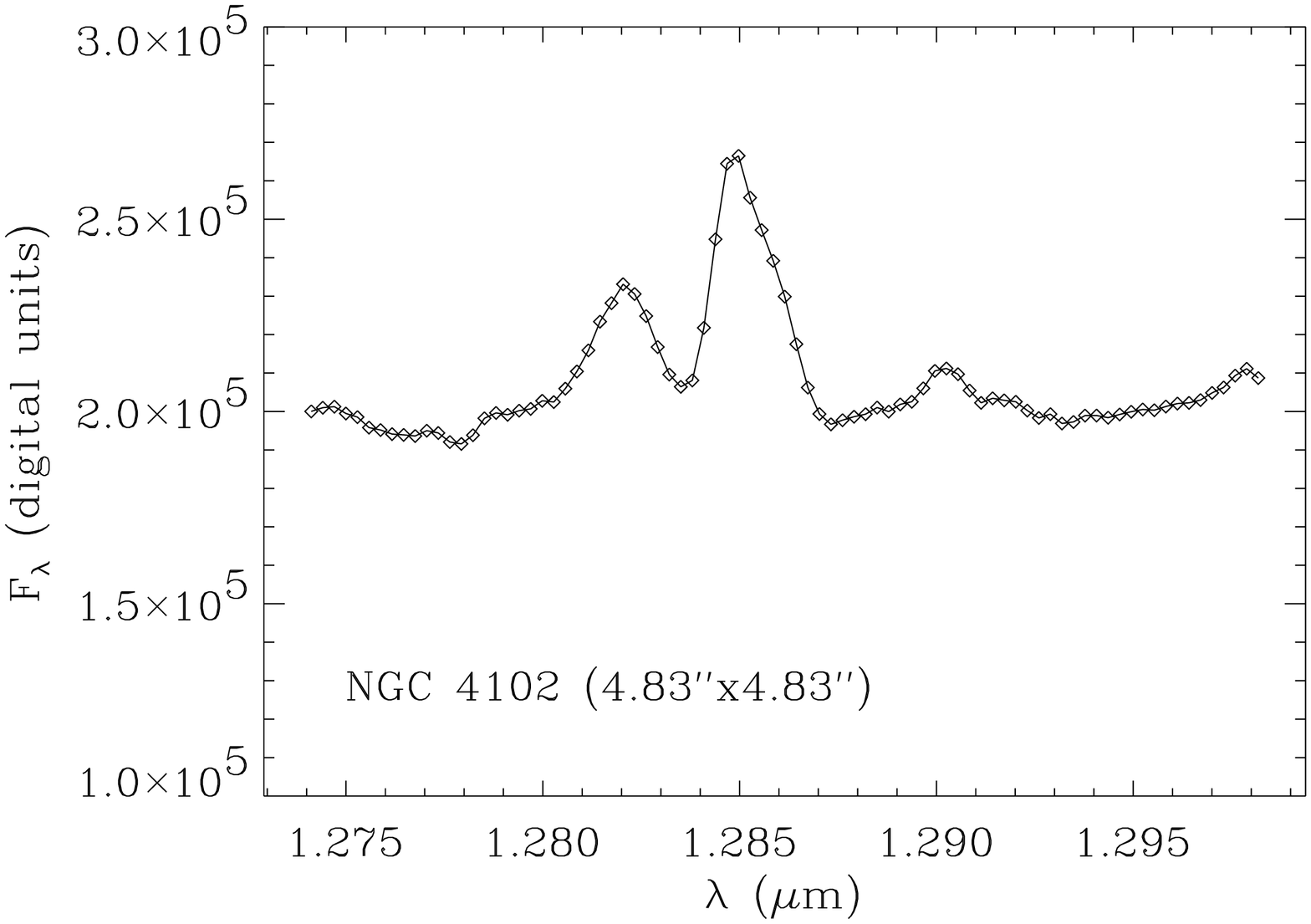}}}
\resizebox{6cm}{!}{\rotatebox{90}{\plotone{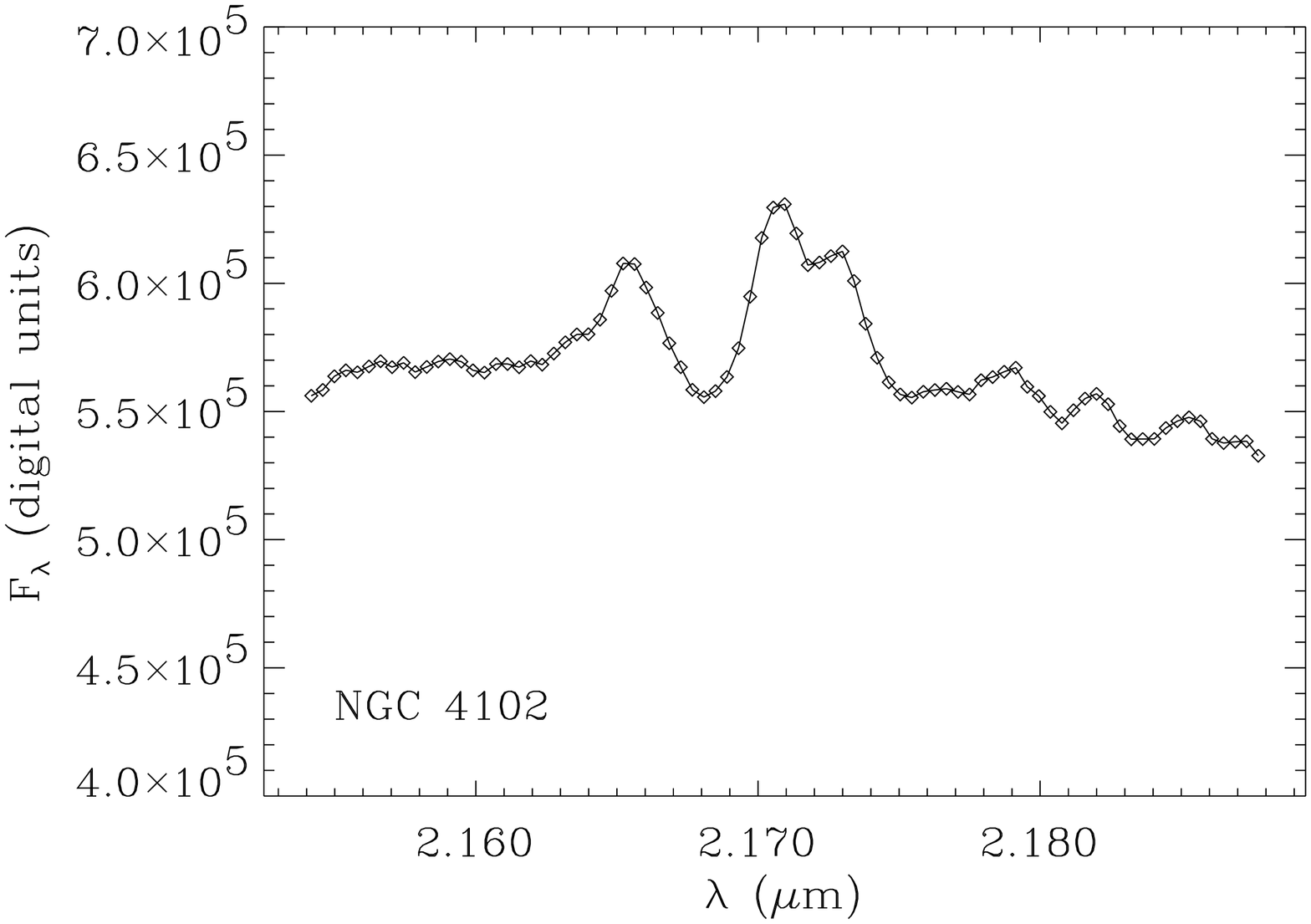}}}
\resizebox{6cm}{!}{\rotatebox{90}{\plotone{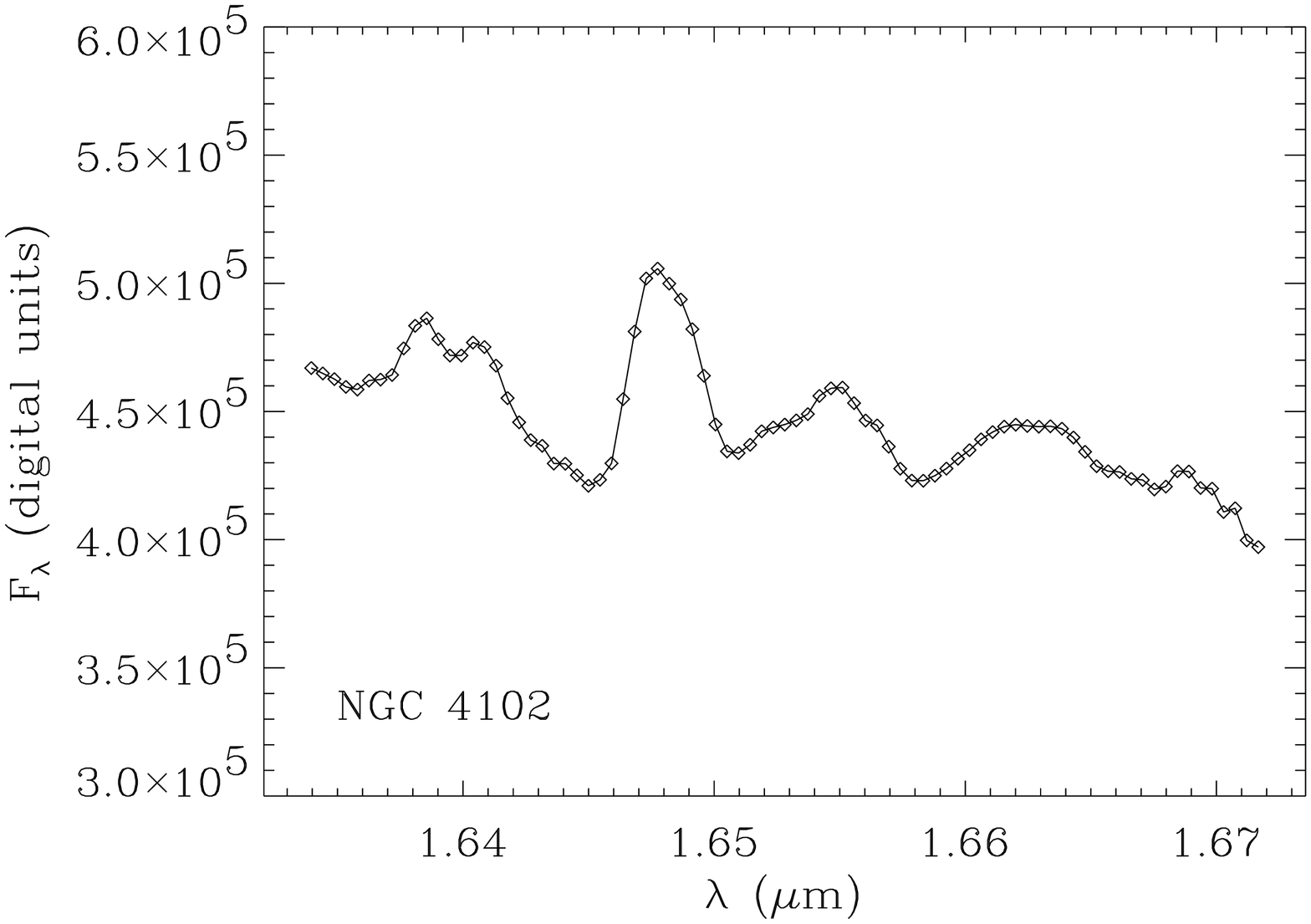}}} \\
\hspace*{-1.8cm}
\resizebox{6cm}{!}{\rotatebox{90}{\plotone{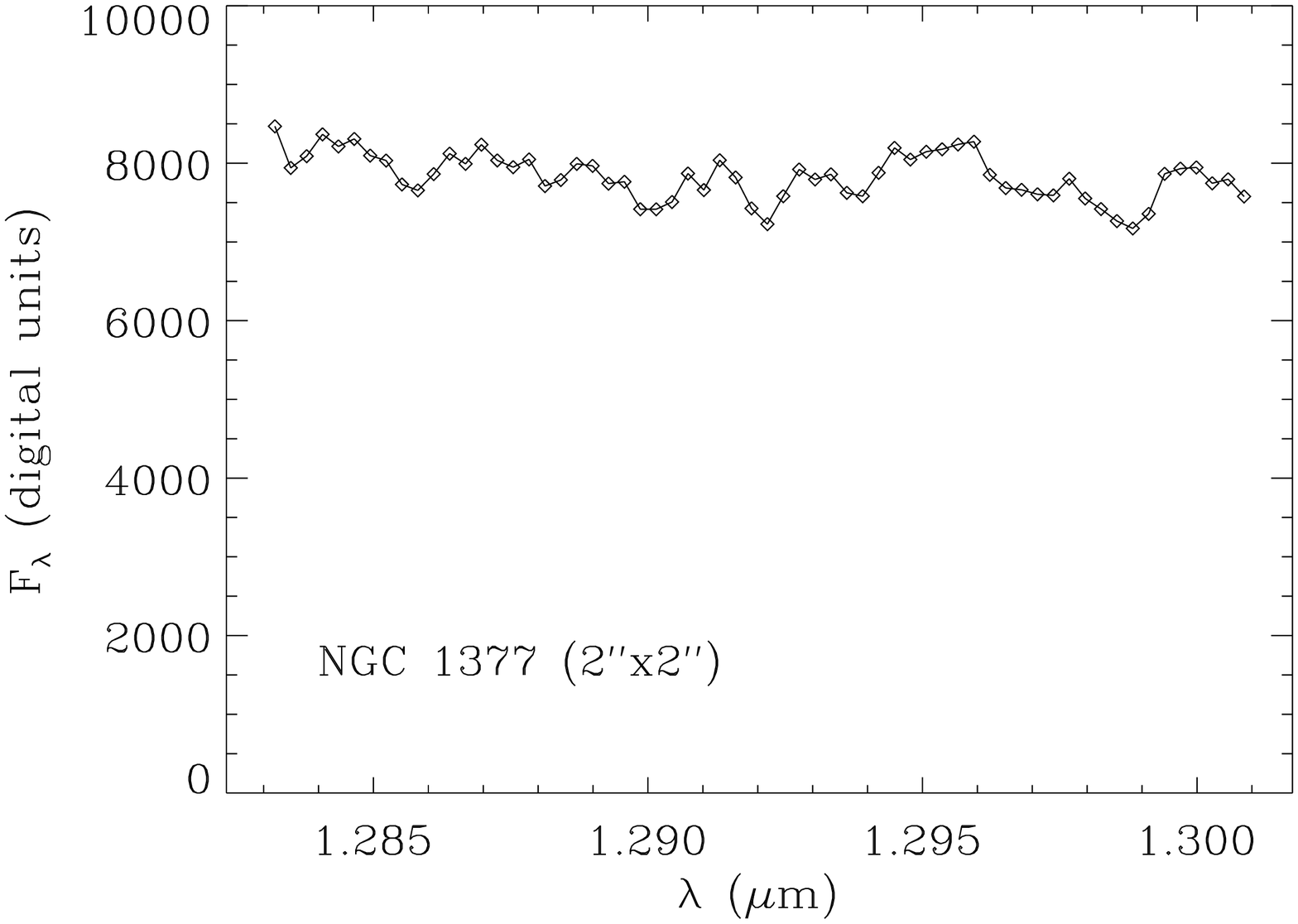}}}
\resizebox{6cm}{!}{\rotatebox{90}{\plotone{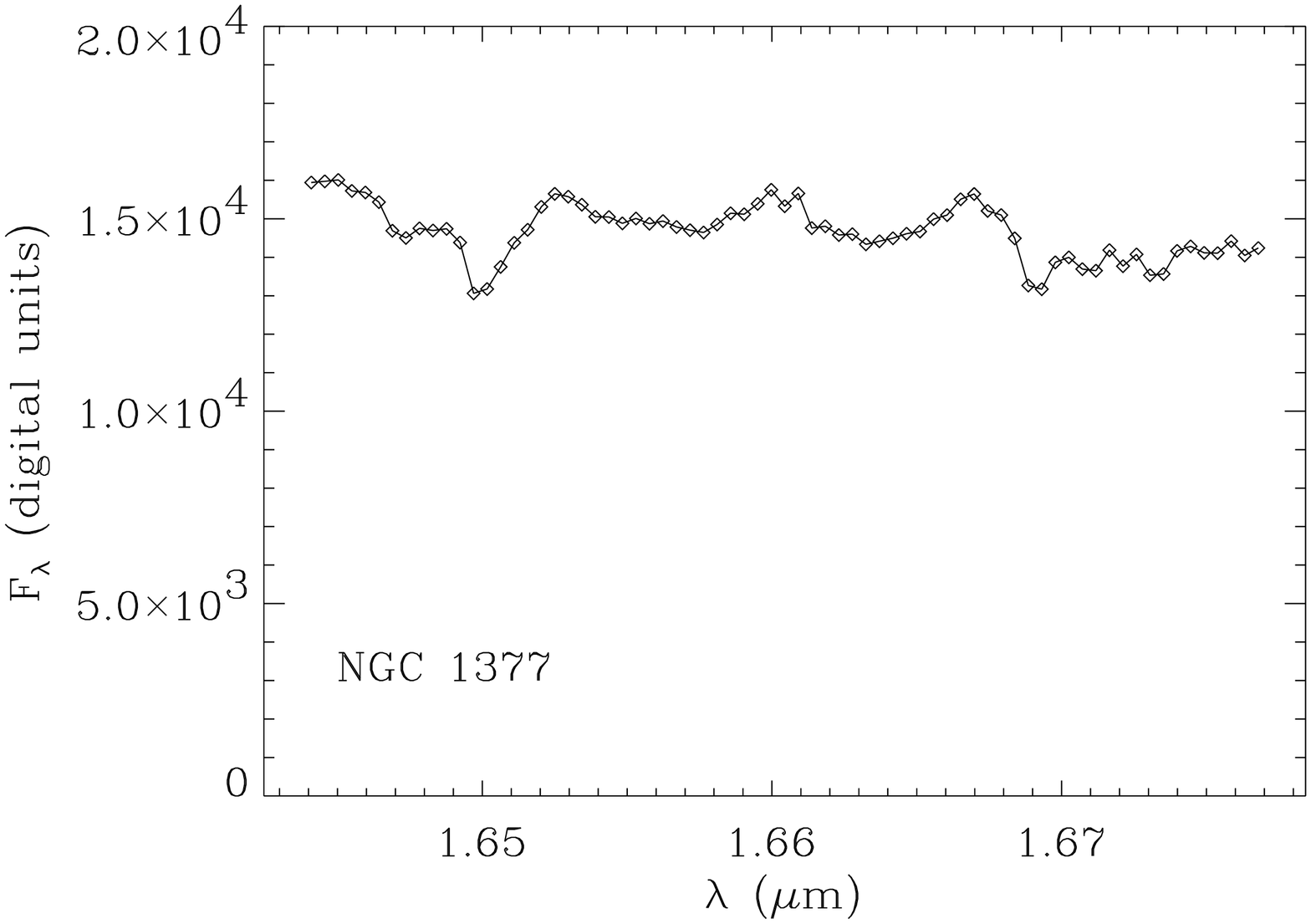}}}
\resizebox{6cm}{!}{\rotatebox{90}{\plotone{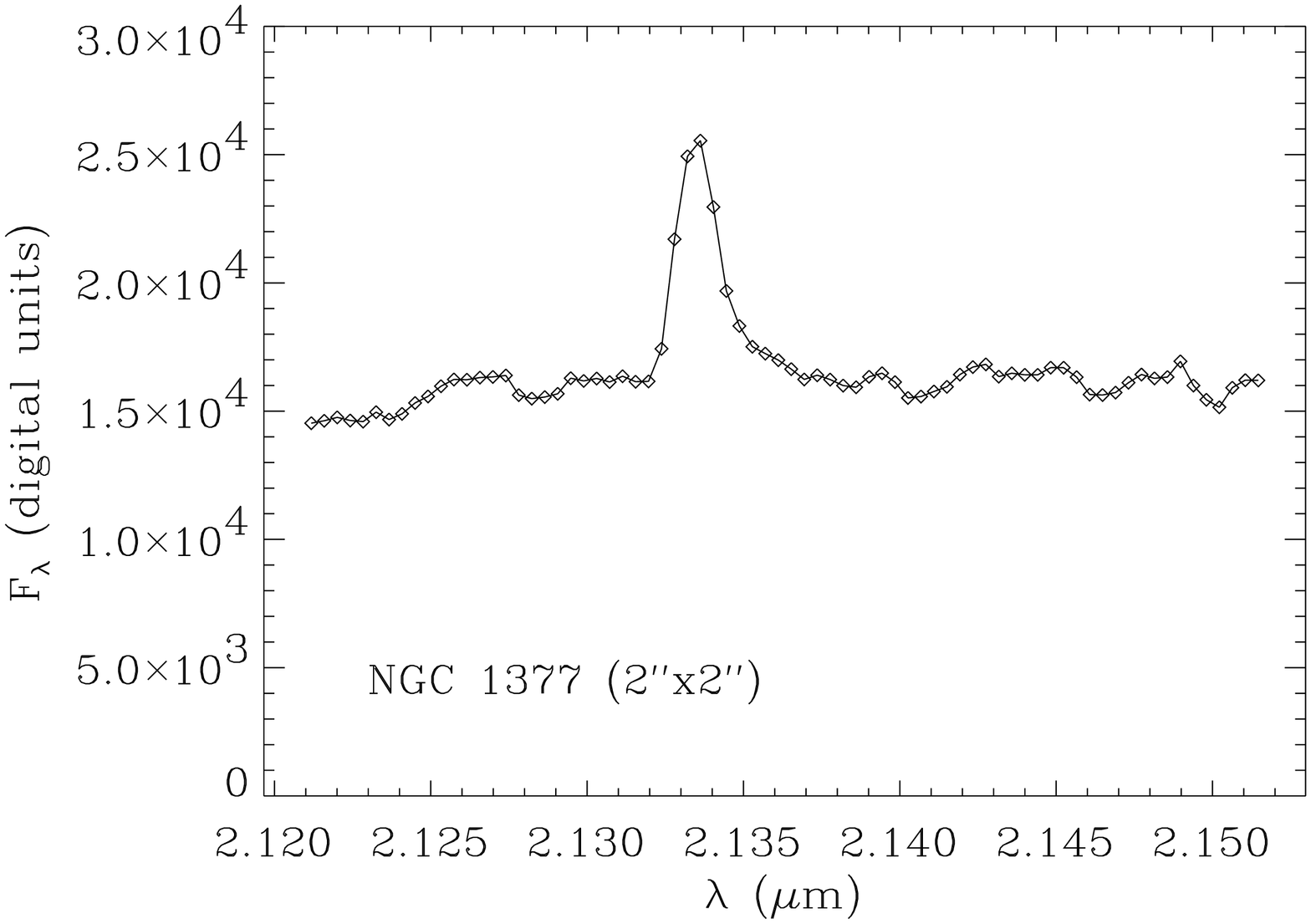}}} \\
\hspace*{-1.8cm}
\resizebox{6cm}{!}{\rotatebox{90}{\plotone{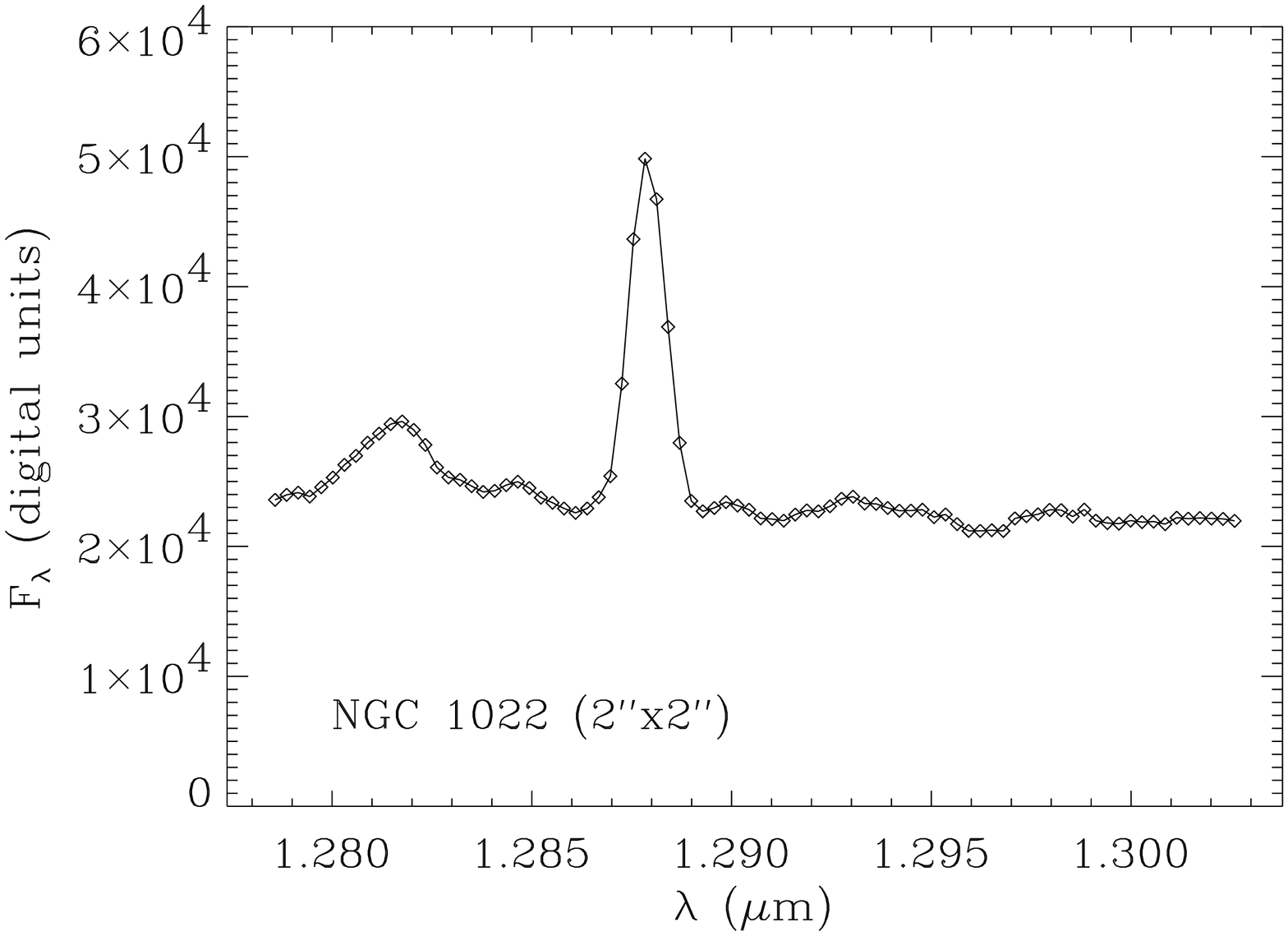}}}
\resizebox{6cm}{!}{\rotatebox{90}{\plotone{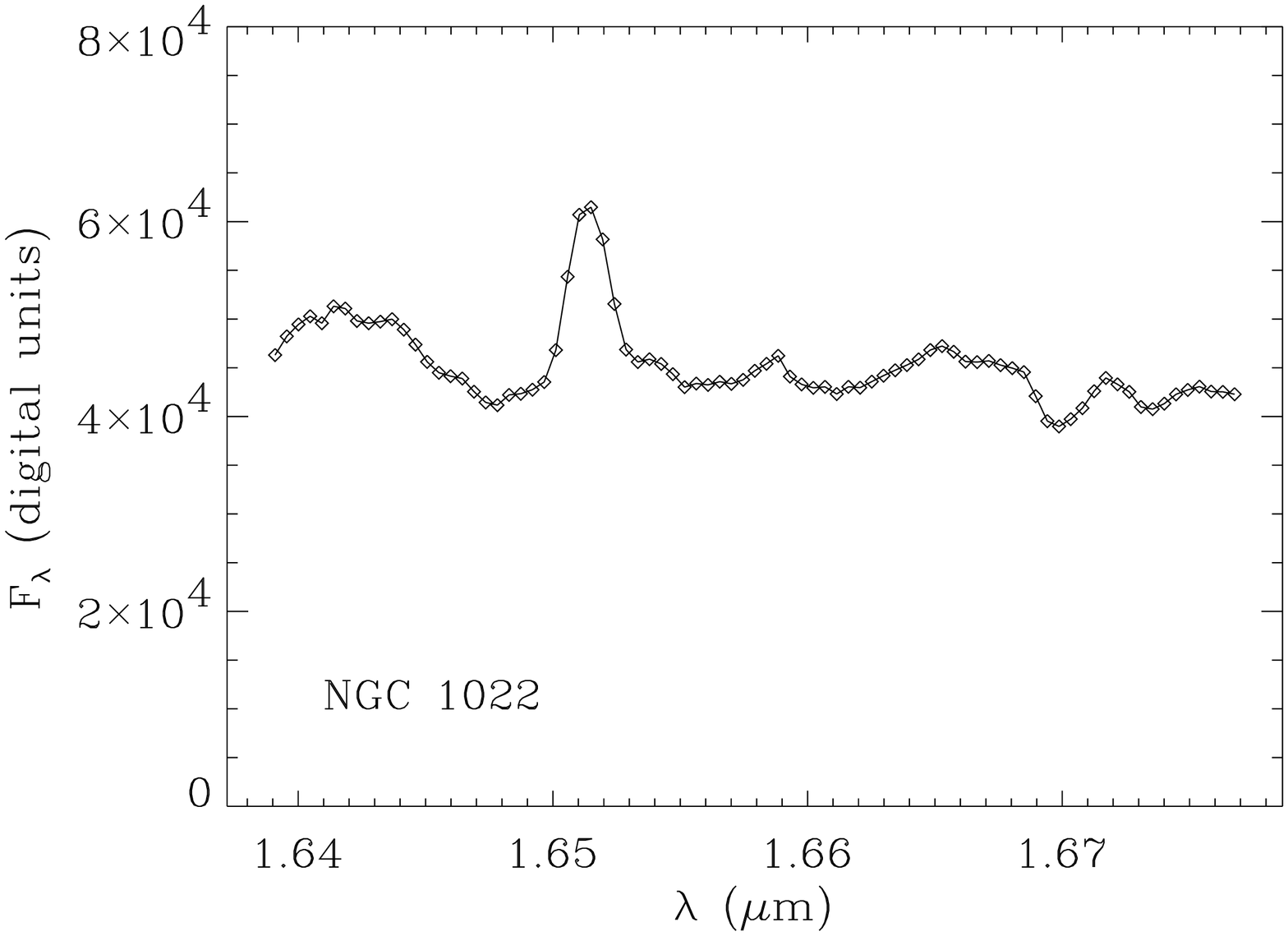}}}
\resizebox{6cm}{!}{\rotatebox{90}{\plotone{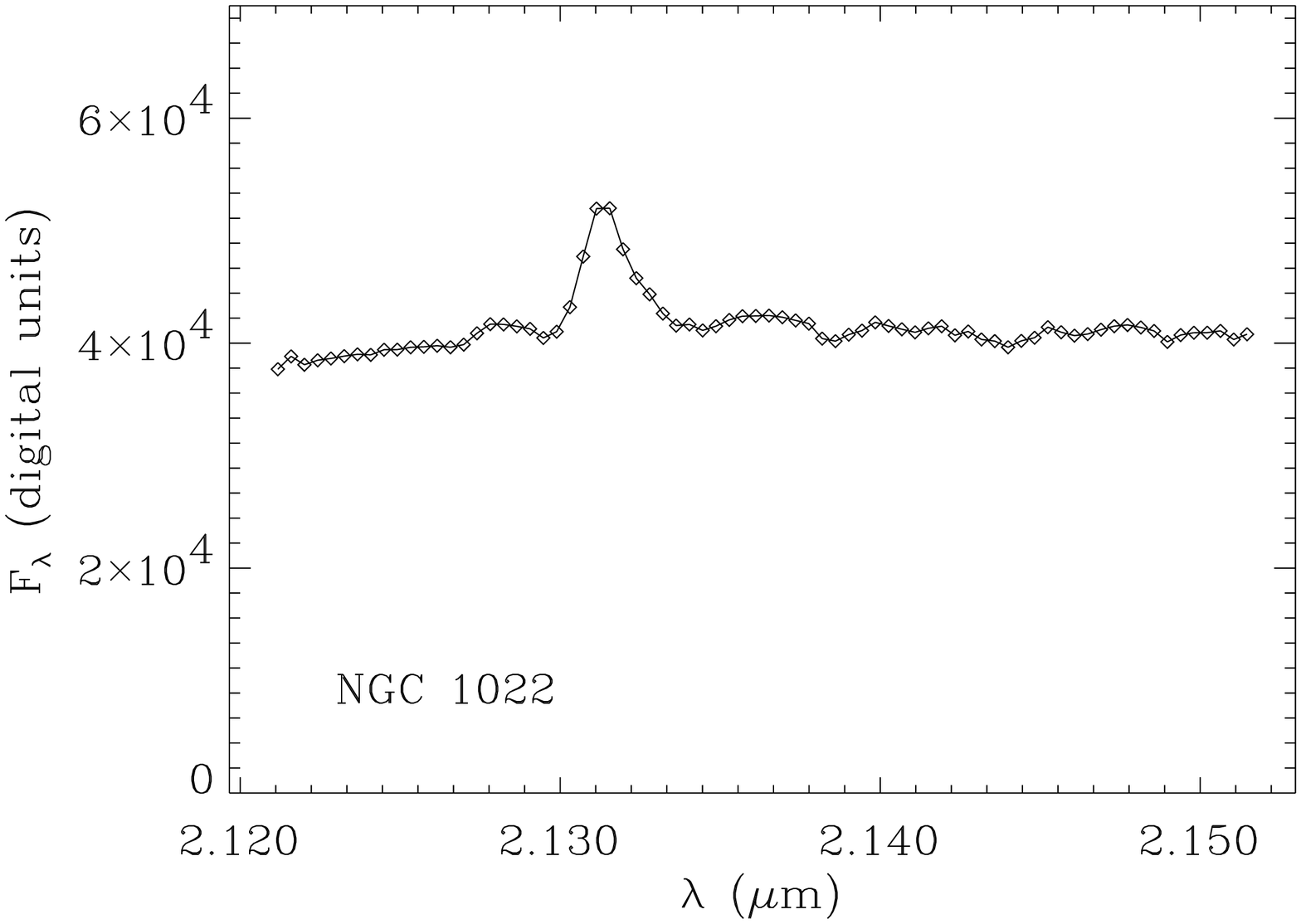}}}
\caption{Near-infrared spectra of NGC\,4491 and the comparison starburst
NGC\,4102 in the Pa$\beta$, Br$\gamma$ and [FeII]\,1.644 lines (in this order
from left to right), and of NGC\,1377 and the comparison starburst NGC\,1022
in the Pa$\beta$, [FeII]\,1.644 and H$_2$\,(1-0)\,S(1) lines.
Since the line emission of NGC\,4102 peaks slightly SE of the nucleus,
the rotation broadening is asymmetric.
\label{spec}}
\end{figure}

\clearpage

\begin{figure}
\vspace*{-1cm}
\resizebox{17cm}{!}{\plotone{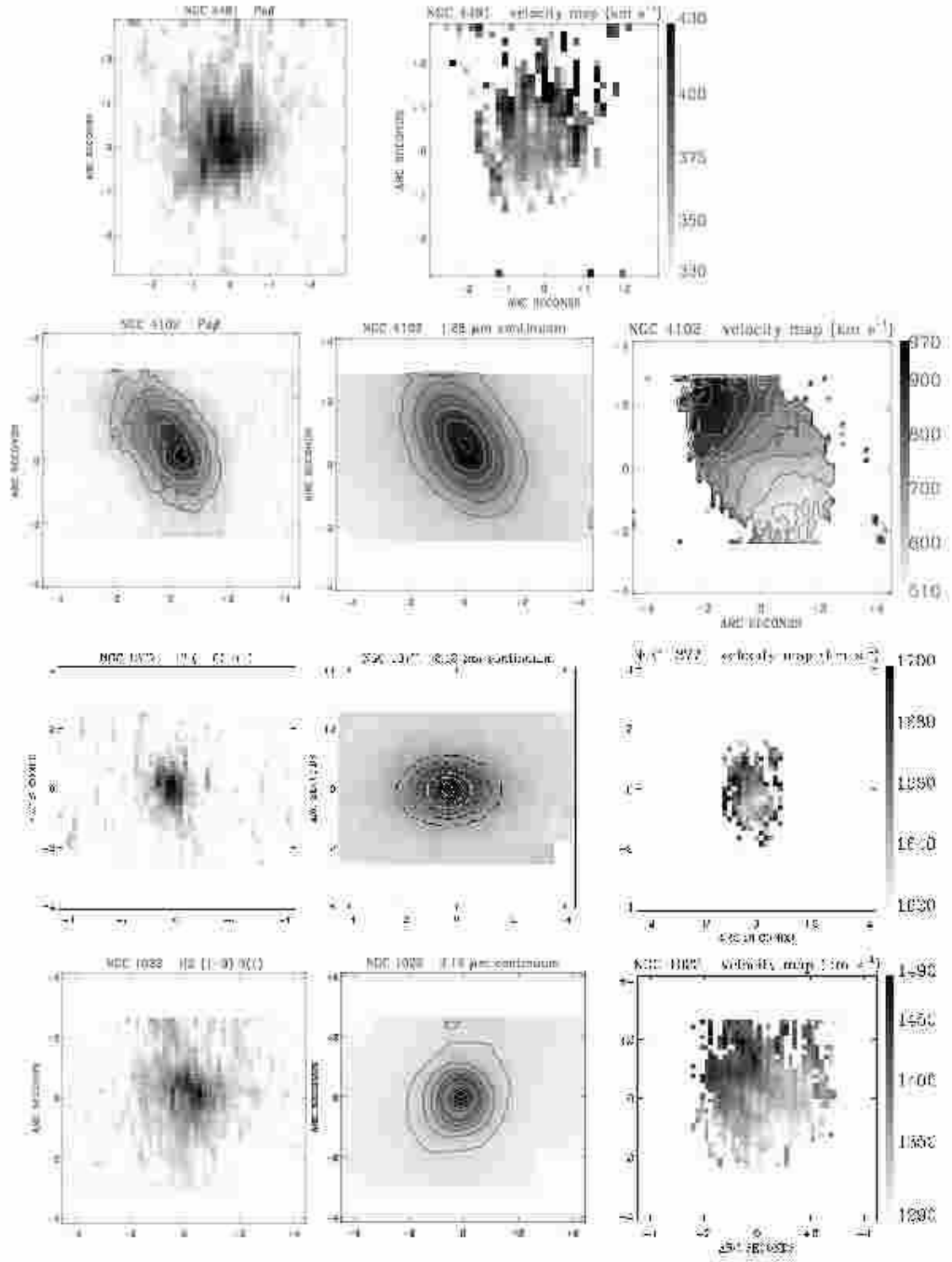}}
\vspace*{-1.5cm}
\caption{Continuum-subtracted images and velocity maps of NGC\,4491 and NGC\,4102
in the Pa$\beta$ line, and of NGC\,1377 and NGC\,1022 in the H$_2$\,(1-0)\,S(1)
line. The intensities range from $1\sigma$ to the maximum flux.
Also shown are maps of the continuum emission around the line.
The contours are equally spaced in flux and in velocity.
\label{pifsima}}
\end{figure}

\clearpage

\begin{figure}
\begin{minipage}[t]{10cm}
\resizebox{16cm}{!}{\rotatebox{-90}{\plotone{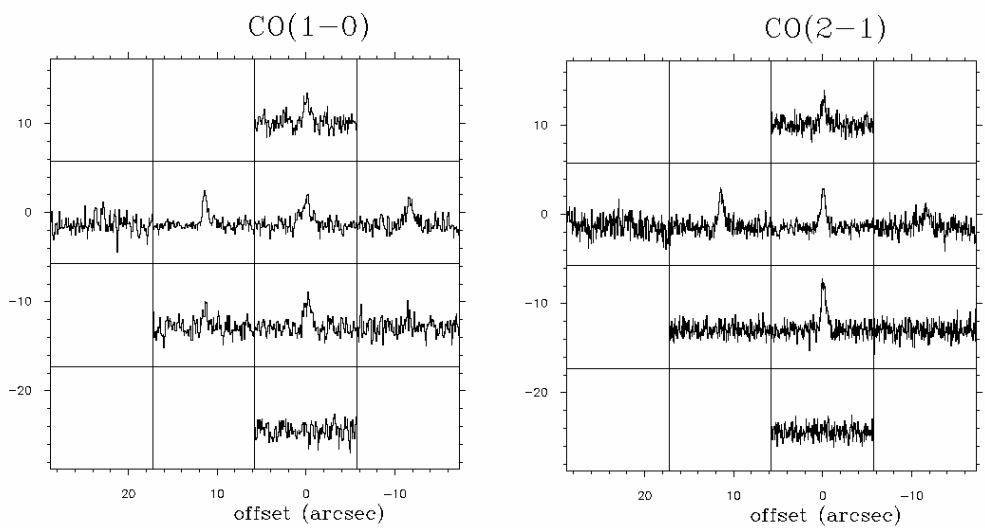}}}
\end{minipage}
\vspace*{-2cm}
\caption{Averaged CO(1-0) and CO(2-1) spectra of NGC\,1377 at the nine
observed positions. The velocity range and temperature scale are the
same for both lines, and the channel width was smoothed to four times
the original width.
\label{co_maps}}
\end{figure}

\begin{figure}
\hspace*{-1cm}
\begin{minipage}[t]{8cm}
\resizebox{9cm}{!}{\rotatebox{-90}{\plotone{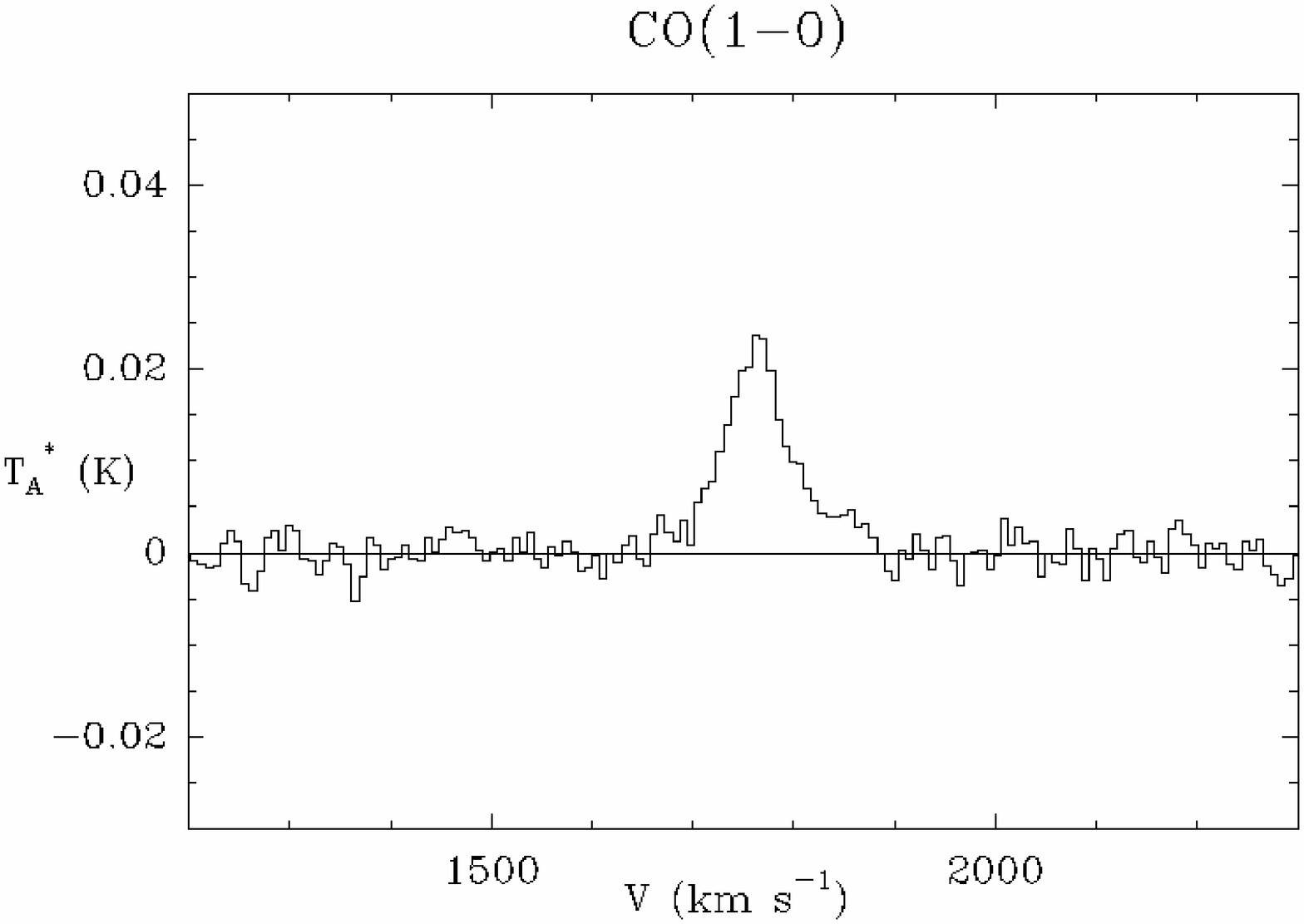}}}
\end{minipage}
\begin{minipage}[t]{8cm}
\resizebox{9cm}{!}{\rotatebox{-90}{\plotone{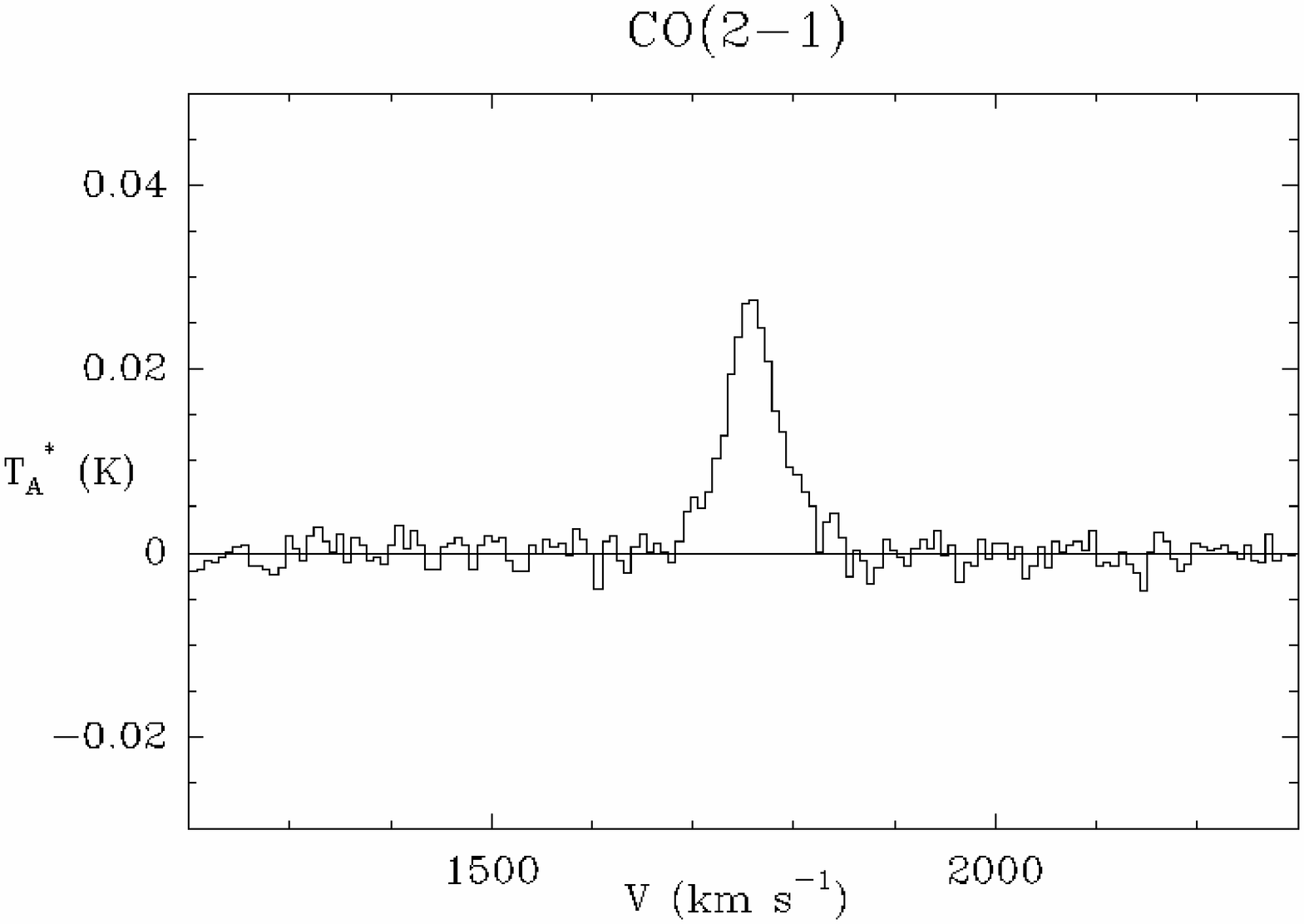}}}
\end{minipage}
\caption{CO(1-0) and CO(2-1) spectra of NGC\,1377 averaged over the five
centermost positions. The channel width was Hanning-smoothed to about
7.3\,km\,s$^{-1}$ for both lines.
\label{co_tot}}
\end{figure}

\end{document}